\documentclass{article}
\usepackage{graphicx,subfigure}
\usepackage{authblk}
\usepackage{tikz}
\usepackage{textcomp}
\usepackage{gensymb}
\usepackage{hyperref}
\usepackage{amsmath,bm}
\usepackage{amssymb}
\usepackage{tablefootnote}
\usepackage{nicefrac}
\usepackage{tabularx}
\usepackage{comment}
\usepackage[section]{placeins}
\usepackage[para]{footmisc}
\usepackage{mathtools}
\usepackage{dsfont}
\usepackage{microtype}
\usepackage{float}
\usepackage{xurl}
\usepackage[title]{appendix}
\usepackage[margin=1in]{geometry}
\usetikzlibrary{positioning, calc}
\usetikzlibrary{shapes.geometric}
\usepackage[backend=biber,style=vancouver]{biblatex}
\begin{document}
\title{\Large Equation Learning for multiscale models of infectious diseases}
\author[1,*]{James W. G. Doran}
\author[1]{Cameron A. Smith}
\author[1]{Christian A. Yates}
\author[1]{Ruth Bowness}
\affil[1]{Department of Mathematical Sciences, University of Bath, Bath, United Kingdom BA2 7AY}
\affil[*]{Corresponding author: James W. G. Doran, jd521@bath.ac.uk}
\maketitle
\begin{abstract}
    Tuberculosis (TB) is an airborne disease caused by the pathogen \textit{Mycobacterium tuberculosis}. In 2023, according to the World Health Organization, it ``probably'' replaced COVID-19 as the leading cause of death from an infectious agent globally; in the nineteenth century, one in seven of all humans deaths were as a result of tuberculosis. More than 10 million people are diagnosed with TB every year. The majority of cases in adults occur in males (62.5\% of all global adult cases in 2023, compared to 37.5\% in females). The main reasons for males suffering from a higher burden of global TB cases, compared to females, is likely to be a combination of within-host factors, such as differences in immune response, and population-scale factors, such as likelihood of completing treatment. To investigate the impact different scales have in determining this higher TB burden in males, we have developed a gender/sex-stratified multiscale framework. We have learnt ordinary differential equations (ODEs) to capture the average output of an agent-based within-host model, and used the resulting equations to describe the within-host scales of the multiscale framework. We evolve the population demographics at the between-host scale using ODEs, and link the scales with stochastic coupling functions. We have considered counterfactual scenarios to elucidate the impact of sex and gender on the infectious disease dynamics of TB. This paper is intended to provide a proof-of-concept for the development and implementation of the presented multiscale framework.
\end{abstract}
\section*{Keywords}
Multiscale model, tuberculosis, equation learning, within-host, between-host
\section{Introduction}
\subsection{Multiscale modelling in infectious diseases}
\begin{figure}
    \centering
    \includegraphics[width=0.9\linewidth]{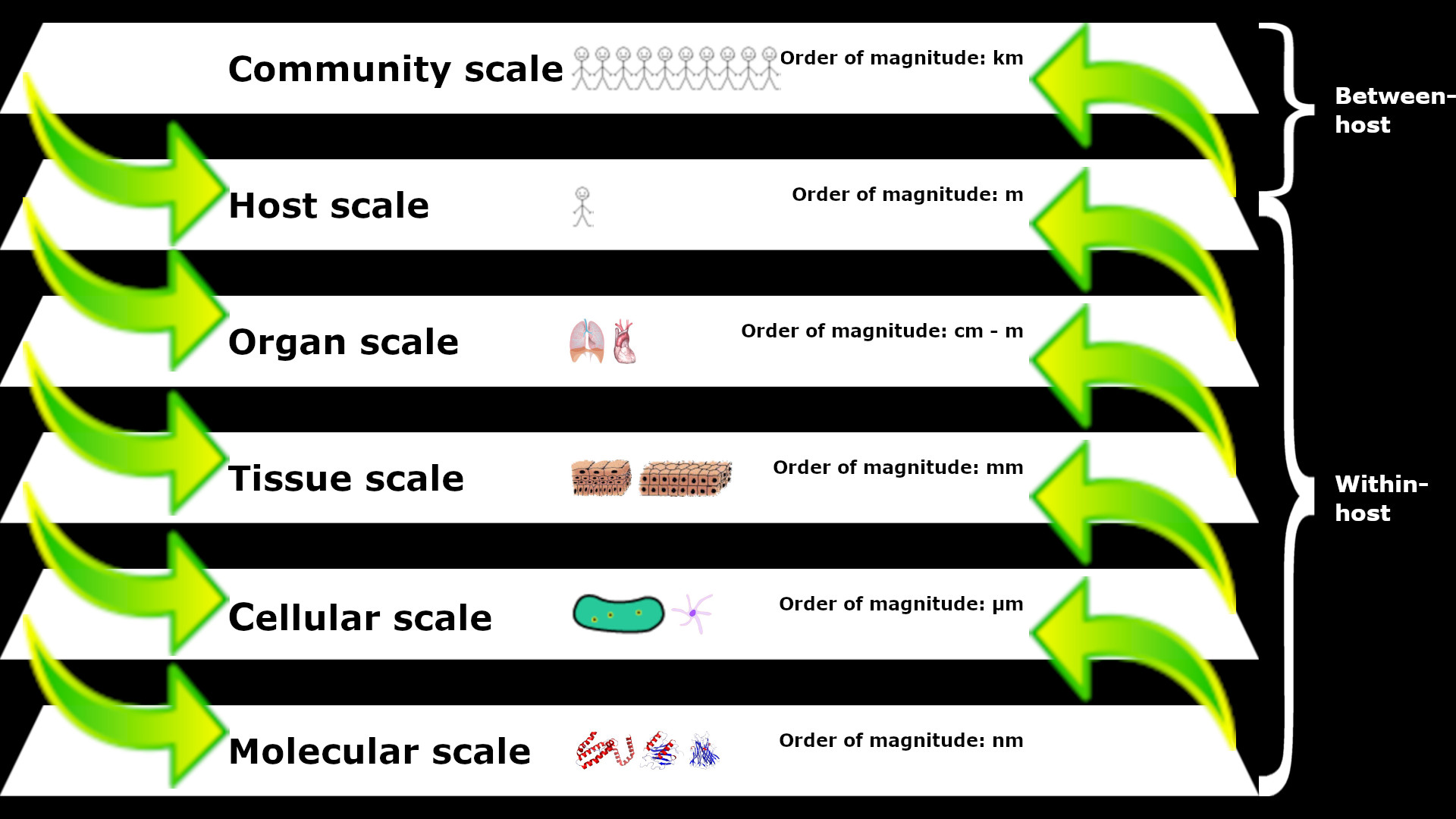}
    \caption{A schematic showing the different organisational levels of infectious disease systems, and their categorisation into the ``within-host'' and ``between-host'' scales. Reprinted from \citetitle{Doran2023} by \citeauthor{Doran2023} \cite{Doran2023} (licensed under \url{https://creativecommons.org/licenses/by/4.0/}).}
    \label{fig:schematic}
\end{figure}

Infectious disease dynamics play out across multiple spatial and temporal scales. Mathematical models can be developed to investigate the impact of different infectious diseases and the effectiveness of public health interventions (e.g. vaccination campaigns) in reducing their impact at each of these different scales. Often, these focus on two main scales: the ``within-host'' scale and the ``between-host'' scale, which themselves can contain multiple scales. Within-host models investigate host-pathogen interactions occurring at the order of nanometres to metres, typically focusing on pathogen replication and treatment, whilst between-host models will explore the dynamics of host-host interactions over several metres to kilometres and are mainly concerned with the transmission of these pathogens \cite{Garira2019} (see Figure \ref{fig:schematic} and Tables \ref{table: within-host} and \ref{table: between-host} for more information). However, it is important to note that these scales are not independent and will have reciprocal feedback on one another \cite{Handel2015}. For example, as pathogens reproduce and pathogen load increases, some of them are expelled into the environment, leading to an increased probability of transmission, while transmission between hosts will lead to infections that allow pathogens to start replicating inside a new host \cite{Garira2019}. As a result, some mathematical models look to combine within-host and between-host dynamics into a single framework that link these scales. We describe these as ``multiscale'' models \cite{Childs2019}. A primer on multiscale models has been published by \citeauthor{Garira2018} \cite{Garira2018}, as well as a comprehensive outline of the methodology to design such models \cite{Garira2020}. A recent review of the current state of this emerging field was published by \citeauthor{Doran2023} in \citeyear{Doran2023} \cite{Doran2023}.
\begin{table}
    \centering
    \resizebox{14.5cm}{!}{
    \begin{tabular}{|l|l|l|}
        \hline
        \textbf{Scale} & \textbf{Entities} & \textbf{Example processes}\\
        \hline
        Molecular & Cytokines, oxygen & Pathogen replication, immune signalling\\
        Cellular & Pathogens, immune cells & Infection of cells, apoptosis, cytokine response\\
        Tissue & Epithelial tissue & Localised inflammation\\
        Organ & Lungs, bladder, liver etc & Organ-specific pathology (e.g. TB granuloma)\\
        Host (individual) & The entire organism & Disease progression, symptom onset\\
        \hline
    \end{tabular}}
    \caption{Terms and processes at the within-host scales. The within-host scales focus on the biological mechanisms inside an individual host, ranging from molecules to entire organs.}
    \label{table: within-host}
\end{table}
\begin{table}
    \centering
    \resizebox{14.5cm}{!}{
    \begin{tabular}{|l|l|l|}
         \hline
         \textbf{Scale} & \textbf{Entities} & \textbf{Example processes}\\
         \hline
         Individual host & Susceptible, infected and recovered individuals & Contact, transmission, recovery\\
         Population & Communities, demographics & Epidemic spread\\
         Metapopulation & Cities, regions & Spatial transmission\\
         \hline
    \end{tabular}}
    \caption{Terms and processes at the between-host scales. The between-host scales capture the spread and dynamics across individuals or populations.}
    \label{table: between-host}
\end{table}
\subsection{Importance of tuberculosis (TB)}
Tuberculosis (TB) is an airborne disease caused by the bacterium \textit{Mycobacterium tuberculosis} (\textit{M. tb}). In 2023, according to the World Health Organization, it ``probably'' regained its position as the leading cause of death from a single infectious agent, after being overtaken by COVID-19 for three years \cite{WHO2024}. That year, TB accounted for nearly twice the number of deaths as HIV/AIDS. Over 10 million people continue to develop TB annually, with case numbers steadily increasing since 2021 \cite{WHO2024}. One challenge that comes with modelling TB is the variability in within-host dynamics between different individuals \cite{Childs2015}. Although approximately $90\pm5\%$ of people exposed to \textit{M. tb} will not develop active TB immediately after exposure \cite{Ahmad2011,Glaziou2018}, a small proportion will become almost instantly infectious. Others who initially developed latent TB will become sick with active TB weeks to years after inhaling the pathogen \cite{Childs2015}. Even after successful treatment with antibiotics, individuals can relapse and start infecting others again months-to-years later \cite{Bowness2018}. Therefore, any epidemiological model of tuberculosis can not simply assume the within-host scale reaches a predictable `steady-state' equilibrium and disregard the dynamics at that scale, as the timing of active disease will vary substantially from person to person. Moreover, it has been observed that males are more likely to develop active TB than females (in 2023, 62.5\% of global TB cases in adults occurred in males, compared to 37.5\% in females \cite{WHO2024}), meaning that a single (albeit stochastic) within-host model does not capture the extrinsic variability in individual disease progression.
\subsubsection{Host-pathogen interactions}
\label{subsubsec:host-pathogen}
The immune response to \textit{M. tb} infection can be split into an initial innate aspect and a subsequent adaptive aspect \cite{OGarra2013}. When the bacteria enter the lung tissue, one of the first groups of immune cells that encounter them are resting macrophages \cite{OGarra2013,Schluger1998}. There is typically a consistent background level of resting macrophages in the lung tissue in anticipation of an invading entity \cite{OGarra2013}. When these resting macrophages discover the \textit{M. tb} bacteria, they initiate the innate immune response by attempting to phagocytose the bacteria and destroy them internally \cite{OGarra2013,Schluger1998}. However, \textit{M. tb} has evolved to withstand the destructive mechanisms of the macrophages, and is able to enter a ``dormant'' state, reducing its rate of replication to increase its chances of survival \cite{Hammond2015,Lipworth2016}\footnote{A number of reasons have been hypothesised for dormancy in \textit{M. tb}, one of which is insufficient oxygen levels to maintain a high replication rate \cite{Lipworth2016}.}. Subsequently, any macrophages that phagocytose \textit{M. tb} become infected \cite{OGarra2013}. At this point, they release a number of chemokines. These signalling chemicals alert the rest of the immune system to the invading pathogen, upregulate the immune response and guide immune cells to the site of infection via chemotaxis \cite{OGarra2013,Schluger1998}. Infected macrophages continue to phagocytose the bacteria they encounter, but once the number of intracellular bacteria they have ingested exceeds a certain threshold, they burst, releasing the \textit{M. tb} bacteria back into the lung tissue and leaving behind necrotic material, also known as caseum \cite{Bowness2018}.\par
As part of the adaptive immune response, dendritic cells travel to the lymph nodes and activate relevant T-cells; these T-cells then head to the part of the lung tissue infected with \textit{M. tb}, following the chemoattractant gradient caused by the diffusing chemokines released by the infected macrophages \cite{OGarra2013}. Once there, T-cells kill infected macrophages, as well as any intracellular bacteria within them, and activate resting macrophages, enabling them to destroy more bacteria without becoming infected \cite{OGarra2013,Schluger1998}. This aggregation of immune cells, bacteria and necrotic material leads to a structure known as a granuloma; infection from \textit{M. tb} typically leads to an average of 10 granulomas \cite{Joslyn2022}.
\label{sec:TB biology}
\subsubsection{Differences seen between sexes/genders in TB}
Both physiological factors, related to sex at the within-host scale, and behavioural factors, related to gender at the between-host scale, have been proposed for why there is a greater TB burden in males \cite{Nhamoyebonde2014}. At the within-host scale, testosterone has been found to negatively impact the immune response against \textit{M. tb} \cite{Gupta2022,Shrivastava2021}. Furthermore, males are more likely than females to develop cavitary TB \cite{Balogun2021}, with cavities forming and causing treatment relapse, higher transmission rates and making the development of drug resistance more likely \cite{Urbanowski2020}. There is also evidence that genetic events linked to the X chromosome reduce the immune response to tuberculosis in males \cite{Nhamoyebonde2014,Gupta2022}, although we have not explicitly considered genetic differences in this paper. At the between-host scale, certain behavioural risk factors that enhance TB transmission have been identified, such as homelessness and substance use, as well as certain occupations being riskier than others, for example working in health care, driving public transport, or working in prisons \cite{Childs2015}. In some countries, these riskier occupations are likely to be male-dominated, which can lead to higher proportions of male TB cases than would otherwise be expected (for example, public transport sector workers in Lima, Peru \cite{HornaCampos2010}). The picture is mixed when it comes to diagnosis delays and adherence to treatment, however: some studies have suggested women may have longer delays \cite{Yang2014,Karim2007} and be more likely to stop their treatment early \cite{Kaona2004}, while others suggest men are more likely to delay seeking treatment and do not adhere to treatment as well as women \cite{Hof2010}. We acknowledge that there may be other studies on diagnosis delays and treatment adherence which we have not considered here that may provide more evidence one way or another.
\subsection{Modelling methods and model structure}
Although there have been attempts to develop multiscale models of TB \cite{Pereira2021}, these have not addressed the differing immune responses between the sexes at the within-host scale, nor gender differences at the between-host scale. Other models have investigated the impact of gender on the epidemiology of TB (e.g. \cite{KisselevskayaBabinina2018,Kubjane2023,Wang2024}), but have not included the within-host dynamics as part of their modelling framework. The within-host representation of any such multiscale model would need to consider the impact of the phenotypes and spatial locations of \textit{M. tb} within lung tissue, as both have been found to be important factors in determining whether an individual develops latent or active TB, or goes on to relapse \cite{Bowness2018}.\par
An agent-based model (ABM), such as the one presented by \citeauthor{Bowness2018} \cite{Bowness2018}, would be a suitable choice for modelling the within-tissue scale of a single infected host; this could either be used once to model the formation of a single representative granuloma or used repeatedly to generate multiple granulomas, as part of a within-host model. However, agent-based models are computationally expensive, so simulating a population in which every infected individual has their own within-host dynamics determined by one or more ABMs would be unfeasible. One alternative approach is to ``learn'' ordinary differential equations (ODEs) which capture the average behaviour of a sufficiently large number of simulations of an ABM accurately whilst remaining computationally efficient \cite{Nardini2021}. It should be noted that this approach sacrifices between-individual variability for computational efficiency. Having said this, individuals in such a model could be initialised with a randomly distributed number of granulomas, allowing some between-individual variability. This will lead to different pathogen loads between individuals whose active infections began at the same time but with differing degrees of severity. Every infected individual can be assigned the same set of learned ODEs to govern their within-host dynamics, with different parameters depending on their sex\footnote{For the purposes of this model, we make a simplifying assumption by considering only two biological sexes (male and female) and two corresponding gender identities (man/boy and woman/girl). We further assume that each individual's gender identity aligns with their assigned biological sex, that is, individuals identifying as men/boys are assigned male at birth, and individuals identifying as women/girls are assigned female at birth.}. The within-host scale can then be linked to the between-host scale to create a multiscale model for TB that can more effectively capture how the mechanisms important at the within-host scale can influence population-level outcomes.\par
In this paper, we present a gender-stratified multiscale model of TB dynamics, adapting the framework developed by \citeauthor{Smith2025} \cite{Smith2025}. Every person who develops active TB is individually tracked, and their within-host dynamics (WHD) models at the within-host scale are stochastically linked to continuous densities of individuals in a compartmental between-host model, also governed by ODEs. The between-host model considers the evolution of the host demographic dynamics (HDD); the disease-associated events are added through coupling functions connecting the within-host and between-host scales. We learnt the average behaviour of an updated version of the within-host ABM presented by \citeauthor{Bowness2018} \cite{Bowness2018} (available at \url{https://github.com/Ruth-Bowness-Group/Equation-Learning-for-multiscale-models-of-infectious-diseases-WHIDM} and described in Section \ref{subsec:WHIDM behaviour}), with different sets of parameters for the two sexes (and a third ``neutral'' set, which did not take sex into account, that was used in some counterfactual scenarios), and embedded the sets of learned within-host ODEs into our ``WHD-HDD'' multiscale framework to simulate a TB epidemic (hereafter referred to as Scenario 0). The results of Scenario 0 are contrasted with three counterfactual scenarios, outlined in Figure \ref{fig:counterfactuals}. In Scenario 1, differences between genders at the between-host scale are ignored: epidemiological parameters are set equal for males and females, while within-host differences between the sexes are maintained. In Scenario 2, the differences between the sexes at the within-host scale are ignored: new learned ODEs are derived using a third averaged set of parameters for the within-host scale and applied to both male and female TB cases while differences at the between-host scale between genders are maintained. In Scenario 3, differences at both scales are ignored. In this way, we aim to improve our understanding of the impacts of the within-host and between-host scales, both reciprocally on each other and on the overall disease dynamics. A schematic summarising the construction of the multiscale modelling framework is shown in Figure \ref{fig:multiscale schematic}.
\begin{figure}
    \centering
    \begin{tikzpicture}
        \node[draw,diamond] (check 1) at (0,6){Within-host differences?};
        \node[draw,diamond] (check 2a) at (-4,0){Between-host differences?};
        \node[draw,ellipse] (scenario 0) at (-6,-6){Scenario 0};
        \node[draw,ellipse] (scenario 1) at (-2,-6){Scenario 1};
        \node[draw,diamond] (check 2b) at (4,0){Between-host differences?};
        \node[draw,ellipse] (scenario 2) at (2,-6){Scenario 2};
        \node[draw,ellipse] (scenario 3) at (6,-6){Scenario 3};
        \draw[-stealth] (check 1.south) -- node[midway, above]{Yes} (check 2a.north);
        \draw[-stealth] (check 2a.south) -- node[midway, left]{Yes} (scenario 0.north);
        \draw[-stealth] (check 2a.south) -- node[midway, right]{No} (scenario 1.north);
        \draw[-stealth] (check 1.south) -- node[midway, above]{No} (check 2b.north);
        \draw[-stealth] (check 2b.south) -- node[midway, left]{Yes} (scenario 2.north);
        \draw[-stealth] (check 2b.south) -- node[midway, right]{No} (scenario 3.north);
    \end{tikzpicture}
    \caption{Summary of the ground truth (Scenario 0) and three counterfactual scenarios (Scenarios 1 to 3) considered in this paper. Within-host differences are between sexes and between-host differences are between genders.}
    \label{fig:counterfactuals}
\end{figure}
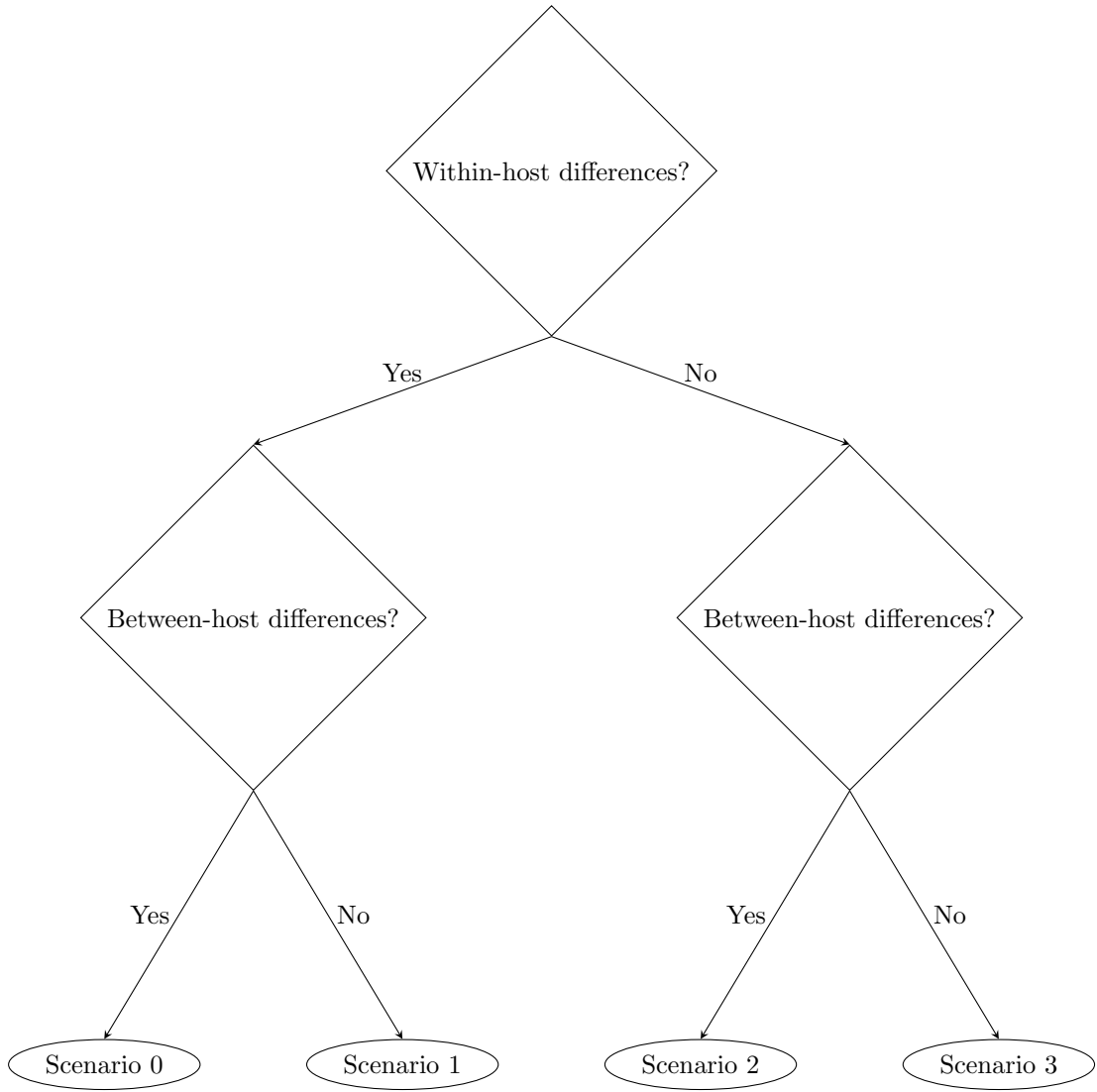
\begin{figure}
    \centering
    \begin{tikzpicture}
        \node[draw, text width=15.5cm, align=center] (step 1) at (0,4){1. Run simulations on agent-based model and generate within-host scale output (\S \ref{subsubsec:WHIDM output})};
        \node[draw, text width=15.5cm, align=center] (step 2) at (0,3){2. Approximate derivatives with respect to time of the variables in the ABM to be learnt (\S \ref{subsubsec:derivative estimation})};
        \node[draw, text width=15.5cm, align=center] (step 3) at (0,2){3. Construct library of possible terms for learnt equations (\S \ref{subsubsec:library})};
        \node[draw, text width=15.5cm, align=center] (step 4) at (0,1){4. Learn equations that align with agent-based model time derivatives from possible terms (\S \ref{subsubsec:equation inference})};
        \node[draw, text width=15.5cm, align=center] (step 5) at (0,0){5. Add learnt ODEs to adapted WHD-HDD framework to represent within-host dynamics (\S \ref{subsubsec:learnt ODEs})};
        \node[draw, text width=15.5cm, align=center] (step 6) at (0,-1){6. Represent host demographics dynamics (births and natural deaths) with ODEs (\S \ref{subsubsec:compartmental model})};
        \node[draw, text width=15.5cm, align=center] (step 7) at (0,-2){7. Link scales with coupling functions (\S \ref{subsubsec:coupling})};
        \draw[-stealth] (step 1.south) -- (step 2.north);
        \draw[-stealth] (step 2.south) -- (step 3.north);
        \draw[-stealth] (step 3.south) -- (step 4.north);
        \draw[-stealth] (step 4.south) -- (step 5.north);
        \draw[-stealth] (step 5.south) -- (step 6.north);
        \draw[-stealth] (step 6.south) -- (step 7.north);        
    \end{tikzpicture}
    \caption{Schematic outlining the model construction. Steps 1 to 4 represent the equation learning pipeline outlined by \citeauthor{Nardini2021} \cite{Nardini2021} and are described in detail in Sections \ref{subsubsec:WHIDM output} to \ref{subsubsec:equation inference}; steps 5 to 7 represent the WHD-HDD framework introduced by \citeauthor{Smith2025} \cite{Smith2025} and are described in detail in Sections \ref{subsubsec:learnt ODEs} to \ref{subsubsec:coupling}.}
    \label{fig:multiscale schematic}
\end{figure}
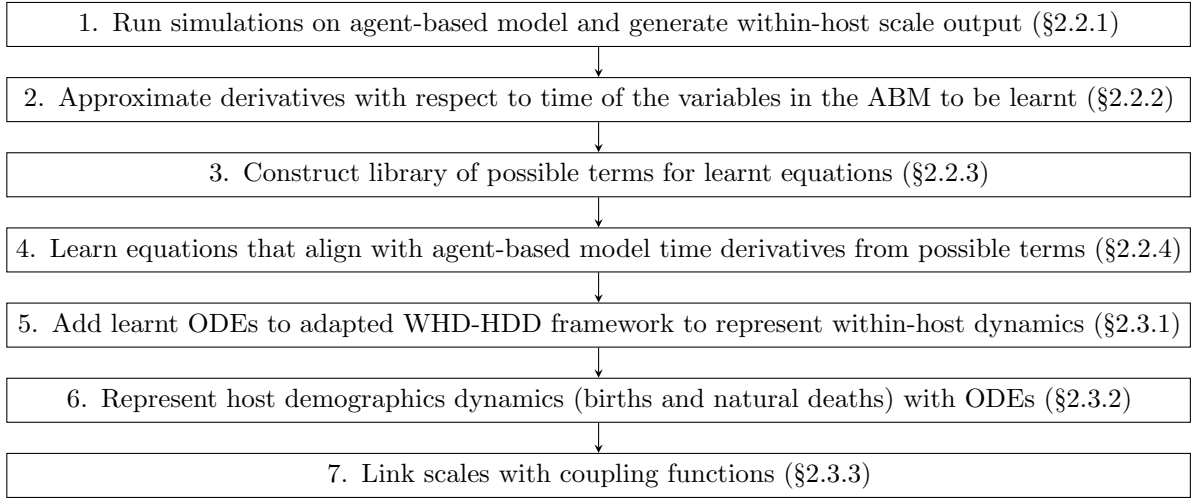
\subsection{Outline for the rest of the paper}
The rest of this paper is structured as follows. In Section \ref{sec:methodology}, we outline the framework of our multiscale model: in Section \ref{subsec:WHIDM behaviour}, we describe the behaviour of the Within-Host Infectious Disease Model (hereafter referred to as WHIDM), the within-host ABM developed by \citeauthor{Bowness2018} that we use in this paper; in Section \ref{subsec:EQL}, we explain how the within-host scale ODEs were learned based on the interactions described in Section \ref{subsubsec:host-pathogen} and the behaviour of WHIDM; in Section \ref{subsec:multiscale model}, we present the host demographic dynamics, the ODEs for evolving the compartments at the between-host scale, and how we bidirectionally linked the within-host dynamics to the host demographic dynamics. In Section \ref{sec:results}, we present the results of our investigations: the evolution of the learned within-host ODEs and the changes over time in the between-host compartments in Scenario 0, and the three counterfactual scenarios. In Section \ref{sec:discussion}, we discuss the implications of our findings, as well as some ideas for future work worth considering as a consequence of this study.
\section{Methods}
In this section, we first discuss the behaviour of WHIDM before outlining how the equation learning algorithm works in deriving the final ODEs. After that, we present the library of possible terms to be included in the ODEs, based on the behaviour of WHIDM, that should govern the within-host dynamics of the multiscale model; how the host demographic dynamics evolve; and finally, the transmission function which links the scales, and the probabilities determining transitions between infectious disease states.
\label{sec:methodology}
\subsection{Behaviour of Within-Host Infectious Disease Model (WHIDM)}
\label{subsec:WHIDM behaviour}
The original version of WHIDM framework was first published by \citeauthor{Bowness2018} \cite{Bowness2018} to capture the evolution of a single TB granuloma; here, we are using the latest iteration of the model, which has been optimised for improved performance but maintains the same model mechanisms (available at \url{https://github.com/Ruth-Bowness-Group/Equation-Learning-for-multiscale-models-of-infectious-diseases-WHIDM}). WHIDM is also used and described by \citeauthor{Doran2026a} \cite{Doran2026a}. WHIDM is an agent-based model, in which fast-growing and slow-growing \textit{M. tb} phenotypes exist as agents within a $2\text{mm} \times 2\text{mm}$ cross-section of lung tissue. The lung tissue is represented in the model as a square lattice, of size $100 \times 100$ (so each lattice site has length and width $0.02\text{mm}$). Immune cells - specifically, macrophages and T-cells - are also present as agents in the model. The movements and interactions of all agents are influenced by chemical fields, specifically oxygen (which is supplied via blood vessels) and chemokines (which are supplied by certain immune cells); these evolve according to partial differential equations (PDEs). It is worth noting here that this model is a simplified version of the host-pathogen interactions described in Section \ref{subsubsec:host-pathogen}. More detail about the rules governing the dynamics of this model are provided in the following section.\par
To incorporate the fact that \textit{M. tb} can enter a dormant state and replicate more slowly due to insufficient oxygen levels \cite{Lipworth2016}, we have considered two phenotypes of extracellular bacteria: fast-growing and slow-growing. Fast-growing and slow-growing extracellular bacteria replicate at different rates ($Rep_f$ and $Rep_s$, respectively), and the extracellular bacteria switch between phenotypes depending on the local oxygen level in the lung tissue. The rate of replication is then impacted by the remaining available space within the lung tissue. We have assumed fast-growing extracellular \textit{M. tb} reduce their rate of replication (i.e. switch to a different phenotypic state) when local oxygen levels are below a given threshold, hereafter referred to as $O_{low}$, and that slow-growing extracellular \textit{M. tb} switch to a different phenotypic state with a greater replication rate when local oxygen levels exceed another given threshold, hereafter referred to as $O_{high}$.\par
Macrophages are modelled as being in one of four states: resting, activated, infected, and chronically infected. Macrophages initially begin simulations in the resting state; once they phagocytose any extracellular bacteria, they become infected, and the extracellular bacteria they ingested become intracellular. Once an infected macrophage has phagocytosed a certain quantity of \textit{M. tb}, this threshold hereafter being referred to as $M_{ici}$, they become chronically infected. Once a chronically infected macrophage has phagocytosed another given quantity of \textit{M. tb}, hereafter referred to as $M_{cib}$, they burst and become caseum (necrotic material). The intracellular bacteria within these bursting macrophages become extracellular, with the slow-growing phenotype; the proportion of the total intracellular bacterial load that survives the macrophage bursting and becomes extracellular depends on the remaining available space within the lung tissue. Macrophages in all four states phagocytose extracellular bacteria at rate $M_{phag}$. Resting macrophages can be recruited to the site of infection via blood vessels at rate $M_{recr}$, which is then modified by the amount of remaining available space within the lung tissue. Resting, infected and chronically infected macrophages die at a rate inversely proportional to their expected lifespan, $M_{life}$. When infected and chronically infected macrophages die, they leave caseum and release their intracellular bacterial load; as with bursting, the proportion of this intracellular bacterial load that survives and becomes extracellular depends on the remaining available space.\par
T-cells are recruited to the site of infection after a minimum period of time has elapsed in the simulation, hereafter referred to as $T_{enter}$, at rate $T_{recr}$. $T_{enter}$ was set to equal 9 days: experimental data suggests that T-cell responses to \textit{M. tb} infection do not start until at least 9 days after initial infection, as it takes this long for live bacteria to be transported to the lung-draining lymph node to begin T-cell priming \cite{Urdahl2011}. This recruitment rate is modified depending on the remaining available space within the lung tissue. Resting macrophages become activated in the presence of T-cells at rate $M_{ra}$. After this occurs, any extracellular bacteria they phagocytose is destroyed. Infected and chronically infected macrophages and any intracellular bacteria within them are killed by T-cells at rate $T_{kill}$. Activated macrophages die at a rate inversely proportional to their expected lifespan, $M_{alife}$. Similarly, T-cells die at a rate inversely proportional to their expected lifespan, $T_{life}$.\par
\subsection{Equation learning algorithm}
\label{subsec:EQL}
We decided to use equation learning to capture the dynamics of WHIDM, rather than derive mean-field differential equations. As stated by \citeauthor{Nardini2021} \cite{Nardini2021}, mean-field differential equations can accurately describe the output of agent-based models when the mean-field assumption (that is, that the occupancies of neighbouring lattice sites are independent) is met. However, it is clear that this assumption will not hold when simulating the within-host dynamics of \textit{M. tb} infection using WHIDM. For example, the initial \textit{M. tb} bacteria are placed in a small cluster uniformly at random within the domain and do not move throughout the simulation, so if a bacterium is found at one lattice site, it is highly probable that another bacterium will be occupying a neighbouring lattice site. For such agent-based models, equation learning can lead to a much more accurate representation of the model output than can be achieved using mean-field differential equations, as shown by \citeauthor{Nardini2021} \cite{Nardini2021}. Furthermore, using learnt ODEs can allow us to predict the emergent behaviour of the model if we perturb some of the parameters in counterfactual scenarios, without having to run more computationally expensive simulations of WHIDM; although we have not done this here, this would be a benefit to using this multiscale framework in future work. We followed the methodology outlined by \citeauthor{Nardini2021} \cite{Nardini2021} to learn the equations governing the within-host dynamics of the multiscale model, albeit with a few differences; we summarise our revised protocol in this section.\par
As shown in steps 1 to 4 in Figure \ref{fig:multiscale schematic}, the equation learning algorithm can be summarised in four steps:
\begin{enumerate}
    \item Generate averaged agent-based model output;
    \item Estimate derivates of the output with respect to time;
    \item Construct a library of possible terms that could be included in the learnt equations;
    \item Perform regression analysis to infer the form that the learnt equations should take \cite{Nardini2021}.
\end{enumerate}
We discuss each of these four steps in turn in Sections \ref{subsubsec:WHIDM output} to \ref{subsubsec:equation inference}.
\subsubsection{WHIDM output}
\label{subsubsec:WHIDM output}
The first step in developing the multiscale model is to generate the averaged WHIDM output. In order to determine how many simulations are required to control for intrinsic noise, we have previously conducted a consistency analysis, using the methodology given by \citeauthor{Hamis2021} in \cite{Hamis2021}. From this, we determined that 300 simulations of WHIDM would be sufficient to give us consistent averages \cite{Doran2026a}. We simulated the model this number of times to get output density vectors, one per simulation for each output of interest in WHIDM (the numerical solutions to Equations (\ref{eq:FGEB}) to (\ref{eq:T}) in Section \ref{subsubsec:learnt ODEs}). Here, the density is equal to the number of cells of interest divided by the total number of lattice sites. From here, we sampled the output at $n$ different time points $t_i = (i-1)\Delta t, i \in \{1,...,n\}$ (where $\Delta t$ is the model time step) and averaged the outputs at each time point over the total number of simulations to arrive at average output density vectors. For example, $\bm{F}^{(x)}(t)$ would be the $n \times 1$ output density vector for fast-growing extracellular bacteria (that is, the number of fast-growing extracellular bacteria divided by $100^2$) in the $x^{\text{th}}$ simulation of 300; and $\bm{F_d}(t) = \frac{1}{300} \sum_{x=1}^{300} \bm{F}^{(x)}(t)$ would be the $n \times 1$ average output density vector for fast-growing extracellular bacteria.\par
\subsubsection{Derivative estimation}
\label{subsubsec:derivative estimation}
The second step in developing the multiscale model is to approximate the temporal derivatives of the quantities of interest. We did this by computing finite differences, specifically the forward difference for the first time point $t_1$, the backward difference for the final time point $t_n$, and central differences at all other time points. For example, for the average output density vector for fast-growing extracellular bacteria $\bm{F_d}$, sampled with time interval $\Delta t$, the approximation of the derivative of $\bm{F_d}$ at time point $t_i, i \in \{1,...,n\}$ was calculated as follows (equivalent calculations were made for the other outputs of interest):
\begin{align}
    \frac{d\bm{F_d}(t_1)}{dt} &\approx \frac{\bm{F_d}(t_2) - \bm{F_d}(t_1)}{\Delta t};\\
    \frac{d\bm{F_d}(t_i)}{dt} &\approx \frac{\bm{F_d}(t_{i+1}) - \bm{F_d}(t_{i-1})}{2\Delta t},\quad \text{for }i \in \{2,...,n-1\};\\
    \frac{d\bm{F_d}(t_n)}{dt} &\approx \frac{\bm{F_d}(t_n) - \bm{F_d}(t_{n-1})}{\Delta t}.
\end{align}
\subsubsection{Library construction}
\label{subsubsec:library}
The third step in developing the multiscale model is to construct a library of the possible terms we would expect to find in the learnt ODE for any one of the possible quantities of interest (see Section \ref{subsubsec:learnt ODEs} for the library of possible terms and the learnt ODEs). After doing this, we constructed a $n \times k$ matrix $\Theta$ for each variable, with each of the $k$ columns containing the values (averaged over the simulations) of a given term that could be in the learnt equation (that is, the terms we considered, as per Equations (\ref{eq:FGEB}) to (\ref{eq:T}) in Section \ref{subsubsec:learnt ODEs}) across all time points, and each of the $n$ rows containing the values (averaged over the simulations) of all the terms at a given time point. Specifically, $\Theta$ would be of the form:
\[
\Theta =
\begin{bmatrix}
    \Theta_1(t_1) & \Theta_2(t_1) & \cdots & \Theta_k(t_1)\\
    \Theta_1(t_2) & \Theta_2(t_2) & \cdots & \Theta_k(t_2)\\
    \vdots & \vdots & \ddots & \vdots\\
    \Theta_1(t_n) & \Theta_2(t_n) & \cdots & \Theta_k(t_n)
\end{bmatrix},
\]
where $t_i, i \in \{1,...,n\}$ is the $i^\text{th}$ time point and  $\Theta_j, j \in \{1,...,k\}$ is the $j^\text{th}$ possible term in the learnt equation. As we considered a different number of terms across Equations (\ref{eq:FGEB}) to (\ref{eq:T}), the value of $k$ will vary for each $\Theta$. The parameters to be learnt were stored in a column vector $\bm{\xi}$, with entry $k$ of this vector corresponding to the coefficient of the term $\Theta_k$; one such $\bm{\xi}$ was generated for each equation to be learnt. For example, the equation to be learnt for estimating the derivative of the average density of fast-growing extracellular bacteria can be summarised as
\begin{equation}
\label{eq:FGEB derivative}
    \frac{d\bm{F_d}}{dt} = \Theta \bm{\xi},
\end{equation}
where
\[
\Theta = 
\begin{bmatrix}
    \bm{F}(t_1)(\bm{M_R}(t_1)+\bm{M_I}(t_1)+\bm{M_{CI}}(t_1)+\bm{M_A}(t_1)) & \bm{F}(t_1) & \bm{F}(t_1)\bm{N}(t_1)\\
    \bm{F}(t_2)(\bm{M_R}(t_2)+\bm{M_I}(t_2)+\bm{M_{CI}}(t_2)+\bm{M_A}(t_2)) & \bm{F}(t_2) & \bm{F}(t_2)\bm{N}(t_2)\\
    \vdots & \vdots & \vdots\\
    \bm{F}(t_n)(\bm{M_R}(t_n)+\bm{M_I}(t_n)+\bm{M_{CI}}(t_n)+\bm{M_A}(t_n)) & \bm{F}(t_n) & \bm{F}(t_n)\bm{N}(t_n)\\
\end{bmatrix}
\]
is an $n \times 3$ matrix and $\bm{\xi} = [\xi_1,\xi_2,\xi_3]^T$.\par
\subsubsection{Equation inference}
\label{subsubsec:equation inference}
The fourth step in developing the multiscale model is to infer the equations. To do this, we used linear regression to find the coefficients in $\bm{\xi}$. Initially, we used the \texttt{lstsq} method in Python's \texttt{numpy.linalg} library (as per the tutorial provided by \citeauthor{Nardini2021} for their algorithm \cite{Nardini2021} - for more information on this method, see \url{https://numpy.org/doc/stable/reference/generated/numpy.linalg.lstsq.html} and Appendix \ref{appendix:regression comparison}). However, this method may lead to overfitting - some of the possible terms that would be logical to include in one differential equation, based on the processes within WHIDM, would not be logical to include in other equations - so we subsequently tried using the Lasso (least absolute shrinkage and selection operator) algorithm from Python's \texttt{sklearn.linear\_model} library (for more information on this method, see \url{https://scikit-learn.org/stable/modules/generated/sklearn.linear_model.Lasso.html} and Appendix \ref{appendix:regression comparison}). By using this method, the accuracy of the learnt equations can be maintained without overfitting, as the sum of the coefficients is forcibly kept small in size: in choosing the value of $\lambda$, the regularisation parameter, we are determining how large a value we will accept for $\|\bm{\xi}\|_1$, that is, the sum of the absolute values of the terms in $\bm{\xi}$. By setting the weighting to be large, we are putting more emphasis on minimising $\|\bm{\xi}\|_1$. Thus, the coefficients of some terms are determined to be zero by the method in order to achieve this.\par
The two methods of regression discussed above were used in examples in the original article presenting equation learning of agent-based model output \cite{Nardini2021}. However, using either of these methods, we could not guarantee that conservation of mass would be incorporated into the equations. Certain coefficients should have been of the same magnitude with opposite signs, to capture one cell type transitioning into another (for example, a resting macrophage transitioning into an infected macrophage). In addition, the signs of the coefficients did not match what would be expected based on the model mechanics: certain terms were expected to have positive coefficients and others to have negative coefficients. As such, we required a method of linear regression that would enforce conservation of mass as closely as possible and give the expected signs of the coefficients correctly. We used the \texttt{lsq\_linear} algorithm from Python's \texttt{scipy.optimize} library (for more information on this method, see \url{https://docs.scipy.org/doc/scipy/reference/generated/scipy.optimize.lsq_linear.html} and Appendix \ref{appendix:regression comparison}) to solve the linear regression with upper and lower bounds imposed on the coefficients of the variables.\par
This regression method, with upper and lower bounds, was not used in any of the examples presented by \citeauthor{Nardini2021} \cite{Nardini2021}. However, the examples demonstrating the algorithm's use in that paper did not use agent-based models as complex as the agent-based model for which we are learning equations in this paper. One agent-based model was a birth-death process with migration for a single cell type; the other agent-based model was a standard susceptible-infected-recovered epidemiological compartmental model with a fixed quantity of individuals in the population \cite{Nardini2021}. By comparison, our agent-based model considers seven cell types, none of which have a fixed quantity, with replications, deaths, migration and more interactions between different cell types. Therefore, it is perhaps not surprising that the two less constrained regression methods of standard least squares and Lasso regression proved unsuitable for fitting equations to our more complex model. Our decision to consider least squares regression with upper and lower bounds is validated by the improvement of the fit of the equations to the data. Using \texttt{lsq\_linear} led to the best fitting equations, in terms of minimising the mean squared error of most variables (see Appendix \ref{appendix:regression comparison} for more details).
\subsection{WHD-HDD framework}
\label{subsec:multiscale model}
The multiscale modelling framework we use in this paper is an adaptation of the framework first presented by \citeauthor{Smith2025} \cite{Smith2025}. This framework was developed to allow efficient computation of multiscale mathematical models of epidemics, by stochastically coupling the within-host dynamics of each infected individual to continuous state variables at the between-host scale \cite{Smith2025}. We have adapted this framework to make it more suitable to model TB infectious disease dynamics. Specifically, we have increased the number of possible infection states of individuals; added gender, cavitation and number of granulomas as properties of individuals; and included more coupling functions to link the two scales to allow for more types of transition between infection states. We discuss the within-host dynamics of our adapted WHD-HDD framework (step 5 in Figure \ref{fig:multiscale schematic}) in Section \ref{subsubsec:learnt ODEs}, the host demographics dynamics (step 6 in Figure \ref{fig:multiscale schematic}) in Section \ref{subsubsec:compartmental model}, and the coupling functions (step 7 in Figure \ref{fig:multiscale schematic}) in Section \ref{subsubsec:coupling}.\par
\subsubsection{Within-host dynamics (WHD)}
\label{subsubsec:learnt ODEs}
The fifth step in developing the multiscale model is to add the learnt ODEs to the WHD-HDD framework to represent the within-host dynamics. Our construction of the library of possible terms to be included in the ODEs was based on the biology outlined in the introduction and the behaviour of WHIDM discussed in the Section \ref{subsec:WHIDM behaviour}. The equations derived by the equation learning algorithm govern the evolution of the immune cells and the extracellular \textit{M. tb} over time. Specifically, the ODEs control the following quantities: fast-growing extracellular \textit{M. tb} bacteria; slow-growing extracellular \textit{M. tb} bacteria; resting macrophages; active macrophages; infected macrophages; chronically infected macrophages; T-cells. In addition, as in WHIDM, we also include blood vessels in our model, as fixed constants in the ODEs for any terms involving recruitment of immune cells from blood vessels (as blood vessels are a fixed number throughout the simulation).\par
Each of the following equations expresses the evolution of the densities of the associated cell types. In the following equations, $N$ is the total density of occupied space in the lung tissue, and $N_{V}$ is the density of blood vessels in the lung tissue. A summary of the variables in the learnt equations, based on quantities from the WHIDM framework, can be found in Table \ref{tab:EQL variables}. A summary of the parameter values used in the learnt equations, based on interactions and behaviours from the WHIDM model, can be found in Table \ref{tab:EQL params}. The names of variables and parameters are based on the naming conventions used in \cite{Bowness2018}.\par
\begin{table}
    \centering
    \begin{tabular}{|l|l|}
        \hline
        \textbf{Variable} & \textbf{Description}\\
        \hline
        $F$ & Fast-growing extracellular \textit{M. tb}\\
        $S$ & Slow-growing extracellular \textit{M. tb}\\
        $M_R$ & Resting macrophages\\
        $M_I$ & Infected macrophages\\
        $M_{CI}$ & Chronically infected macrophages\\
        $M_A$ & Activated macrophages\\
        $T$ & T-cells\\
        $N_V$ & Blood vessels\\
        $N$ & Total occupied space\\
        \hline
    \end{tabular}
    \caption{Variables from the learnt equations used in the within-host dynamics, based on quantities from the WHIDM framework.}
    \label{tab:EQL variables}
\end{table}
\begin{table}
    \centering
    \resizebox{15.5cm}{!}{
    \begin{tabular}{|l|l|l|}
        \hline
        \textbf{EQL parameter} & \textbf{Related WHIDM parameter(s)} & \textbf{Description of the EQL parameter}\\
        \hline
        $\xi_1$ & $M_{phag}$ & Macrophage phagocytosis rate in (\ref{eq:FGEB})\\
        $\xi_2$ & $Rep_f$ & Fast-growing extracellular \textit{M. tb} replication rate in (\ref{eq:FGEB})\\
        $\xi_3$ & None & Clustering effect on fast-growing extracellular bacteria replication in (\ref{eq:FGEB})\\
        $\xi_4$ & $M_{phag}$ & Macrophage phagocytosis rate in (\ref{eq:SGEB})\\
        $\xi_5$ & $Rep_s$ & Slow-growing extracellular \textit{M. tb} replication rate in (\ref{eq:SGEB})\\
        $\xi_6$ & None & Clustering effect on slow-growing extracellular bacteria replication in (\ref{eq:SGEB})\\
        $\xi_7$ & $M_{life}$ & Transition rate from $B_I$ to $S$ after $M_I$ death in (\ref{eq:SGEB})\\
        $\xi_8$ & $M_{life}$/$M_{phag}$ & Transition rate from $B_I$ to $S$ after $M_{CI}$ death/bursting in (\ref{eq:SGEB})\\
        $\xi_9$ & $M_{recr}$ & Recruitment rate of resting macrophages from blood vessels in (\ref{eq:MR})\\
        $\xi_{10}$ & $M_{phag}$ & Macrophage transition from resting to infected in (\ref{eq:MR})\\
        $\xi_{11}$ & $M_{life}$ & Resting macrophage death rate in (\ref{eq:MR})\\
        $\xi_{12}$ & $M_{ra}$ & Macrophage transition rate from resting to activated in (\ref{eq:MR})\\
        $\xi_{13}$ & $M_{phag}$ & Macrophage transition from resting to infected in (\ref{eq:MI})\\
        $\xi_{14}$ & $M_{phag}$ & Transition rate from $M_I$ to $M_{CI}$ due to phagocytosis in (\ref{eq:MI})\\
        $\xi_{15}$ & $M_{life}$ & Infected macrophage natural death rate in (\ref{eq:MI})\\
        $\xi_{16}$ & $T_{kill}$ & Death rate of infected macrophages due to T-cells in (\ref{eq:MI})\\
        $\xi_{17}$ & $M_{phag}$ & Transition from $M_I$ to $M_{CI}$ due to phagocytosis in (\ref{eq:MCI})\\
        $\xi_{18}$ & $M_{phag}$/$M_{life}$ & Death rate of $M_{CI}$ due to bursting or naturally dying in (\ref{eq:MCI})\\
        $\xi_{19}$ & $T_{kill}$ & Death rate of $M_{CI}$ due to T-cells in (\ref{eq:MCI})\\
        $\xi_{20}$ & $M_{ra}$ & Macrophage transition rate from resting to activated in (\ref{eq:MA})\\
        $\xi_{21}$ & $M_{alife}$ & Activated macrophage death rate in (\ref{eq:MA})\\
        $\xi_{22}$ & $T_{recr}$ & Recruitment rate of T-cells from blood vessels in (\ref{eq:T})\\        
        $\xi_{23}$ & $T_{life}$ & Death rate of T-cells in (\ref{eq:T})\\        
        \hline
    \end{tabular}}
    \caption{Parameters in the learnt equations used in the within-host dynamics, derived from interactions and behaviours from the WHIDM framework. For descriptions of the variables, see Table \ref{tab:EQL variables}. Abbreviations: EQL, equation learning; $B_I$, intracellular bacteria; $S$, slow-growing extracellular bacteria; $M_I$, infected macrophage; $M_{CI}$, chronically infected macrophage.}
    \label{tab:EQL params}
\end{table}
Note that we are learning the equations by inferring the values of the parameters $\xi_k,k\in\{1,...,23\}$ that lead to equations that provide the best estimation of the average WHIDM output, as determined by the optimisation function being used. The ODEs to be learnt that govern the evolution of the within-host dynamics are as follows.\par
\newpage
The density of the fast-growing extracellular bacteria, $F$, evolved over time according to the following equation:
\begin{equation}
\label{eq:FGEB}
    \frac{dF}{dt} = \xi_1F(M_{R} + M_{I} + M_{CI} + M_{A}) + \xi_2F + \xi_3FN.
\end{equation}
The density of the slow-growing extracellular bacteria, $S$, evolved over time according to the following equation:
\begin{equation}
\label{eq:SGEB}
    \begin{aligned}
    \frac{dS}{dt} = &\xi_4S(M_{R} + M_{I} + M_{CI} + M_{A}) + \xi_5S + \xi_6SN\\
    &+ \xi_7M_I(1-N) + \xi_8M_{CI}(1-N).
    \end{aligned}
\end{equation}
The density of the resting macrophages, $M_R$, evolved over time according to the following equation:
\begin{equation}
\label{eq:MR}
    \frac{dM_{R}}{dt} = \xi_9N_{V}(1 - N) + \xi_{10}M_{R}(F + S) + \xi_{11}M_{R} + \xi_{12}M_{R}T.
\end{equation}
The density of the infected macrophages, $M_I$, evolved over time according to the following equation:
\begin{equation}
\label{eq:MI}
    \frac{dM_I}{dt} = \xi_{13}M_{R}(F+S) + \xi_{14}M_{I}(F+S) + \xi_{15}M_{I} + \xi_{16}TM_{I}.
\end{equation}
The density of the chronically infected macrophages, $M_{CI}$, evolved over time according to the following equation:
\begin{equation}
\label{eq:MCI}
    \frac{dM_{CI}}{dt} = \xi_{17}M_{I}(F+S) + \xi_{18}M_{CI} + \xi_{19}TM_{CI}.
\end{equation}
The density of the activated macrophages, $M_A$, evolved over time according to the following equation:
\begin{equation}
\label{eq:MA}
    \frac{dM_{A}}{dt} = \xi_{20}M_{R}T + \xi_{21}M_{A}.
\end{equation}
The density of T-cells, $T$, evolved over time according to the following equation:
\begin{equation}
\label{eq:T}
    \frac{dT}{dt} = \xi_{22}\mathbf{1}_{t \ge T_{enter}}N_{V}(1-N) + \xi_{23}T.
\end{equation}
 The upper and lower bounds for each coefficient and justification for these bounds are shown in Table \ref{tab:bounds}.
\begin{table}
    \centering
    \begin{tabular}{|l|l|l|}
        \hline
        \textbf{EQL parameter} & \textbf{Bounds} & \textbf{Justification}\\
        \hline
        $\xi_1$ & $(-\infty,0]$ & Expected to be negative\\
        $\xi_2$ & $[0.02167,0.021671]$ & \cite{Bowness2018},\cite{Shorten2013}\\
        $\xi_3$ & $[-0.02,0]$ (female, neutral) & Expected to be negative;\\
         & $[-0.07,0]$ (male) & Improved fit by using different bounds\\
        $\xi_4$ & $(-\infty,-0.00001]$ & Expected to be negative\\
        $\xi_5$ & $[0.00723,0.008]$ & \cite{Bowness2018},\cite{HendonDunn2016}\\
        $\xi_6$ & $(-\infty,-0.00001]$ & Expected to be negative\\
        $\xi_7$ & $[0.0004,\infty)$ & \cite{Bowness2018},\cite{VanFurth1973}\\
        $\xi_8$ & $[0.00001,\infty)$ & Expected to be positive\\
        $\xi_9$ & $[0,\infty) $& Expected to be positive\\
        $\xi_{10}$ & $(-\infty,-0.525]$ (female) & Expected to be negative and conserve mass;\\
         & $(-\infty,-0.45]$ (male) & Improved fit by using different bounds\\
         & $(-\infty,-0.4875]$ (neutral) & \\        
        $\xi_{11}$ & $(-\infty,-0.0004]$ & \cite{Bowness2018},\cite{VanFurth1973}\\
        $\xi_{12}$ & $[-58.5,0]$ (female) & Expected to be negative and conserve mass;\\
         & $[-46.5,0]$ (male) & Improved fit by using different bounds\\
         & $[-47.5,0]$ (neutral) & \\
        $\xi_{13}$ & $[0.525,\infty)$ (female) & Expected to be positive and conserve mass;\\
         & $[0.45,\infty)$ (male) & Improved fit by using different bounds\\
         & $[0.4875,\infty)$ (neutral) & \\
        $\xi_{14}$ & $[-0.002,0]$ (female, neutral) & Expected to be negative and conserve mass;\\
         & $[-0.0015,0]$ (male) & Improved fit by using different bounds\\
        $\xi_{15}$ & $(-\infty,-0.0004]$ & \cite{Bowness2018},\cite{VanFurth1973}\\
        $\xi_{16}$ & $(-\infty,0]$ & Expected to be negative\\
        $\xi_{17}$ & $[0,\infty)$ & Expected to be positive\\
        $\xi_{18}$ & $(-\infty,-0.0004]$ & \cite{Bowness2018},\cite{VanFurth1973}\\
        $\xi_{19}$ & $(-\infty,0]$ & Expected to be negative\\
        $\xi_{20}$ & $[0,\infty)$ & Expected to be positive\\
        $\xi_{21}$ & $[-0.00438,-0.00397]$ & \cite{Bowness2018},\cite{SegoviaJuarez2004}\\
        $\xi_{22}$ & $[0,\infty)$ & Expected to be positive\\
        $\xi_{23}$ & $(-\infty,-0.01]$ & \cite{Bowness2018},\cite{Sprent1993}\\
        \hline
    \end{tabular}
    \caption{Upper and lower bounds for the coefficients in the learnt ODEs, along with justifications.}
    \label{tab:bounds}
\end{table}
\subsubsection{Host demographic dynamics (HDD)}
\label{subsubsec:compartmental model}
The sixth step in developing the multiscale model is to represent the host demographics dynamics with ODEs. This section outlines the host demographic equations used to evolve the compartment densities at the between-host scale in the absence of TB within-host dynamics. Section \ref{subsubsec:coupling} presents the coupling functions we use to link the scales in order to model TB infectious disease dynamics across both scales.\par
The between-host dynamics determine the evolution of the host population demographics, due to processes such as births and background mortality. We have assumed individuals belong to one of ten compartments, after stratifying the population based on gender and infection status. Specifically, the current infection status of individuals can be either susceptible; suffering from latent TB (i.e. exposed to \textit{M. tb} but not actively infected nor infectious to others); suffering from active TB; undergoing treatment and no longer infectious; recovered but with potential to relapse.\par
The list of compartments is summarised in Table \ref{tab:hdd}, along with the initial values used in generating the results. The population we have modelled is hypothetical, so initial values in each compartment were assumed. In this hypothetical population, we have assumed that one in four individuals have previously been infected with \textit{M. tb}, in keeping with estimates by \citeauthor{Houben2016} \cite{Houben2016}, and that 1.6 times more males have been infected than females, based on the males-to-females TB case ratio \cite{WHO2024}. Of individuals infected by \textit{M. tb}, we have assumed 10\% have gone on to develop active TB at some point after initial infection, whilst the rest have remained latently infected, in keeping with real-world estimates \cite{Kiazyk2017}. Finally, we have assumed that the population is approximately meeting the WHO's targets of treating 90\% of individuals diagnosed with active TB and curing 90\% of individuals diagnosed with TB \cite{Suthar2016}. Hence, we have set the initial conditions such that there are more people with active TB diagnosed and undergoing treatment than undiagnosed, and more people who have completed treatment than are still undergoing treatment. The exact numbers have been tuned so that TB is endemic in the hypothetical population (that is, the densities of males with active TB and females with active TB are maintained at baseline levels and do not stochastically die out) in each simulation with both within-host and between-host differences implemented between males and females.\par
\begin{table}
    \centering
    \begin{tabular}{|l|l|r|}
        \hline
        \textbf{Variable} & \textbf{Description} & \textbf{Initial condition}\\
        \hline
        $S_F$ & Density of susceptible females & 28,269\\
        $S_M$ & Density of susceptible males & 24,231\\
        $L_F$ & Density of latently infected females & 6,058\\
        $L_M$ & Density of latently infected males & 9,692\\
        $I_F$ & Density of females with active TB & 11\\
        $I_M$ & Density of males with active TB & 18\\
        $T_F$ & Density of females undergoing treatment & 117\\
        $T_M$ & Density of males undergoing treatment & 187\\
        $R_F$ & Density of recovered females who could relapse & 545\\
        $R_M$ & Density of recovered males who could relapse & 872\\
        \hline
    \end{tabular}
    \caption{Compartments from the host demographic dynamics. The initial conditions are given as the number of individuals in the compartment out of an initial population of size 70,000 (i.e. the initial densities were the values in the `Initial condition' column divided by 70,000). As the population is hypothetical, all values have been assumed, with the initial number of males with active TB set to be 1.6 times the initial number of females with active TB \cite{WHO2024} and one in four individuals has been infected with \textit{M. tb} at some point in time \cite{Houben2016}. For more details, see Section \ref{subsubsec:compartmental model}. For results with an equal initial number of males and females in each compartment, see Appendix \ref{appendix:same IC}.}
    \label{tab:hdd}
\end{table}
\begin{table}
    \centering
    \begin{tabular}{|l|l|l|}
        \hline
        \textbf{Parameter} & \textbf{Description} & \textbf{Value}\\
        \hline
        $b$ & Birth rate & 0.000012\\
        $\mu$ & Background mortality rate & 0.00001\\
        $f$ & Proportion of female births & 0.5\\
        $m$ & Proportion of male births & 0.5\\   
        \hline
    \end{tabular}
    \caption{Host demographics dynamics parameters. The birth rate is equal to the fertility rate of women of reproductive age (15-49) multiplied by the proportion of females whose ages fall within that range. As the population is hypothetical, all values have been assumed. In Equation (\ref{eq:fem susc}), $b_f = b \times f$; in Equation (\ref{eq:male susc}), $b_m = b \times m$.}
    \label{tab:hdd params}
\end{table}
Individuals are assumed to identify with one of two genders: women/girls (hereafter referred to as ``females'') or men/boys (hereafter referred to as ``males''). Individuals are born into the susceptible females compartment at rate $b_f$ and into the susceptible males compartment at rate $b_m$. It is assumed the amount of births is proportional to the density of total females in the population, as we have set the birth rate based on the total fertility rate (that is, the total number of births per woman of reproductive age). We have assumed the background mortality rate (i.e. non-TB-associated death rate), $\mu$, is equal for both females and males, regardless of current infection status. Furthermore, we assume no migration occurs into, or out of, the population.\par
Based on these assumptions, the host demographic dynamics evolve according to the following ODEs. In Equations (\ref{eq:fem susc}) and (\ref{eq:male susc}), $N_F$ is the density of total females in the population. The list of parameter values used is given in Table \ref{tab:hdd params}.
\begin{align}
    \frac{dS_F}{dt} &= b_fN_F - {\mu}S_F, \label{eq:fem susc}\\
    \frac{dS_M}{dt} &= b_mN_F - {\mu}S_M, \label{eq:male susc}\\
    \frac{dL_F}{dt} &= -{\mu}L_F,\\
    \frac{dL_M}{dt} &= -{\mu}L_M,\\
    \frac{dI_F}{dt} &= -{\mu}I_F,\\
    \frac{dI_M}{dt} &= -{\mu}I_M,\\
    \frac{dT_F}{dt} &= -{\mu}T_F,\\
    \frac{dT_M}{dt} &= -{\mu}T_M,\\
    \frac{dR_F}{dt} &= -{\mu}R_F,\\
    \frac{dR_M}{dt} &= -{\mu}R_M.
\end{align}
\subsubsection{Coupling}
\label{subsubsec:coupling}
The seventh and final step in developing the multisale model is to stochastically link the within-host dynamics to the host demographic dynamics with coupling functions. Each individual suffering from active TB has their within-host dynamics modelled individually, using the learnt ODEs described in Section \ref{subsubsec:learnt ODEs} with a set of parameters that will depend on the individual's sex\footnote{It is worth remembering that we are making the simplifying assumptions that there are only two sexes and two genders, and that an individual's gender identity always aligns with their assigned biological sex, that is, men/boys are assigned the male sex and women/girls are assigned the female sex}. Unless stated otherwise (in certain counterfactual scenarios), the set of parameters that is used in the learnt ODEs to model an individual's within-host dynamics will depend on the sex the individual is assigned.\par
The within-host dynamics for these individuals are stochastically linked to the host demographic dynamics occurring at the between-host scale to produce our multiscale model of tuberculosis. Some of the linking functions consider tuberculosis-associated events in order to characterise the influence of the within-host scale on the between-host scale by changing compartment densities to reflect transitions from one infection status to another. Other functions model non-tuberculosis-associated events (i.e. natural mortality) in order to characterise the influence of the between-host scale on the within-host scale by considering the evolving host demographics to stochastically remove individuals' representations at the within-host scale. Whenever a tuberculosis-associated linking event occurs, the number of individuals being modelled in that compartment increases or decreases by one, and the compartment densities are increased or decreased by the density of one individual at the between-host scale to update the host disease states at the host demographics dynamics level. Whenever a non-tuberculosis-associated linking event occurs, a Poisson-distributed number of individuals are removed to bring the number of modelled individuals in a given compartment closer to the amount one would expect to see based on the current density for that compartment. For example, if we are modelling $N_I(t)$ individuals with Active TB at time $t$ and we expect $AI(t)$ individuals to have Active TB at time $t$ based on the current density of the Active TB compartment, $I(t)$ (where $A$ is a scaling constant), then the mean of the Poisson distribution would be $|AI(t) - N_I|$. There are ten linking events included in our model: the first nine are tuberculosis-associated where the within-host dynamics influence the host demographic dynamics, whilst the tenth is non-tuberculosis-associated where the host demographic dynamics influence the within-host dynamics. The number of times each event occurs in a given time step is determined using the tau-leaping method \cite{Gillespie2001}. More details on how the linking events occur are given in the following sections.\par
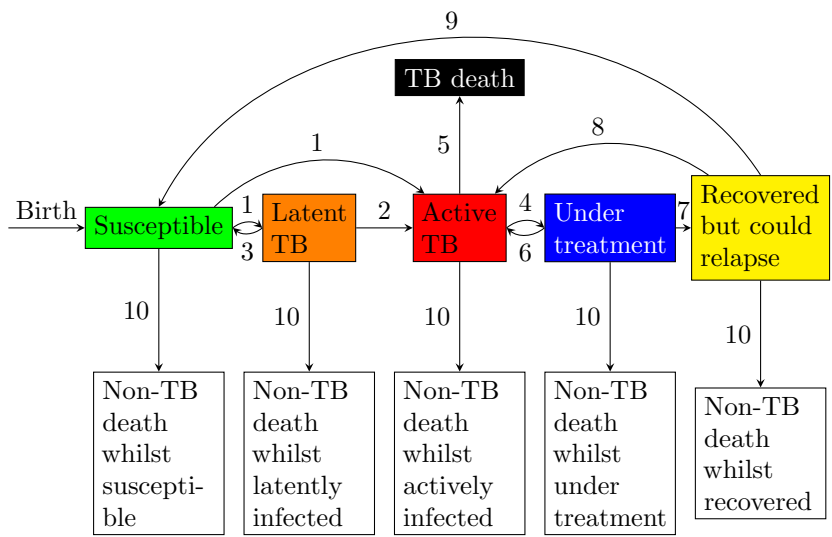
\begin{figure}[H]
    \centering
    \begin{tikzpicture}
        \node[draw, fill = green] (healthy) at (-3,0){\raggedright Susceptible};
        \node[draw, text width = 1.5cm] (death healthy) at (-3,-3){\raggedright Non-TB death whilst susceptible};
        \node[draw, fill = orange, text width = 1cm] (latent) at (-1,0){\raggedright Latent TB};
        \node[draw, text width = 1.5cm] (death latent) at (-1,-3){\raggedright Non-TB death whilst latently infected};
        \node[draw, fill = red, text width = 1cm] (sick) at (1,0){\raggedright Active TB};
        \node[draw, text width = 1.5cm] (death sick) at (1,-3){\raggedright Non-TB death whilst actively infected};
        \node[draw, fill = black, text = white] (dead) at (1, 2){\raggedright TB death};
        \node[draw, fill = blue, text = white, text width = 1.5cm] (treat) at (3,0){\raggedright Under\\treatment};
        \node[draw, text width = 1.5cm] (death treat) at (3,-3){\raggedright Non-TB death whilst under\\treatment};
        \node[draw, fill = yellow, text width = 1.6cm] (recover) at (5,0){\raggedright Recovered but could relapse};
        \node[draw, text width = 1.5cm] (death recover) at (5,-3){\raggedright Non-TB death whilst recovered};
        \draw[-stealth] (healthy.east) to[bend left=40] node[midway,above]{1} (latent.west);
        \draw[-stealth] ($(healthy.north)!0.75!(healthy.north east)$) to[bend left=40] node[midway,above]{1} ($(sick.north west)!0.25!(sick.north)$);
        \draw[-stealth] (latent.west) to[bend left=40] node[midway,below]{3} (healthy.east);
        \draw[-stealth] (-5,0) -- node[midway,above]{Birth} (healthy.west);
        \draw[-stealth] (healthy.south) -- node[midway,left]{10} (death healthy.north);
        \draw[-stealth] (latent.east) -- node[midway,above]{2} (sick.west);
        \draw[-stealth] (latent.south) -- node[midway,left]{10} (death latent.north);
        \draw[-stealth] (sick.north) -- node[midway,left]{5} (dead.south);
        \draw[-stealth] (sick.east) to[bend left=40] node[midway,above]{4} (treat.west);
        \draw[-stealth] (sick.south) -- node[midway,left]{10} (death sick.north);
        \draw[-stealth] (treat.east) -- node[midway,above]{7} (recover.west);
        \draw[-stealth] (treat.west) to[bend left=40] node[midway,below]{6} (sick.east);
        \draw[-stealth] (treat.south) -- node[midway,left]{10} (death treat.north);
        \draw[-stealth] ($(recover.north)!0.75!(recover.north west)$) to[bend right=40] node[midway, above]{8} ($(sick.north east)!0.25!(sick.north)$);
        \draw[-stealth] (recover.north) to[bend right=60] node[midway,above]{9} (healthy.north);
        \draw[-stealth] (recover.south) -- node[midway,left]{10} (death recover.north);
    \end{tikzpicture}
    \caption{Schematic showing the transitions between compartments at the between-host scale. The numbers adjacent to the arrows correspond to the linking events listed previously. The number of times each event takes place over a time step is determined using the tau-leaping method \cite{Gillespie2001}. Newborns enter the Susceptible compartment. It is worth noting that there are two sub-compartments for each of these compartments when gender is taken into account: a males sub-compartment and a females sub-compartment. Adapted from \citetitle{Doran2026} by \citeauthor{Doran2026} \cite{Doran2026} (licensed under \url{https://creativecommons.org/licenses/by/4.0/}).}
    \label{fig:transitions}
\end{figure}

The linking events are as follows:
\begin{enumerate} 
    \item transmission from individuals suffering active TB to susceptible individuals;
    \item progression to active TB for individuals suffering from latent TB;
    \item spontaneous recovery for individuals suffering with latent TB;
    \item diagnosis for individuals suffering with active TB;
    \item disease-associated mortality for individuals suffering with active TB;
    \item abandoning treatment early for individuals undergoing treatment;
    \item successful completion of treatment for individuals undergoing treatment;
    \item relapse for individuals who have recovered but could potentially relapse;
    \item returning to being susceptible for individuals who have recovered but could potentially relapse;
    \item removal of infectious individuals (and the index numbers of individuals with other infection states) due to natural mortality.
\end{enumerate}
All links, as well as births and natural mortality, are shown in Figure \ref{fig:transitions}. Unless stated otherwise (in certain counterfactual scenarios), each of these events is further sub-divided to reflect gender. For example, transmission events between two individuals with no specified gender can be sub-divided into transmission from females to females, from females to males, from males to females and from males to males. Each of these events is discussed further in the following sections.\par
At any given time, there are $N_I$ individuals with active TB in our model, all of whom have associated within-host dynamics. The number of transmission events in a given time step caused by individual $i \in \{1,...,N_I\}$ suffering from active TB depends on the current infectiousness of the potential infector (which is in turn dependent on their current within-host state, stored in their specific state vector $\bm{W}_i$, transmission-specific parameters, $\theta_C$ - which may or may not depend on gender - and the time since their infection began, $T_i$) and the density of susceptible individuals in the population. Each infector is initiated with a number of granulomas, $G$, drawn from a Poisson distribution with a mean equal to 10 (based on real-world data) \cite{Joslyn2022}; each granuloma is assumed to evolve in an identical manner. Additionally, each infector is initiated with an indicator variable, $C$, to determine if the individual is suffering from cavitary TB. Cavitary TB is known to make TB cases more infectious \cite{Urbanowski2020}, so infectors with $C$ equal to 1 have their transmission function increased by a scalar multiplier to make infection events more likely. The probability of developing cavitary TB is set equal to 0.32 for females and 0.48 for males, to reflect the fact that males are more likely to develop cavitary TB than females \cite{Balogun2021} and cavitary TB affects around 40\% of individuals with active TB \cite{Zhang2016}.\par
The probability of a single transmission from potential infector $i$ is determined by the transmission function $\beta(\bm{W}_i,\theta_C,T_i)$. The number of new infections caused by infectious individual $i$, $n_i$, is a Poisson distributed random variable with mean $\beta(\bm{W}_i,\theta_{C},T_i)S(t)\Delta t$, where $S(t)$ is the density of susceptible individuals at time $t$ and $\Delta t$ is a single model time step. This can be further sub-divided into the number of new female infections and new male infections by replacing $S$ with $S_F$ and $S_M$, respectively. This is to reflect the fact that sex assortativity in contacts appears to influence TB transmission probabilities \cite{Horton2020}, so the amount of transmissions from infector $i$ to infectees of a certain gender may differ from the amount of transmissions from infector $i$ to infectees from the other gender. The difference in transmission probabilities depending on the genders of the potential infector and infectee are also captured by adding in a scalar multiplier to the transmission coupling function, which is dependent upon the genders of the two individuals involved in a potential transmission event.\par
A proportion of individuals with these new infections will develop active TB immediately. This proportion has previously been found to be around $10 \pm 5\%$ \cite{Ahmad2011,Kiazyk2017}, and we have set this proportion to 15\%, in keeping with this real-world data, to ensure the number of individuals with active TB is unlikely to drop to zero through stochastic fluctuations of the model. For each such individual, a new representation of the within-host dynamics is created, with the ODEs depending on the sex of the individual (unless stated otherwise in certain counterfactual scenarios); these individuals are added to a list of currently infectious individuals. In the host demographic part of the model, the density of susceptible individuals, $S$, is reduced by an amount equal to the density of $n_i$ individuals, equal to $\frac{n_i}{A}$, where $A$ is ``a positive constant representing the area over which individuals interact'' \cite{Smith2025} and thus $\frac{1}{A}$ is the density of one individual. A density equal to the proportion of the $n_i$ transmissions that lead to active TB, $n_i^I$ is added to the density of infectious individuals, $I$. For the remaining transmissions, $n_i^L = n_i - n_i^I$, that led to latent TB, the individual is added to a list of latently infected individuals. The remaining density that did not lead to active TB is added to the density of the Latent TB compartment, $L$ in the host demographic part of the model. This process is repeated to calculate the number of transmissions caused by each of the $N_I$ individuals with active TB.\par
The densities of the Susceptible compartment, $S$, the Latent TB compartment, $L$, and the Active TB compartment, $I$, change in a single time step due to transmission events as follows:
\begin{align}
    S(t+{\Delta t}) &= S(t) - \frac{1}{A}\sum_{i=1}^{N_I(t)} n_i,\\
    L(t+{\Delta t}) &= L(t) + \frac{1}{A}\sum_{i=1}^{N_I(t)} n_i^L,\\
    I(t+{\Delta t}) &= I(t) + \frac{1}{A}\sum_{i=1}^{N_I(t)} n_i^I.
\end{align}
The function used to model transmission from individual $i$ in our model was a Hill function of the form:
\begin{equation}
\label{eq:transmission}
    \beta(F_i,S_i,G,C,T_i) = (1+(\theta_1-1)C)\theta_2\theta_3\frac{(100^2G(F_i+S_i))^{\theta_4}}{\theta_5^{\theta_4} + (100^2G(F_i+S_i))^{\theta_4}},
\end{equation}
where $F_i$ is the density of fast-growing extracellular bacterial load in individual $i$ at time $T_i$, $S_i$ is the density of slow-growing extracellular bacterial load in individual $i$ at time $T_i$, $G$ is the number of granulomas within individual $i$, $C$ is an indicator variable equal to 1 if individual $i$ has cavitary TB and 0 otherwise, $\theta_1$ is the infectiousness multiplier if individual $i$ has cavitary TB, $\theta_2$ is a  susceptibility multiplier based on the genders of the potential infector and potential infectee, $\theta_3$ is the maximum possible infectiousness, $\theta_4$ is the Hill coefficient, and $\theta_5$ is the bacterial load at which infectiousness is half of the maximum. This equation is based on the transmission function used in a similar multiscale model of tuberculosis, an agent-based model with within-host dynamics controlled by ODEs, but with neither sex nor gender considered \cite{Pereira2021}. The value of $\theta_1$ was chosen based on research conducted by \citeauthor{Melsew2018} \cite{Melsew2018}. The values for $\theta_2$ (given in Table \ref{tab:coupling params}) are based on the assumption that there is strong sex assortativity between contacts in our hypothetical population, which has been observed in multiple countries with varying degrees of TB burden \cite{Horton2020}. The other values were taken from \cite{Pereira2021}.\par
The coupling functions used for the other transitions between compartments that were not due to an infection event are described in the following paragraphs. For each individual, a number, $r$, was generated uniformly at random between 0 and 1 for each coupling function related to their current infection state. A transition event occurred if $r$ was less than the value generated by the coupling function corresponding to that transition event (given by Equations (\ref{eq:trans func 1}) to (\ref{eq:trans func 8}), which we will meet later in this section). Whenever a transition event occurs, the individual is removed from the list of people in their previous infection state and added to the list of people in their new infection state (with the exception of disease-induced mortality, where the individual is removed from the model). If their previous infection state was Active TB, the representation of their within-host dynamics is removed. If their new infection state is Active TB, a representation of the within-host dynamics is instantiated for the individual. Simultaneously, the density of the compartment corresponding to their previous infection state decreases by the density of one individual, and the density of the compartment corresponding to their new infection state increases by the density of one individual (with the exception of disease-induced mortality, where the density of one individual is removed from the model). $\Delta t$ represents the length of the time step of the model in each of the following functions in which it occurs. In scenarios where differences in sex and/or gender were taken into account, some of the parameter values in the following coupling functions may have been allowed to differ between sexes/genders (see Table \ref{tab:coupling params} for the parameter values used in each scenario).\par
A proportion of latently infected individuals will see their infection status change to active. The coupling function used to determine whether progression from Latent TB to Active TB occurred in a given time step was:
\begin{equation}
\label{eq:trans func 1}
    1 - \exp(-\theta_6\Delta t),
\end{equation}
where $\theta_6$ is the average rate of progression from Latent TB to Active TB. The value for $\theta_6$ was chosen to equal the reciprocal of the average time before transitioning from Latent TB to Active TB (assumed to be $2.\dot{3}$ years, one-third of the time before spontaneous recovery is guaranteed to have occurred in a similar agent-based model of tuberculosis transmission dynamics by \citeauthor{Prats2016} \cite{Prats2016}\footnote{In a similar model of tuberculosis transmission \cite{Prats2016}, any latently infected individual that had not progressed to Active TB after 7 years was considered to have spontaneously recovered. As we are assuming the time to progress from Latent TB to Active TB is exponentially distributed, we have assumed the reciprocal of the average time until progression from Latent TB to Active TB is such that, if the time until progression from Latent TB to Active TB is longer than 7 years, such a time would be defined as an outlier according to \citeauthor{Tukey1977} \cite{Tukey1977}. For an exponential distribution with rate $\lambda$, an outlier would be larger than $Q_3 + 1.5(Q_3-Q_1)$, where $Q_1 = \frac{\ln{(4/3)}}{\lambda}$ is the lower quartile and $Q_3 = \frac{\ln{(4)}}{\lambda}$ is the upper quartile. Thus, 7 years should equal $\frac{\ln{(4)}}{\lambda} + 1.5\frac{\ln{(3)}}{\lambda}$. This can be approximately achieved by setting $\lambda$ such that $\frac{1}{\lambda}$ is one-third of 7 years.}), multiplied by the probability of this transition occurring (based on the WHIDM simulation results: see Section \ref{sec:WHIDM results}).\par
All other latently infected individuals will eventually recover from their infection and return to the Susceptible compartment, that is, they eventually ``spontaneously recover''. The coupling function used to determine whether spontaneous recovery occurred in a given time step was:
\begin{equation}
    1 - \exp(-\theta_7\Delta t),
\end{equation}
where $\theta_7$ is the average rate of spontaneous recovery occurrence. The value of $\theta_7$ was chosen to equal the reciprocal of the average time before spontaneous recovery occurred (assumed to be 7 years, the time by which spontaneous recovery is guaranteed to have occurred in a similar agent-based model of tuberculosis transmission dynamics by \citeauthor{Prats2016} \cite{Prats2016}), multiplied by the probability of spontaneous recovery occurring (assumed to be 90\%, as there is a 10\% chance of reactivation \cite{Ahmad2011}).\par
A proportion of infectious individuals will be diagnosed in a given time step, and will transition from the Active TB compartment, $I$, to the Under Treatment compartment, $T$. The coupling function used to determine whether diagnosis occurred in a given time step was:
\begin{equation}
    1 - \exp(-\theta_8\Delta t),
\end{equation}
where $\theta_8$ is the average rate of diagnosis. The value of $\theta_8$ was chosen to equal the reciprocal of the average time before diagnosis occurred (assumed to be 90 days, typical in some countries, e.g. Ghana \cite{Osei2015}), multiplied by the probability of diagnosis occurring (assumed to be 90\% as we assume the population is meeting the WHO's End TB targets \cite{Suthar2016}). It should be noted that, due to computational restrictions when running WHIDM on the Nimbus cloud supercomputer, we had within-host simulation data from WHIDM covering only the first 15 days post-infection. Although this is not an unreasonable length, as shorter diagnosis delays have been observed in some countries (e.g. Ethiopia \cite{Asefa2014}), it does not allow for the longer diagnosis delays that are observed in some populations \cite{Osei2015}. As such, we specified that, if an individual had been in the Active TB compartment for at least 15 days, they should automatically transition to the Under Treatment compartment. Hence, individuals transitioned from Active TB to Under Treatment if $r < 1 - \exp(-\theta_8\Delta t)$ or $T_i \ge 15$, where $r$ is a uniformly distributed random number between 0 and 1 and $T_i$ is the time (in days) since individual $i$ was infected.\par
Other infectious individuals will die due to the disease in the same time step: this is hereafter referred to as disease-induced mortality. The coupling function used to determine whether disease-induced mortality occurred for individual $i$ in a given time step was:
\begin{equation}
    \theta_9\frac{100^2G(F_i+S_i)}{\theta_{10} + 100^2G(F_i+S_i)},
\end{equation}
where $\theta_9$ is the maximum probability of disease-induced mortality, $\theta_{10}$ is the bacterial load at which the probability of disease-induced mortality is half of the maximum probability, and $F_i$, $S_i$ and $G$ are defined in Equation (\ref{eq:transmission}). The value of $\theta_9$ was chosen to equal the average time before disease-induced mortality occurred (assumed to be 10 years, although evidence suggests individuals can survive much longer than this even without treatment \cite{Rodriguez2023}), multiplied by the probability of disease-induced mortality occurring (assumed to be 10\%, as before \cite{Suthar2016}). The value of $\theta_{10}$ was taken from \cite{Pereira2021}.\par
In each time step, some individuals that are undergoing treatment will abandon the treatment early and return to being infectious. It should be noted that the within-host dynamics for these individuals are not a continuation of their previous within-host dynamics: to save on computational memory, we do not continue to store the within-host dynamics for any individual once they leave the Active TB compartment, so a new representation of the individual's within-host scale dynamics need to be instantiated if an individual returns to the compartment. The coupling function used to determine whether abandonment of treatment occurred in a given time step was:
\begin{equation}
    1 - \exp(-\theta_{11}\Delta t),
\end{equation}
where $\theta_{11}$ is the average rate of abandonment of treatment. The value of $\theta_{11}$ was chosen to equal the reciprocal of the average time until abandonment of treatment occurred (assumed to be 60 days, one-third of the typical treatment length of 180 days \cite{WHO2010}\footnote{As we are assuming the time to abandon treatment is exponentially distributed, we have assumed the reciprocal of the average time until treatment abandonment is such that, if the time until treatment abandonment is longer than 180 days, such a time would be defined as an outlier according to \citeauthor{Tukey1977} \cite{Tukey1977}. For an exponential distribution with rate $\lambda$, an outlier would be larger than $Q_3 + 1.5(Q_3-Q_1)$, where $Q_1 = \frac{\ln{(4/3)}}{\lambda}$ is the lower quartile and $Q_3 = \frac{\ln{(4)}}{\lambda}$ is the upper quartile. Thus, 180 days should equal $\frac{\ln{(4)}}{\lambda} + 1.5\frac{\ln{(3)}}{\lambda}$. This can be approximately achieved by setting $\lambda$ such that $\frac{1}{\lambda}$ is 60 days.}), multiplied by the probability of treatment abandonment occurring (assumed to be 10\%, as before \cite{Suthar2016}).\par
Other individuals undergoing treatment will successfully complete their treatment in the same time step, and will transition to the ``Recovered but could relapse'' compartment, $R$. The coupling function used to determine whether completion of treatment occurred in a given time step was:
\begin{equation}
    1 - \exp(-\theta_{12}\Delta t),
\end{equation}
where $\theta_{12}$ is the average rate of completion of treatment. The value of $\theta_{12}$ was chosen to equal the reciprocal of the average time until treatment completion occurred (assumed to be 180 days, the full standard treatment regimen length \cite{WHO2010}), multiplied by the probability of treatment completion occurring (assumed to be 90\% if the population is meeting the WHO's End TB targets as assumed \cite{Suthar2016}).\par
A proportion of the individuals that have recovered will relapse in a given time step, and transition back to the Active TB compartment. The coupling function used to determine whether relapse occurred in a given time step was:
\begin{equation}
    1 - \exp(-\theta_{13}\Delta t),
\end{equation}
where $\theta_{13}$ is the average rate of relapse. The value of $\theta_{13}$ was chosen to equal the reciprocal of the average time until relapse occurred (assumed to be $0.\dot{6}$ years, one-third of the total time an individual was assumed to be at risk of TB reactivation in the model presented by \citeauthor{Prats2016} \cite{Prats2016}\footnote{In a similar model of tuberculosis transmission \cite{Prats2016}, any individual in the Recovered but could relapse compartment that had not relapsed after 2 years was considered to have returned to the Susceptible compartment. As we are assuming the time to relapse is exponentially distributed, we have assumed the reciprocal of the average time until relapse is such that, if the time until relapse is longer than 2 years, such a time would be defined as an outlier according to \citeauthor{Tukey1977} \cite{Tukey1977}. For an exponential distribution with rate $\lambda$, an outlier would be larger than $Q_3 + 1.5(Q_3-Q_1)$, where $Q_1 = \frac{\ln{(4/3)}}{\lambda}$ is the lower quartile and $Q_3 = \frac{\ln{(4)}}{\lambda}$ is the upper quartile. Thus, 2 years should equal $\frac{\ln{(4)}}{\lambda} + 1.5\frac{\ln{(3)}}{\lambda}$. This can be approximately achieved by setting $\lambda$ such that $\frac{1}{\lambda}$ is one-third of 2 years.}), multiplied by the probability of relapse occurring (assumed to be 10\%, as before \cite{Suthar2016}).\par
Other recovered individuals will return to being susceptible from a new infection in the same time step, and transition to the Susceptible compartment; this is hereafter referred to as ``waning immunity''. The coupling function used to determine whether waning immunity occurred in a given time step was:
\begin{equation}
\label{eq:trans func 8}
    1 - \exp(-\theta_{14}\Delta t),
\end{equation}
where $\theta_{14}$ is the average rate of waning immunity. The value of $\theta_{14}$ was chosen to equal the reciprocal of the average time until waning immunity occurred (assumed to be 2 years, the total time an individual was assumed to be at risk of TB reactivation in the model presented by \citeauthor{Prats2016} \cite{Prats2016}), multiplied by the probability of waning immunity occurring (assumed to be 90\% if the population is meeting the WHO's End TB targets as assumed \cite{Suthar2016}).\par
All of the previous events are tuberculosis-associated; it is worth noting that the compartments are further sub-divided by gender, and thus these events occur for both males and females separately. Each of these causes a change in densities to at least one compartment in the host demographic part of the model. The last event, natural mortality, links the host demographics to the within-host part of the model by finding the difference between the number of infectious individuals we are currently modelling, $N_I$, and the expected number of individuals in the Active TB compartment according to its density, $AI(t)$. As the solutions to the host demographic dynamics equations evolve over time, the densities of the compartments may decrease due to natural mortality, and eventually may have decreased by a density corresponding to more than one individual; without a link to the within-host scale, we would continue to model too many individuals in a given compartment than the between-host scale densities suggest should be modelled. If the absolute value of the difference between the number of infectious individuals with within-host dynamics, $N_I$, and the expected number of individuals in the Active TB compartment according to its density, $AI(t)$, is non-zero, we draw a Poisson-distributed random variable with mean $|{\delta}I|$ (where ${\delta}I=AI(t)-N_I(t)$), and this amount of infectious individuals are removed (along with their within-host dynamics) to account for the fact that they died due to non-tuberculosis-associated mortality. This removal process is also done to individuals dying from natural mortality in the other compartments.\par
\begin{table}
    \centering
    \begin{tabular}{|l|l|l|l|}
        \hline
        \textbf{Parameter} & \textbf{Description} & \textbf{Value} & \textbf{Justification}\\
        \hline
        $\theta_1$ & Cavitary TB infectiousness multiplier & 2.5 & \cite{Melsew2018}\\
        $\theta_2$ & Susceptibility multiplier (female infector, female infectee) & 0.005 & \cite{Horton2020},\cite{Miller2021}\\
         & Susceptibility multiplier (female infector, male infectee) & 0.004 & \\
         & Susceptibility multiplier (male infector, female infectee) & 0.0045 & \\
         & Susceptibility multiplier (male infector, male infectee) & 0.0065 & \\
        $\theta_3$ & Maximum infectiousness & 15 & \cite{Pereira2021}\\
        $\theta_4$ & Hill coefficient of transmission function & 18 & \cite{Pereira2021}\\
        $\theta_5$ & Half-max of transmission function & 111 & \cite{Pereira2021}\\
        $\theta_6$ & Rate of progression to active TB (female) & 0.00006 & \cite{Prats2016}\\
         & Rate of progression to active TB (male) & 0.00013 & \\
        $\theta_7$ & Rate of spontaneous recovery & 0.00035 & \cite{Bowness2018},\cite{Prats2016}\\
        $\theta_8$ & Diagnosis rate & 0.01 & \cite{Bowness2018},\cite{Suthar2016}\\
        $\theta_9$ & Maximum probability of disease-induced mortality & 0.00002 & \cite{Rodriguez2023},\cite{Suthar2016}\\
        $\theta_{10}$ & Half-max of disease-induced mortality function & 111 & \cite{Pereira2021}\\
        $\theta_{11}$ & Treatment abandonment rate & 0.00167 & \cite{Prats2016},\cite{Suthar2016}\\
        $\theta_{12}$ & Treatment completion rate & 0.005 & \cite{Prats2016},\cite{Suthar2016}\\
        $\theta_{13}$ & Relapse rate & 0.00041 & \cite{Prats2016},\cite{Suthar2016}\\
        $\theta_{14}$ & Waning immunity rate & 0.00123 & \cite{Prats2016},\cite{Suthar2016}\\
        \hline
    \end{tabular}
    \caption{Parameters used in the coupling functions and their assigned values, along with justifications for the assignment of these values. Note that the parameter values given in this table were used when both sex and gender stratification were in effect, for the results presented in Section \ref{sec:WHBH}. When differences between sexes and/or genders were removed, the values of some of these parameters were altered to reflect this. These changes are stated in Sections \ref{sec:WHnoBH} to \ref{sec:noWHnoBH}.}
    \label{tab:coupling params}
\end{table}
\section{Results}
\label{sec:results}
In this section, we first present the results of the simulations of WHIDM, then the three sets of learnt ODEs that were subsequently used for the within-host part of the multiscale model, and finally, a summary of the results generated by the multiscale model for the four scenarios considered.
\subsection{WHIDM results}
\label{sec:WHIDM results}
To capture the differences in immune response between males and females, we set three parameters to have unique values in each of the three parameter sets: the probability of T-cell recruitment per time step, the probability of a T-cell activating a resting macrophage, and the probability of a macrophage phagocytosing an extracellular \textit{M. tb}. This was based on research that suggested, during the immune response against \textit{M. tb} in males, that there is reduced phagocytosis by macrophages, fewer $\text{CD4}^+$ T-cells (which help with activation of macrophages) and fewer $\text{CD8}^+$ T-cells \cite{Nhamoyebonde2014}. The list of the parameter values used in WHIDM simulations can be found in Appendix \ref{appendix:WHIDM param sets}.\par
\begin{figure}
    \centering
    \subfigure[Female cell numbers]{\includegraphics[width=0.4\linewidth]{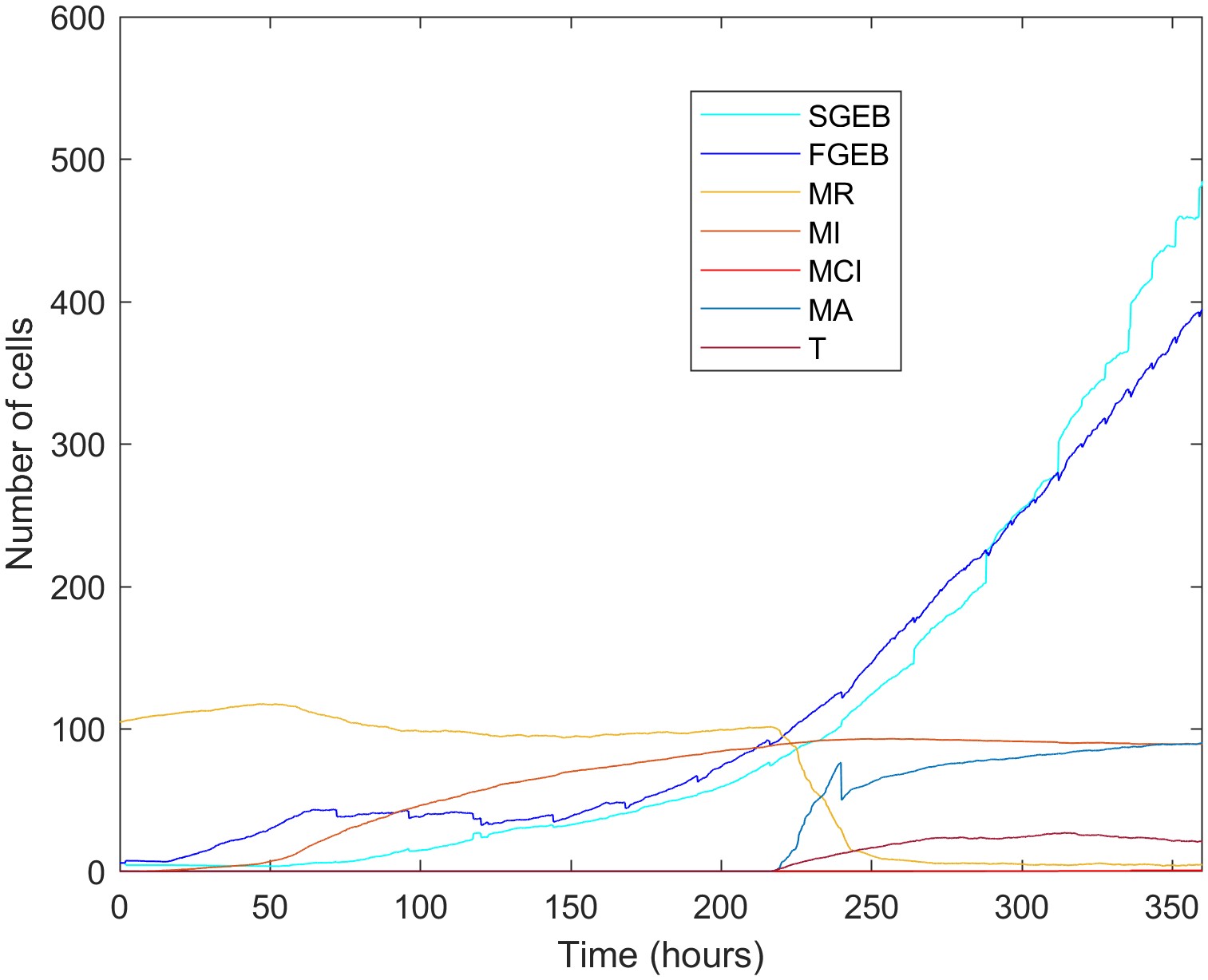}\label{fig:cell numbers female}}
    \hfill
    \subfigure[Female cell densities and learnt ODEs] {\includegraphics[width=0.4\linewidth]{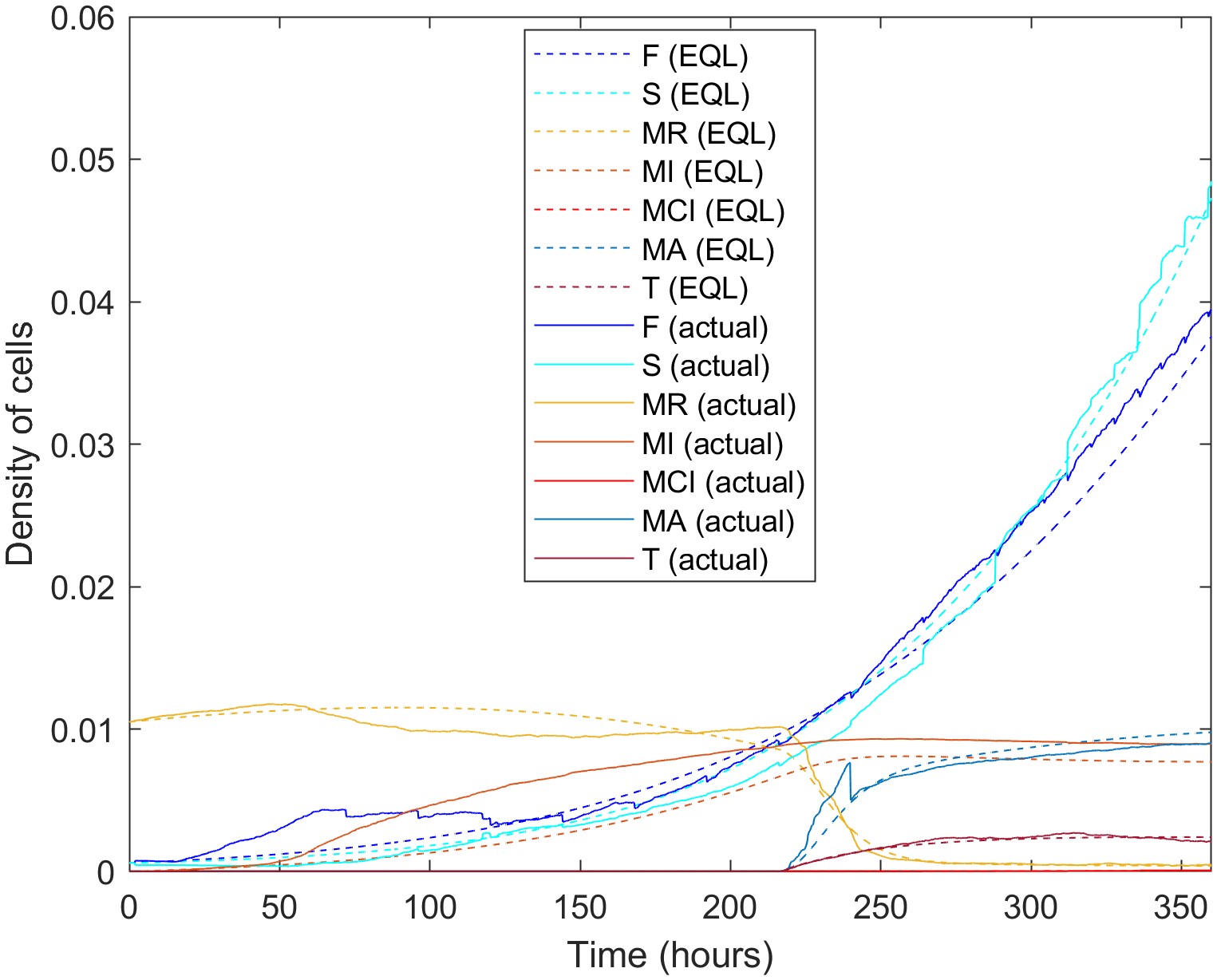}\label{fig:EQL female}}
    \subfigure[Male cell numbers] {\includegraphics[width=0.4\linewidth]{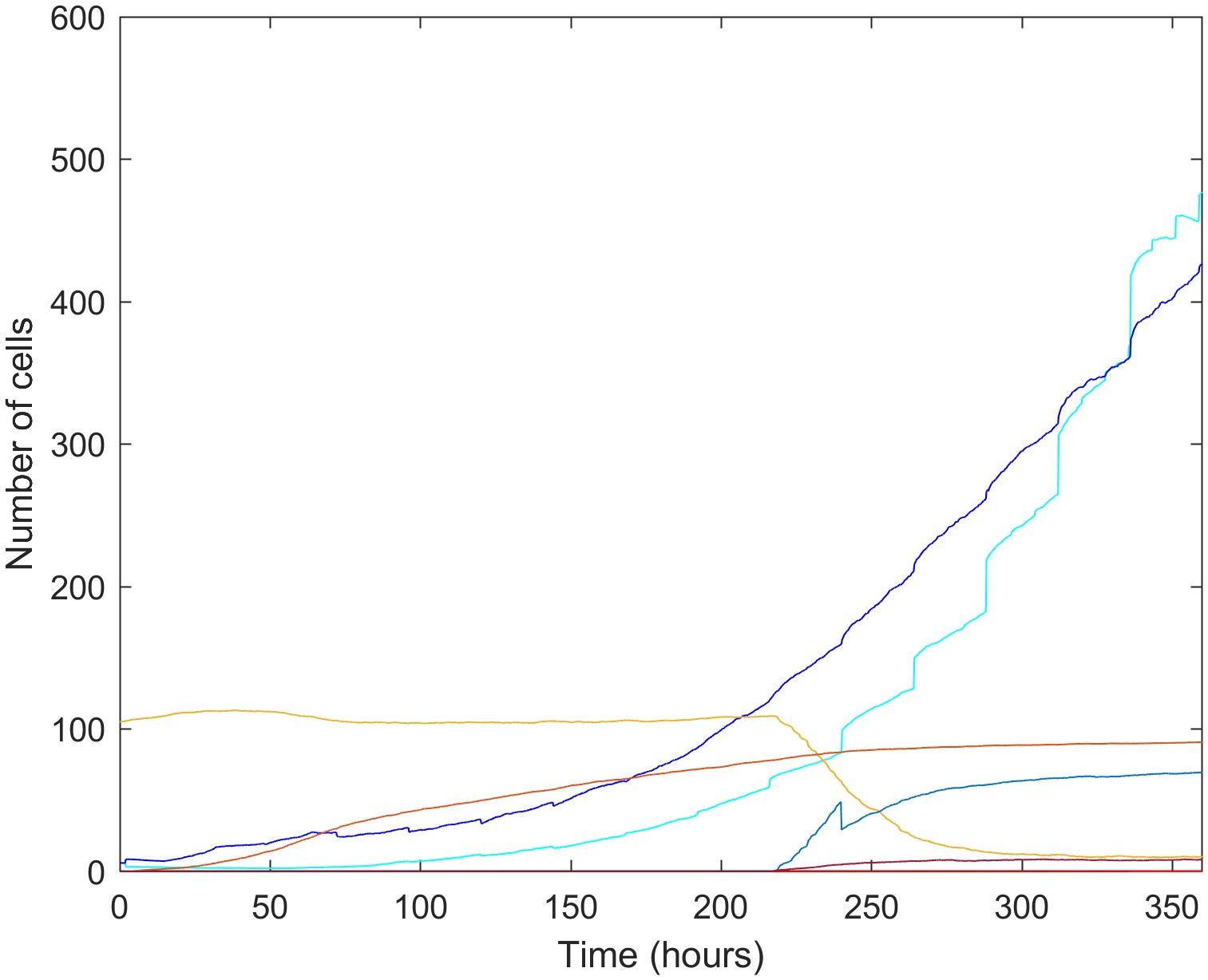}\label{fig:cell numbers male}}
    \hfill
    \subfigure[Male cell densities and learnt ODEs] {\includegraphics[width=0.4\linewidth]{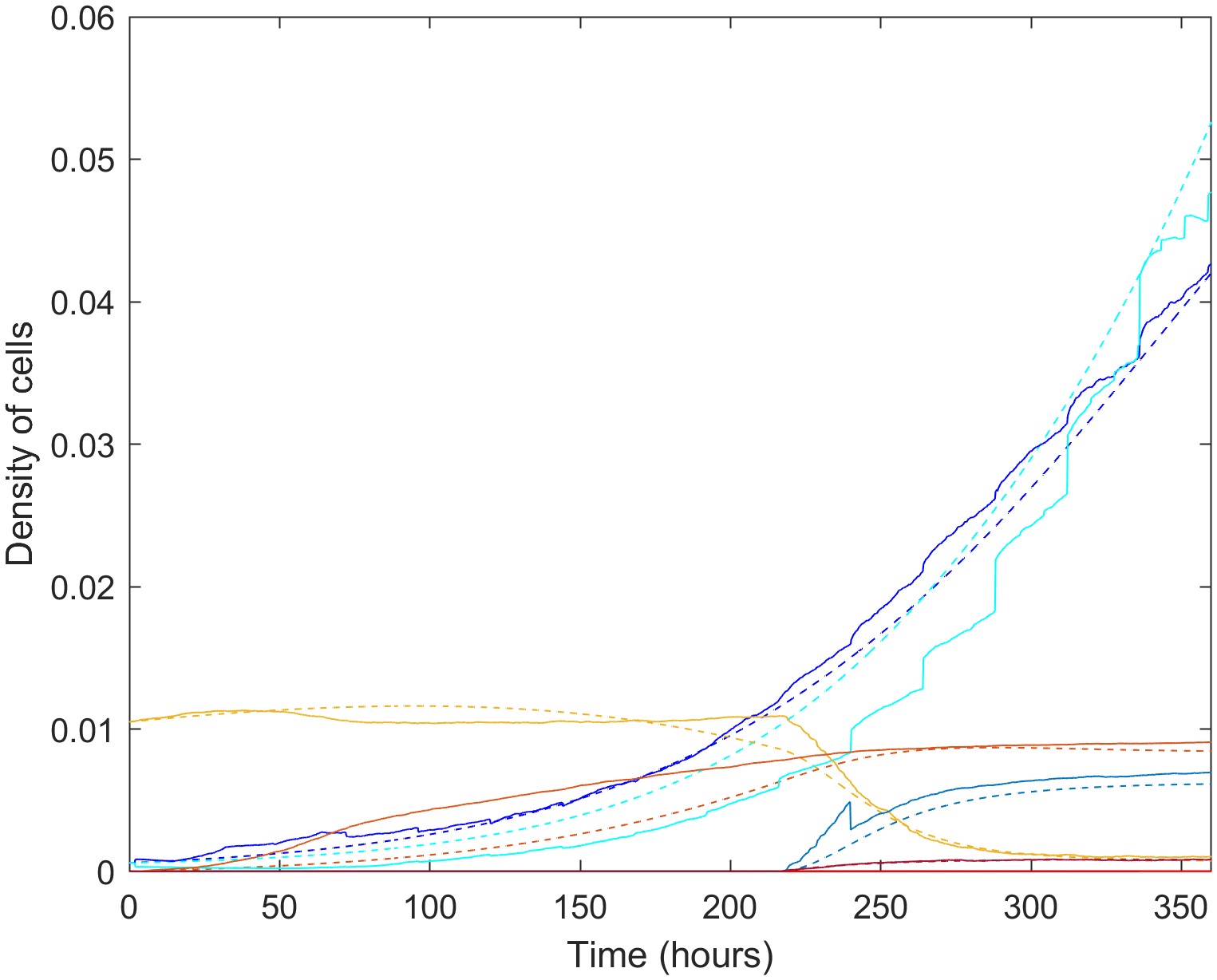}\label{fig:EQL male}}
    \subfigure[Neutral cell numbers] {\includegraphics[width=0.4\linewidth]{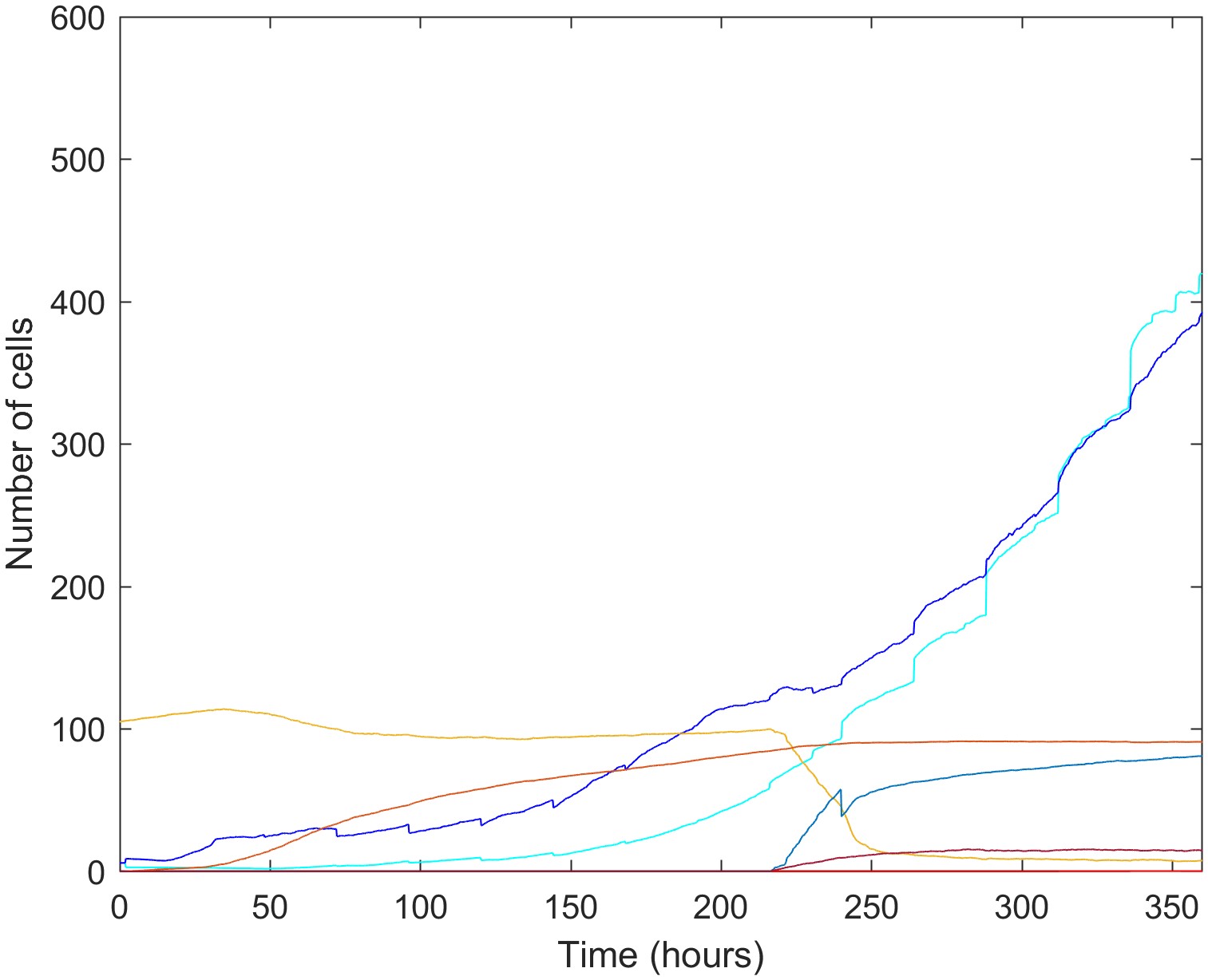}\label{fig:cell numbers neutral}}
    \hfill
    \subfigure[Neutral cell densities and learnt ODEs] {\includegraphics[width=0.4\linewidth]{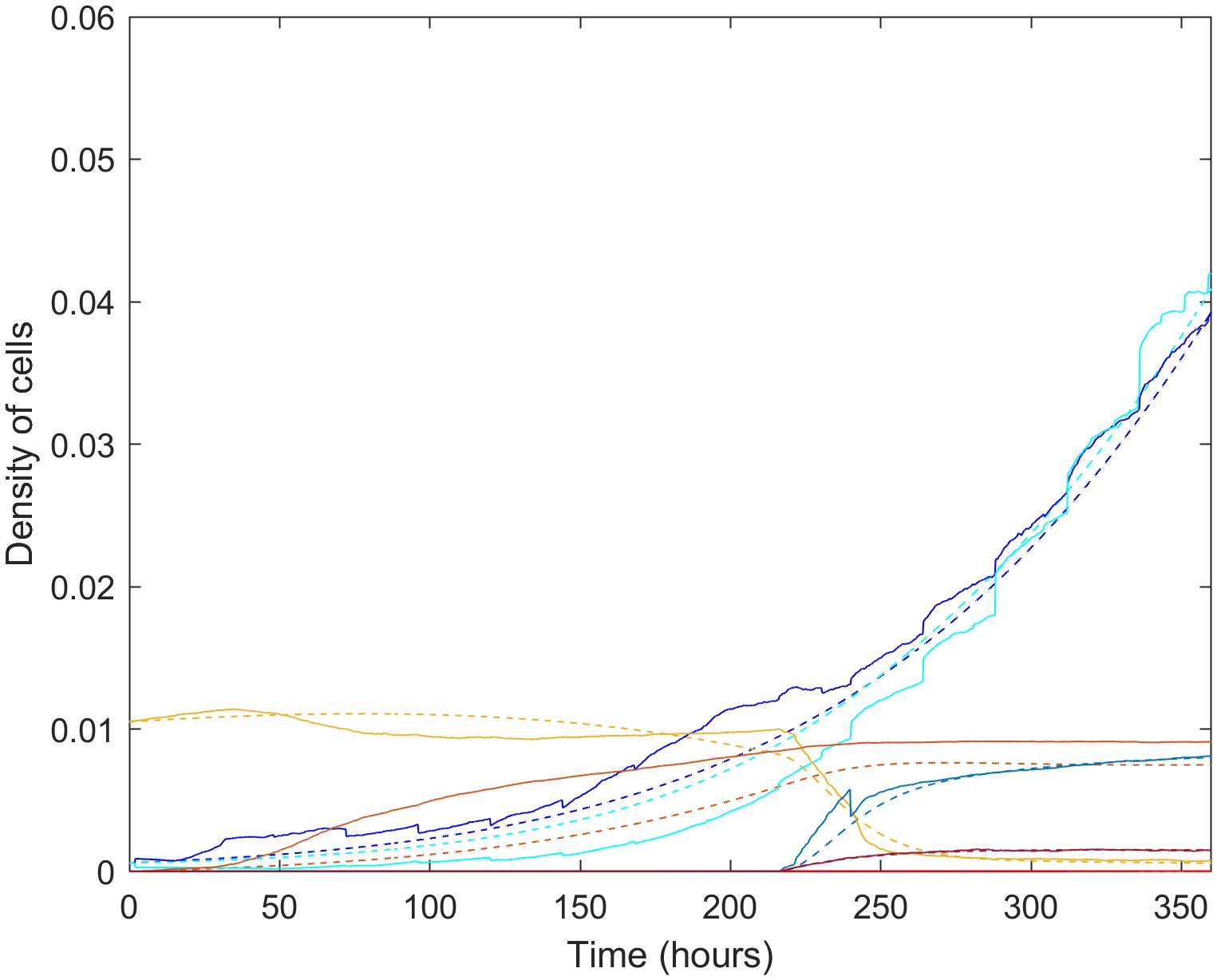}\label{fig:EQL neutral}}
    \caption{Average cell numbers observed in simulations ran using WHIDM that led to dissemination and comparison of average cell densities and the learnt ODEs from WHIDM simulations that led to dissemination. \subref{fig:cell numbers female} cell numbers with female parameters; \subref{fig:cell numbers male} cell numbers with male parameters; \subref{fig:cell numbers neutral} cell numbers with neutral parameters. \subref{fig:EQL female} learnt equations and cell densities with female parameters; \subref{fig:EQL male} learnt equations and cell densities with male parameters; \subref{fig:EQL neutral} learnt equations and cell densities with neutral parameters. Abbreviations: SGEB, slow-growing extracellular bacteria; FGEB, fast-growing extracellular bacteria; MR, resting macrophages; MI, infected macrophages; MCI, chronically infected macrophages; MA, activated macrophages; T, T-cells; EQL, equation learning.}
    \label{fig:cell numbers dissemination}
\end{figure}

Based on a consistency analysis of WHIDM conducted in previous work \cite{Doran2026a}, we ran 300 simulations of WHIDM for each of the three parameter sets, for a total of 900 simulations. Of these, 14 ($4.\dot{6}\%$) of the 300 simulations ran using the female parameters led to dissemination (which we have defined as at least 10 extracellular bacteria remaining at the end of a simulation, to account for similar ``detection limits'' in experimental measures); 29 ($9.\dot{6}\%$) of the 300 simulations ran using the male parameters led to dissemination; and 23 ($7.\dot{6}\%$) of the 300 simulations ran using the neutral parameters (an average of the male and female parameters) led to dissemination. This is in keeping with expected real-world proportions of individuals developing active TB after infection with \textit{M. tb} \cite{Kiazyk2017}. Summary plots of the average evolution of the numbers of each cell type in simulations that led to dissemination can be seen in Figures \ref{fig:cell numbers female}, \ref{fig:cell numbers male} and \ref{fig:cell numbers neutral}. For summary plots of the average evolution of the numbers of each cell type in simulations that led to containment, please see Appendix \ref{appendix:WHIDM containment}.
\subsection{Learnt ODEs}
We used each of the simulations that led to dissemination to learn the three sets of ODEs which best described the host-pathogen dynamics during an infection from \textit{M. tb} that led to active TB. The set of ODEs that a simulation was used towards depended on whether sex was taken into account or not and, if so, the sex for which the within-host dynamics were being simulated. For example, if a simulation that led to dissemination used the female parameter values, the simulation output was used towards the inference of the female set of ODEs. The learnt equations for each of the parameter sets were as follows.\par
The learnt within-host ODEs were of the form shown in Section \ref{subsubsec:learnt ODEs}. The parameter values learnt from the WHIDM simulations run using the female, male and neutral parameters are shown in Table \ref{tab:EQL results}.
\begin{table}[H]
    \centering
    \begin{tabular}{|l|l|l|l|}
        \hline
        \textbf{Parameter} & \textbf{Female} & \textbf{Neutral} & \textbf{Male}\\
        \hline
        $\xi_1$ & -0.65264 & -0.69167 & -0.51328\\
        $\xi_2$ & 0.02167 & 0.02167 & 0.02167\\
        $\xi_3$ & -0.02 & -0.02 & -0.07\\
        $\xi_4$ & -0.00001 & -0.00001 & -0.00001\\
        $\xi_5$ & 0.008 & 0.008 & 0.008\\
        $\xi_6$ & -0.00001 & -0.01079 & -0.0013\\
        $\xi_7$ & 0.0072 & 0.00866 & 0.01016\\
        $\xi_8$ & 0.68043 & 0.00002 & 0.00001\\
        $\xi_9$ & 0.01645 & 0.01487 & 0.01368\\
        $\xi_{10}$ & -0.525 & -0.4875 & -0.45\\
        $\xi_{11}$ & -0.00497 & -0.00497 & -0.00377\\
        $\xi_{12}$ & -58.5 & -47.5 & -46.5\\
        $\xi_{13}$ & 0.525 & 0.4875 & 0.45\\
        $\xi_{14}$ & -0.002 & -0.002 & -0.0015\\
        $\xi_{15}$ & -0.00164 & -0.00138 & -0.00216\\
        $\xi_{16}$ & -0.1891 & -0.82445 & -1.9994\\
        $\xi_{17}$ & 0.00193 & 0.0015 & 0.00131\\
        $\xi_{18}$ & -0.00642 & -0.03117 & -0.07891\\
        $\xi_{19}$ & -0.8483 & -20.37359 & -23.71824\\
        $\xi_{20}$ & 58.08771 & 47.01748 & 46.4828\\
        $\xi_{21}$ & -0.00438 & -0.00438 & -0.00438\\
        $\xi_{22}$ & 0.01579 & 0.01228 & 0.00598\\
        $\xi_{23}$ & -0.02891 & -0.03631 & -0.03173\\
        \hline
    \end{tabular}
    \caption{Lists of parameter values in the learnt ODEs used to simulate the within-host dynamics. The equations are of the same form as those shown in Section \ref{subsubsec:learnt ODEs}.}
    \label{tab:EQL results}
\end{table}
The learnt ODEs are compared against the average evolution of the densities of each cell type in Figures \ref{fig:EQL female}, \ref{fig:EQL male} and \ref{fig:EQL neutral}.
\subsection{Summary of multiscale model results}
\label{subsec:results}
We ran 300 simulations of the multiscale model for each of the following four scenarios: sex and gender stratification at both scales (Scenario 0); sex stratification at within-host scale only (Scenario 1); gender stratification at between-host scale only (Scenario 2); and no stratification by sex at the within-host scale or gender at the between-host scale (Scenario 3). In this section we give a summary of the number of TB cases for both males and females, and the ratios of these averages, for each of these four scenarios.
\subsubsection{Scenario 0}
\label{sec:WHBH}
\begin{figure}[H]
    \centering
    \subfigure[Female TB cases.]{\includegraphics[width=0.49\linewidth]{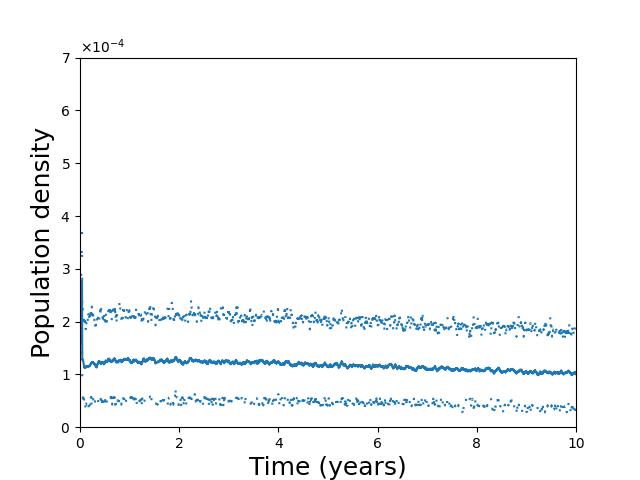} \label{fig:IF WHBH}}
    \subfigure[Male TB cases.]{\includegraphics[width=0.49\linewidth]{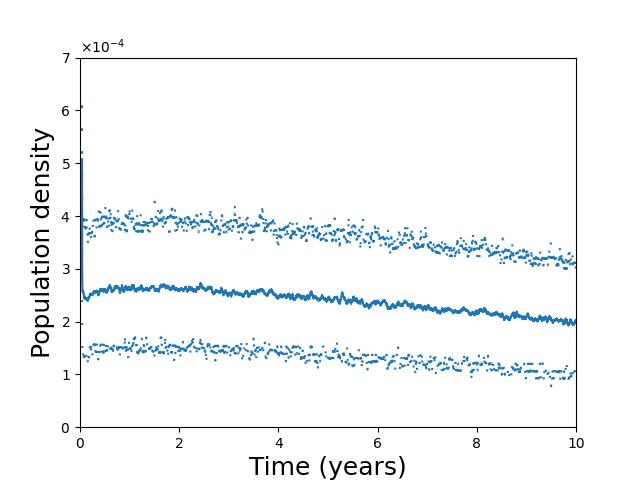}\label{fig:IM WHBH}}
    \subfigure[Male-to-female TB case ratio.]{\includegraphics[width=0.49\linewidth]{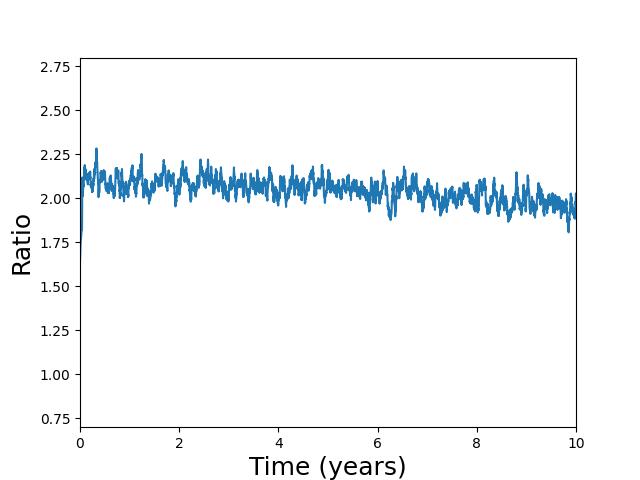}\label{fig:Ratio WHBH}}
    \caption{Plots showing a summary of the average densities of the Males With Active TB and Females With Active TB compartments, and the ratio of the average male cases to average female cases over a 10-years simulated time period, for Scenario 0. In Figures \subref{fig:IF WHBH} and \subref{fig:IM WHBH}, the solid line indicates the mean, and the dotted lines indicate the upper and lower bounds of the 95\% confidence interval. In Figure \subref{fig:Ratio WHBH}, the solid lines indicates the mean number of male TB cases divided by the mean number of female TB cases.}
    \label{fig:WHBH}
\end{figure}

A summary of the evolution of the densities for the Females With Active TB and Males With Active TB compartments, as well as the ratio of these two compartments, is shown in Figure \ref{fig:WHBH}. After an initial spike in the densities of both compartments, both quickly return to sizes close in value to their initial conditions (although the density of males with active TB slowly declines over time). The ratio of males with active TB to females with active TB jumps up to around 2 initially, then slowly declines over time. We ran additional simulations of 20-year and 40-year time periods to check whether this ratio would continue to decline over time. The ratio levels off and is substantially different from 1 after the simulated 20 years (see Appendix \ref{appendix:20 years}) and this pattern is maintained over 40 years (see Appendix \ref{appendix:40 years}). Hence, when we stratify for sex at the within-host scale and gender at the between-host scale, we do observe a clear difference between males and females, with males more likely than females to have active TB, as expected. We also ran additional simulations with equal numbers of males and females in each infection state initially, to check that the initial conditions weren't the primary reason for this occurring; in these simulations, the ratio quickly moved up from 1 and remained above 1 for the remaining simulation time, confirming our hypothesis (see Appendix \ref{appendix:same IC} for more details).
\subsubsection{Scenario 1}
\label{sec:WHnoBH}
\begin{figure}[H]
    \centering
    \subfigure[Female TB cases.]{\includegraphics[width=0.49\linewidth]{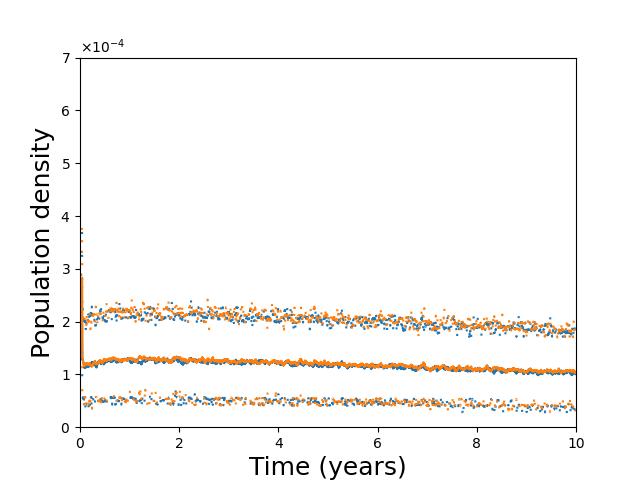}\label{fig:IF noBH}}
    \subfigure[Male TB cases.]{\includegraphics[width=0.49\linewidth]{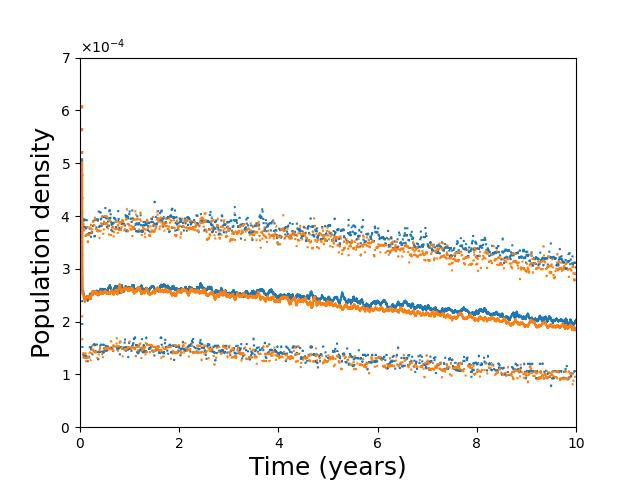}\label{fig:IM noBH}}
    \subfigure[Male-to-female TB case ratio.]{\includegraphics[width=0.49\linewidth]{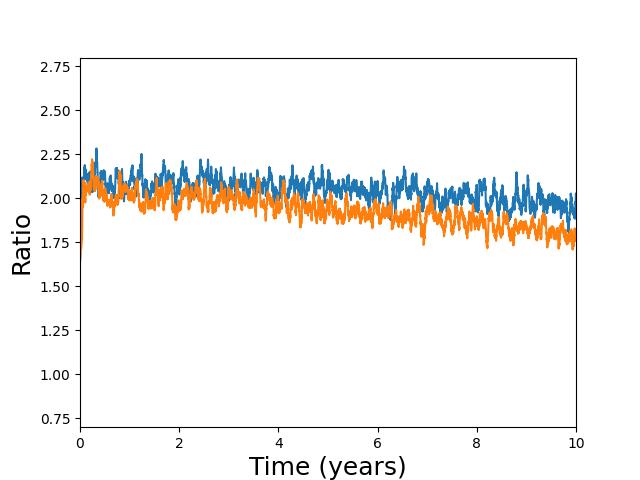}\label{fig:Ratio noBH}}
    \caption{Plots showing a summary of the average densities of the Males With Active TB and Females With Active TB compartments, and the ratio of the average male cases to average female cases over a 10-years simulated time period, for Scenario 1. In Figures \subref{fig:IF noBH} and \subref{fig:IM noBH}, the solid line indicates the mean, and the dotted lines indicate the upper and lower bounds of the 95\% confidence interval. In Figure \subref{fig:Ratio noBH}, the solid lines indicates the mean number of male TB cases divided by the mean number of female TB cases. In all sub-figures, the blue lines indicate Scenario 0, where differences are incorporated between males and females at both scales, for comparison; the orange lines indicate Scenario 1.}
    \label{fig:noBH}
\end{figure}

A summary of the evolution of the densities for the Females With Active TB and Males With Active TB compartments, as well as the ratio of these two compartments, is shown in Figure \ref{fig:noBH}. To remove the between-host scale differences, we set the value of $\theta_2$ (the susceptibility multiplier based on the genders of the potential infector and infectee) equal to 0.005, the average value used for $\theta_2$ for all infector-infectee gender combinations in Section \ref{sec:WHBH}, for female-to-female transmission, female-to-male transmission, male-to-female transmission and male-to-male transmission. This was to ensure that sex assortativity of contacts would not impact upon transmission probability in this counterfactual scenario. Compared to Scenario 0, a decrease in the ratio of the mean males with active TB density to the mean females with active TB density is observed In Scenario 1. We would expect this: if female and male infectors are equally likely to transmit to either gender, then male cases should decrease and female cases should increase, as we observed. The ratio in Scenario 1 continues to be less than the ratio in Scenario 0 if the simulations are run for 20 years or for 40 years instead of 10 (see Appendices \ref{appendix:20 years} and \ref{appendix:40 years}), and when equal numbers of males and females were in each infection state initially (see Appendix \ref{appendix:same IC}). However, the ratio does not drop to 1 over either 20 years or 40 years, and increases above 1 with equal initial conditions, suggesting differences at the within-host scale only are not entirely responsible for the male TB burden.
\subsubsection{Scenario 2}
\label{sec:noWHBH}
\begin{figure}[H]
    \centering
    \subfigure[Female TB cases.]{\includegraphics[width=0.49\linewidth]{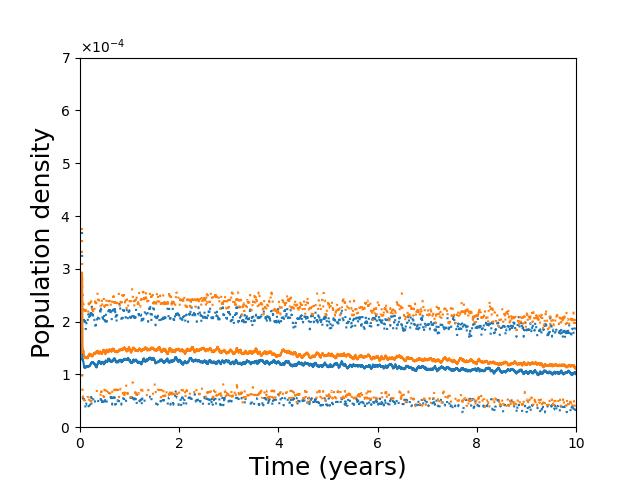}\label{fig:IF noWH}}
    \subfigure[Male TB cases.]{\includegraphics[width=0.49\linewidth]{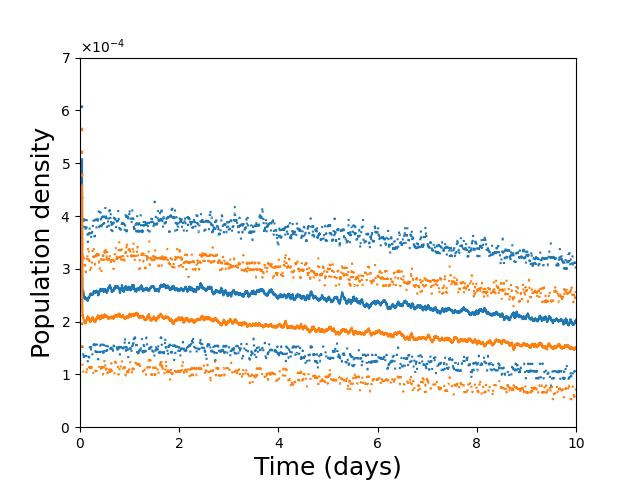}\label{fig:IM noWH}}
    \subfigure[Male-to-female TB case ratio.]{\includegraphics[width=0.49\linewidth]{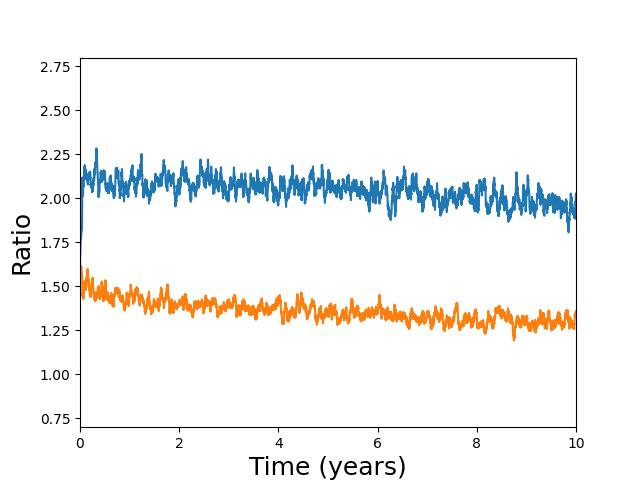}\label{fig:Ratio noWH}}
    \caption{Plots showing a summary of the average densities of the Males With Active TB and Females With Active TB compartments, and the ratio of the average male cases to average female cases over a 10-years simulated time period, for Scenario 2. In Figures \subref{fig:IF noWH} and \subref{fig:IM noWH}, the solid line indicates the mean, and the dotted lines indicate the upper and lower bounds of the 95\% confidence interval. In Figure \subref{fig:Ratio noWH}, the solid lines indicates the mean number of male TB cases divided by the mean number of female TB cases. In all sub-figures, the blue lines indicate Scenario 0, where differences are incorporated between males and females at both scales, for comparison; the orange lines indicate Scenario 2.}
    \label{fig:noWH}
\end{figure}

A summary of the evolution of the densities for the Females With Active TB and Males With Active TB compartments, as well as the ratio of these two compartments, is shown in Figure \ref{fig:noWH}. To remove the within-host scale differences, we used the ODEs learnt from the WHIDM dissemination simulations generated using the neutral parameter set for both males and females; we set the probability of cavitary TB (given an individual has active TB) equal to 0.4, the average probability of cavitary TB for males and females in Section \ref{sec:WHBH}, for both males and females. In addition, we set $\theta_6$ (the rate of transitioning from latent TB to active TB) equal to 0.00009 units, the average of the rates of transitioning from Latent TB to Active TB for males and females in Section \ref{sec:WHBH}, for both males and females. Compared to Scenario 0, a decrease in the ratio of the mean males with active TB density to the mean females with active TB density is observed in Scenario 2 (see Figure \ref{fig:Ratio noWH}). We would expect this: if the typical immune response, the probability of cavitation and the probability of progressing from Latent TB to Active TB are equivalent for males and females, then male cases should decrease and female cases should increase in this counterfactual, as we observed. The ratio in Scenario 2 continues to be less than the ratio in Scenario 0 if the simulations are run for 20 years or 40 years instead of 10 (see Appendices \ref{appendix:20 years} and \ref{appendix:40 years}), and when equal numbers of males and females were in each infection state initially (see Appendix \ref{appendix:same IC}). Although the ratio drops close to 1 over 20 years, it does not appear to reach 1; that continues to be true over 40 years. Similarly, although the ratio stays close to 1 with equal initial conditions, it tends to remain slightly above 1 for most of the simulated time period. These results suggest differences at the between-host scale only are not entirely responsible for the male TB burden.
\subsubsection{Scenario 3}
\label{sec:noWHnoBH}
\begin{figure}[H]
    \centering
    \subfigure[Female TB cases.]{\includegraphics[width=0.49\linewidth]{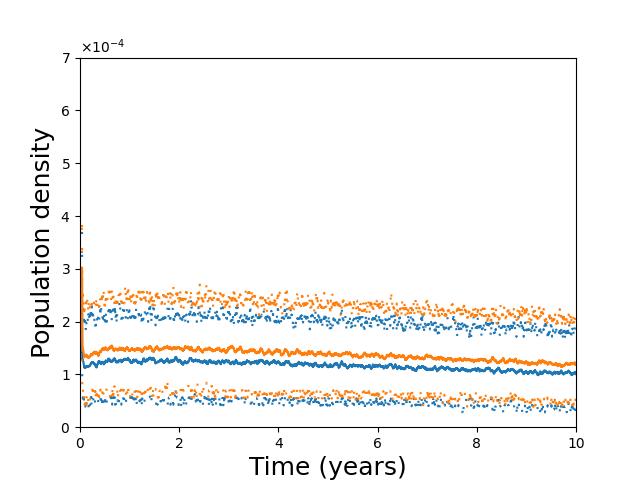}\label{fig:IF noWHnoBH}}
    \subfigure[Male TB cases.]{\includegraphics[width=0.49\linewidth]{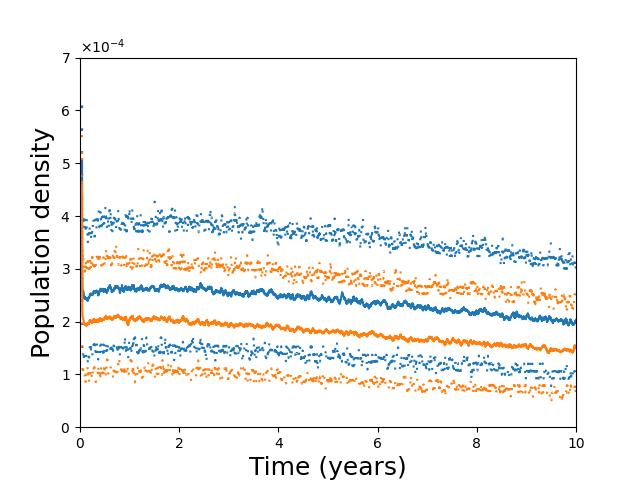}\label{fig:IM noWHnoBH}}
    \subfigure[Male-to-female TB case ratio.]{\includegraphics[width=0.49\linewidth]{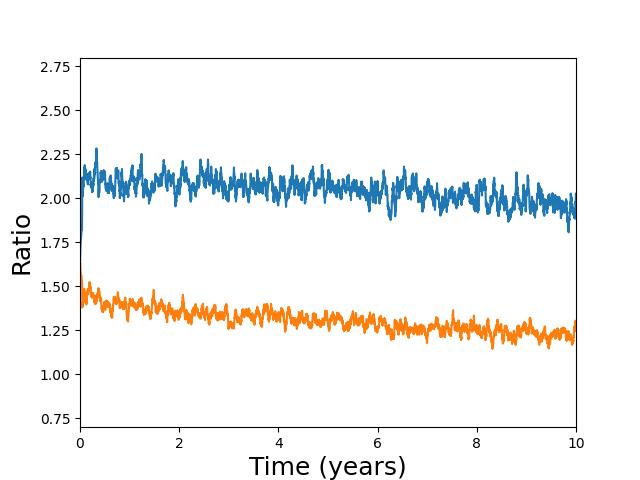}\label{fig:Ratio noWHnoBH}}
    \caption{Plots showing a summary of the average densities of the Males With Active TB and Females With Active TB compartments, and the ratio of the average male cases to average female cases over a 10-years simulated time period, for Scenario 3. In Figures \subref{fig:IF noWHnoBH} and \subref{fig:IM noWHnoBH}, the solid line indicates the mean, and the dotted lines indicate the upper and lower bounds of the 95\% confidence interval. In Figure \subref{fig:Ratio noWHnoBH}, the solid lines indicates the mean number of male TB cases divided by the mean number of female TB cases. In all sub-figures, the blue lines indicate Scenario 0, where differences are incorporated between males and females at both scales, for comparison; the orange lines indicate Scenario 3.}
    \label{fig:noWHnoBH}
\end{figure}

A summary of the evolution of the densities for the Females With Active TB and Males With Active TB compartments, as well as the ratio of these two compartments, is shown in Figure \ref{fig:noWHnoBH}. To remove differences at both scales, we incorporated all the changes outlined in Sections \ref{sec:WHnoBH} and \ref{sec:noWHBH}. Compared to Scenario 0, a decrease in the ratio of the mean males with active TB density to the mean females with active TB density is observed in Scenario 3. We would expect this for the same reasons as outlined in Sections \ref{sec:WHnoBH} and \ref{sec:noWHBH}. The ratio drops and remains close to 1 over 20 years and over 40 years (see Appendices \ref{appendix:20 years} and \ref{appendix:40 years}), and stays around 1 when equal numbers of males and females were in each infection state initially (see Appendix \ref{appendix:same IC}), as we would expect. 
\section{Discussion}
\label{sec:discussion}
In this paper, we have provided a proof-of-concept for a multiscale modelling approach that allows us to both capture the output of within-host agent-based models with the equation learning algorithm, and efficiently model infectious disease systems that capture transient within-host dynamics and stochastic transmission events with the adapted WHD-HDD multiscale framework. This framework provides a powerful tool for modelling multiscale infectious disease dynamics, by successfully combining multiple useful but complex modelling ideas into one unified approach, and could be adapted to represent other infectious diseases besides tuberculosis.\par
We have attempted to determine the impact that differences between sexes (at the within-host scale) and differences between genders (at the between-host scale) can have in causing males to suffer from a higher burden of TB than females. To do this, we have developed an approach to model TB infectious disease dynamics across multiple scales in seven steps. First, we have run simulations of the within-host dynamics of TB using an agent-based model, capturing host-pathogen interactions based on the immune responses of both sexes (as well as an average immune response for counterfactual scenarios). Second, we have approximated the derivatives of the variables in WHIDM to be learnt. Third, we have constructed a library of possible terms to include in learnt ODEs that capture the average behaviour of this model. Fourth, we have inferred these learnt ODEs using regression with upper and lower bounds. Fifth, we have incorporated these learnt ODEs into a multiscale modelling framework, which we have adapted to be suitable for typical TB infectious disease dynamics. Sixth, we have represented the host demographic dynamics using ODEs to capture births and non-disease-induced mortality. Finally, we have stochastically linked the within-host dynamics to the host demographic dynamics with coupling functions. We have subsequently simulated four scenarios (with differences at the two scales retained in some scenarios and discarded in others - see Figure \ref{fig:counterfactuals}) and compared the results found in these four scenarios.\par
Our results suggest that differences at both scales are important in causing the higher male TB burden. At the within-host scale, differences in immune response, the probability of cavitation, and the probability of progressing from latent infection to active disease all appear to contribute towards this burden. At the between-host scale, incorporating sex assortativity in contacts led to a greater male-to-female TB case ratio than when it was removed, which would imply it is a factor in causing the higher TB burden in males.\par 
Our results are in agreement with observations made by \citeauthor{Doran2026} in an earlier study of the impact of gender on tuberculosis transmission in Uganda \cite{Doran2026}. In that article, it was found that setting within-host factors, such as probability of progression from latent infection to active disease and probability of cavitary TB, equal between males and females led to the greatest reduction in the male-to-female TB case ratio. Similarly, in this paper, although it appears that both within-host and between-host factors play a part in causing the higher burden of TB cases in males, it seems that the within-host factors have a larger impact. One reason for this might be that we do not have as many between-host factors incorporated into our modelling framework. The only way we influenced the between-host scale was by altering the susceptibility multiplier in transmission events. It should be noted that the population we have simulated is hypothetical (and loosely based on the population of Kampala, Uganda, which we have investigated in previous work \cite{Doran2026}). Therefore, although our results agree with previous research by \citeauthor{Doran2026} \cite{Doran2026}, it may be that certain assumptions we have made are unrealistic for other populations. For instance, assuming an equal diagnosis rate or an equal treatment completion rate between genders may not hold in other populations with a significant TB burden. If we were to model a real population affected by TB and incorporate real-world data, having different parameter values for some of these factors as well as the susceptibility multiplier may lead to larger differences between the scenario where both within-host and between-host differences are implemented, and the scenario with only within-host differences implemented. Future work will involve refining the estimates of the parameters we have used here, but it should be noted that the primary aim of this paper was to provide proof-of-concept for this multiscale modelling framework.\par
In determining the coupling functions in this model, we have made multiple assumptions about TB infectious disease dynamics. We have assumed that the transition times are exponentially distributed for all of the non-transmission and non-disease-induced mortality transitions between compartments, that infectiousness is a Hill function of pathogen load, and that disease-induced mortality is another Hill function of pathogen load. In reality, there is little data available to support these assumptions, and our coupling functions may need to be reconsidered if additional data comes to light. The issue of how best to link the scales in multiscale infectious disease modelling has been an ongoing problem in recent years \cite{Handel2015}.\par
We have previously conduced a consistency analysis for WHIDM \cite{Doran2026a}, but not for the whole multiscale framework presented in this paper. Therefore, although we have assumed that 300 simulations should be sufficient for the averages and confidence intervals in our results to be consistent, a formal consistency analysis could be conducted to determine if this assumption was valid in this full framework. We would follow the methodology presented by \citeauthor{Hamis2021} when completing this analysis \cite{Hamis2021}, as in our previous work \cite{Doran2026a,Doran2026}. In addition, we could add additional features to the model to allow for further insights into the potential factors leading to a higher burden of TB cases in males. This could include considering more immune cell types in WHIDM simulations - for example, differentiating T-cells and not just using a generic T-cell - and further stratification by age as well as gender. Alternatives to the equation learning algorithm first presented by \citeauthor{Nardini2021} \cite{Nardini2021} and used in this paper could also be considered. For example, \citeauthor{Burrage2025} performed equation learning on their agent-based cancer model using sets of chemical reactions that follow the Law of Mass Action \cite{Burrage2025}. In their paper, conservation of mass was enforced by using the Lasso regression approach and adding an extra loss term. Although we were satisfied that using least squares regression with upper and lower bounds led to good-fitting learnt equations, we could utilise this approach in future work and see if it leads to different equations being learnt that provide a better fit. Finally, our intention would be to generalise the approach taken in this work and apply it to other similar infectious diseases: we believe this framework could be used to model other disease spread via airborne bacteria or viruses.
\section*{Acknowledgements}
JD thanks John Nardini for his help in understanding the equation learning code. This work makes use of the Nimbus cloud computer, and the authors gratefully acknowledge the University of Bath's Research Computing Group (\url{https://doi.org/10.15125/b6cd-s854}) for their support in this work.
\section*{Funding}
This work was supported by the Medical Research Council, United Kingdom [grant number MR/Y010124/1].
\section*{CRediT authorship contribution statement}
\begin{itemize}
    \item \textbf{James W. G. Doran:} Conceptualization, Formal analysis, Investigation, Software, Validation, Visualization, Writing - original draft, Writing - review \& editing.
    \item \textbf{Cameron A. Smith:} Software, Writing - review \& editing.
    \item \textbf{Christian A. Yates:} Conceptualization, Supervision, Writing - review \& editing.
    \item \textbf{Ruth Bowness:} Conceptualization, Funding acquisition, Supervision, Writing - review \& editing.
\end{itemize}
\section*{Declaration of interests}
The authors declare the following financial interests/personal relationships which may be considered as potential competing interests: Ruth Bowness reports a relationship with Medical Research Council that includes: funding grants (grant number MR/Y010124/1).
\section*{Data statement}
The model code can be found at the following links: \url{https://github.com/jwgd93/Equation-Learning-for-multiscale-models-of-infectious-diseases} for the adapted WHD-HDD code used to simulate the multiscale infectious disease dynamics; \url{https://github.com/Ruth-Bowness-Group/Equation-Learning-for-multiscale-models-of-infectious-diseases-WHIDM} for WHIDM. The data presented in this article can be found at the following links: \url{https://doi.org/10.5281/zenodo.18970202} for the WHIDM simulation data; \url{https://doi.org/10.5281/zenodo.18970485} for the WHD-HDD simulation data. These links will become live after publication.
\raggedright \printbibliography

@Article{Nhamoyebonde2014,
  author    = {Nhamoyebonde, Shepherd and Leslie, Alasdair},
  journal   = {The Journal of infectious diseases},
  title     = {Biological differences between the sexes and susceptibility to tuberculosis},
  year      = {2014},
  publisher = {JSTOR},
}

@Article{Gupta2022,
  author    = {Gupta, Manish and Srikrishna, Geetha and Klein, Sabra L. and Bishai, William R.},
  journal   = {Trends in immunology},
  title     = {Genetic and hormonal mechanisms underlying sex-specific immune responses in tuberculosis},
  year      = {2022},
  number    = {8},
  volume    = {43},
  publisher = {Elsevier},
}

@Article{Shrivastava2021,
  author  = {Shrivastava, Parul and Bagchi, Tamishraha},
  journal = {Chemical Biology Letters},
  title   = {Testosterone in pathogenesis of tuberculosis},
  year    = {2021},
  number  = {4},
  volume  = {8},
}

@Article{Smith2025,
  author    = {Smith, Cameron A. and Ashby, Ben},
  journal   = {Journal of Theoretical Biology},
  title     = {Efficient coupling of within-and between-host infectious disease dynamics},
  year      = {2025},
  publisher = {Elsevier},
}

@Article{Nardini2021,
  author    = {Nardini, John T. and Baker, Ruth E. and Simpson, Matthew J. and Flores, Kevin B.},
  journal   = {Journal of the Royal Society Interface},
  title     = {Learning differential equation models from stochastic agent-based model simulations},
  year      = {2021},
  number    = {176},
  volume    = {18},
  publisher = {The Royal Society},
}

@Article{Urdahl2011,
  author    = {Urdahl, K. B. and Shafiani, S. and Ernst, J. D.},
  journal   = {Mucosal immunology},
  title     = {Initiation and regulation of T-cell responses in tuberculosis},
  year      = {2011},
  number    = {3},
  volume    = {4},
  publisher = {Elsevier},
}

@Article{Garira2018,
  author    = {Garira, Winston},
  journal   = {Infectious Disease Modelling},
  title     = {A primer on multiscale modelling of infectious disease systems},
  year      = {2018},
  volume    = {3},
  publisher = {Elsevier},
}

@Article{Doran2023,
  author    = {Doran, James W. G. and Thompson, Robin N. and Yates, Christian A. and Bowness, Ruth},
  journal   = {Epidemics},
  title     = {Mathematical methods for scaling from within-host to population-scale in infectious disease systems},
  year      = {2023},
  publisher = {Elsevier},
}

@Article{Garira2019,
  author    = {Garira, Winston},
  journal   = {Scientific Reports},
  title     = {The replication-transmission relativity theory for multiscale modelling of infectious disease systems},
  year      = {2019},
  number    = {1},
  volume    = {9},
  publisher = {Nature Publishing Group},
}

@Article{Handel2015,
  author    = {Handel, Andreas and Rohani, Pejman},
  journal   = {Philosophical Transactions of the Royal Society B: Biological Sciences},
  title     = {Crossing the scale from within-host infection dynamics to between-host transmission fitness: a discussion of current assumptions and knowledge},
  year      = {2015},
  number    = {1675},
  volume    = {370},
  publisher = {The Royal Society},
}

@Article{Childs2019,
  author    = {Childs, Lauren M. and El Moustaid, Fadoua and Gajewski, Zachary and Kadelka, Sarah and Nikin-Beers, Ryan and Smith Jr, John W. and Walker, Melody and Johnson, Leah R.},
  journal   = {PeerJ},
  title     = {Linked within-host and between-host models and data for infectious diseases: a systematic review},
  year      = {2019},
  volume    = {7},
  publisher = {PeerJ Inc.},
}

@Article{WHO2024,
  author = {{World Health Organization}},
  title  = {Global tuberculosis report 2024},
  year   = {2024},
  url    = {https://iris.who.int/bitstream/handle/10665/379339/9789240101531-eng.pdf?sequence=1},
}

@Article{Childs2015,
  author    = {Childs, Lauren M. and Abuelezam, Nadia N. and Dye, Christopher and Gupta, Sunetra and Murray, Megan B. and Williams, Brian G. and Buckee, Caroline O.},
  journal   = {Epidemics},
  title     = {Modelling challenges in context: lessons from malaria, HIV, and tuberculosis},
  year      = {2015},
  volume    = {10},
  publisher = {Elsevier},
}

@Article{Ahmad2011,
  author    = {Ahmad, Suhail},
  journal   = {Clinical and Developmental Immunology},
  title     = {Pathogenesis, immunology, and diagnosis of latent Mycobacterium tuberculosis infection},
  year      = {2011},
  volume    = {2011},
  publisher = {Hindawi},
}

@InProceedings{Glaziou2018,
  author    = {Glaziou, P. and Floyd, K. and Raviglione, M. C.},
  booktitle = {Seminars in Respiratory and Critical Care Medicine},
  title     = {Global Epidemiology of Tuberculosis.},
  year      = {2018},
  number    = {3},
  volume    = {39},
}

@Article{Bowness2018,
  author   = {Ruth Bowness and Mark A.J. Chaplain and Gibin G. Powathil and Stephen H. Gillespie},
  journal  = {Journal of Theoretical Biology},
  title    = {Modelling the effects of bacterial cell state and spatial location on tuberculosis treatment: Insights from a hybrid multiscale cellular automaton model},
  year     = {2018},
  issn     = {0022-5193},
  volume   = {446},
  url      = {https://www.sciencedirect.com/science/article/pii/S0022519318301206},
}

@Article{Pereira2021,
  author    = {Pereira, Rafael S. and Bauch, Chris T. and Penna, Thadeu J. P. and Espíndola, Aquino L.},
  journal   = {International Journal of Modern Physics C},
  title     = {A nested model for tuberculosis: Combining within-host and between-host processes in a single framework},
  year      = {2021},
  number    = {12},
  volume    = {32},
  publisher = {World Scientific},
}

@Article{KisselevskayaBabinina2018,
  author    = {Kisselevskaya-Babinina, Viktoriya Yaroslavovna and Sannikova, Tat'yana Evgen'evna and Romanyukha, Aleksei Alekseevich and Karkach, Arsenii Sergeevich},
  journal   = {Matematicheskaya Biologiya i Bioinformatika},
  title     = {Modeling of gender differences in tuberculosis prevalence},
  year      = {2018},
  number    = {2},
  volume    = {13},
  publisher = {Institute of Mathematical Problems of Biology, Russian Academy of Sciences},
}

@Article{Kubjane2023,
  author    = {Kubjane, Mmamapudi and Cornell, Morna and Osman, Muhammad and Boulle, Andrew and Johnson, Leigh F.},
  journal   = {Scientific Reports},
  title     = {Drivers of sex differences in the South African adult tuberculosis incidence and mortality trends, 1990–2019},
  year      = {2023},
  number    = {1},
  volume    = {13},
  publisher = {Nature Publishing Group UK London},
}

@Article{Wang2024,
  author    = {Wang, Si and Cao, Hui},
  journal   = {Journal of Biological Dynamics},
  title     = {The dynamics of tuberculosis transmission model with different genders},
  year      = {2024},
  number    = {1},
  volume    = {18},
  publisher = {Taylor & Francis},
}

@Article{Garira2020,
  author    = {Garira, Winston},
  journal   = {PLoS computational biology},
  title     = {The research and development process for multiscale models of infectious disease systems},
  year      = {2020},
  number    = {4},
  volume    = {16},
  publisher = {Public Library of Science San Francisco, CA USA},
}

@Article{Hamis2021,
  author    = {Hamis, Sara and Stratiev, Stanislav and Powathil, Gibin G.},
  journal   = {THE PHYSICS OF CANCER: Research Advances},
  title     = {Uncertainty and sensitivity analyses methods for agent-based mathematical models: An introductory review},
  year      = {2021},
  publisher = {World Scientific},
}

@Article{OGarra2013,
  author    = {O'Garra, Anne and Redford, Paul S. and McNab, Finlay W. and Bloom, Chloe I. and Wilkinson, Robert J. and Berry, Matthew P. R.},
  journal   = {Annual review of immunology},
  title     = {The immune response in tuberculosis},
  year      = {2013},
  number    = {1},
  volume    = {31},
  publisher = {Annual Reviews},
}

@Article{Schluger1998,
  author    = {Schluger, Neil W. and Rom, William N.},
  journal   = {American journal of respiratory and critical care medicine},
  title     = {The host immune response to tuberculosis},
  year      = {1998},
  number    = {3},
  volume    = {157},
  publisher = {American Thoracic Society New York, NY},
}

@Article{Lipworth2016,
  author   = {S. Lipworth and R.J.H. Hammond and V.O. Baron and Yanmin. Hu and A. Coates and S.H. Gillespie},
  journal  = {Tuberculosis},
  title    = {Defining dormancy in mycobacterial disease},
  year     = {2016},
  issn     = {1472-9792},
  volume   = {99},
  url      = {https://www.sciencedirect.com/science/article/pii/S1472979216300658},
}

@Article{Hammond2015,
  author    = {Hammond, Robert J. H. and Baron, Vincent O. and Oravcova, Katarina and Lipworth, Sam and Gillespie, Stephen H.},
  journal   = {Journal of Antimicrobial Chemotherapy},
  title     = {Phenotypic resistance in mycobacteria: is it because I am old or fat that I resist you?},
  year      = {2015},
  number    = {10},
  volume    = {70},
  publisher = {Oxford University Press},
}

@Article{Balogun2021,
  author  = {Balogun, O. O. and Fawole, A. and Osemwinyen, E. and Balogun, B.},
  journal = {J Infect Dis Epidemiol},
  title   = {Predictors of Pulmonary Cavitation among Tuberculosis Patients},
  year    = {2021},
  volume  = {7},
}

@Article{Urbanowski2020,
  author    = {Urbanowski, Michael E. and Ordonez, Alvaro A. and Ruiz-Bedoya, Camilo A. and Jain, Sanjay K. and Bishai, William R.},
  journal   = {The Lancet Infectious Diseases},
  title     = {Cavitary tuberculosis: the gateway of disease transmission},
  year      = {2020},
  number    = {6},
  volume    = {20},
  publisher = {Elsevier},
}

@Article{Kaona2004,
  author    = {Kaona, Frederick A. D. and Tuba, Mary and Siziya, Seter and Sikaona, Lenganji},
  journal   = {BMC Public health},
  title     = {An assessment of factors contributing to treatment adherence and knowledge of TB transmission among patients on TB treatment},
  year      = {2004},
  number    = {1},
  volume    = {4},
  publisher = {Springer},
}

@Article{Yang2014,
  author    = {Yang, Wei-Teng and Gounder, Celine R. and Akande, Tokunbo and De Neve, Jan-Walter and McIntire, Katherine N. and Chandrasekhar, Aditya and de Lima Pereira, Alan and Gummadi, Naveen and Samanta, Santanu and Gupta, Amita},
  journal   = {Tuberculosis research and treatment},
  title     = {Barriers and delays in tuberculosis diagnosis and treatment services: does gender matter?},
  year      = {2014},
  number    = {1},
  volume    = {2014},
  publisher = {Wiley Online Library},
}

@Article{Karim2007,
  author    = {Karim, Fazlul and Islam, Md Akramul and Chowdhury, A. M. R. and Johansson, Eva and Diwan, Vinod K.},
  journal   = {Health policy and planning},
  title     = {Gender differences in delays in diagnosis and treatment of tuberculosis},
  year      = {2007},
  number    = {5},
  volume    = {22},
  publisher = {Oxford University Press},
}

@Article{Hof2010,
  author  = {van den Hof, Susan and Najlis, Camilo Antillon and Bloss, Emily and Straetemans, Masja},
  journal = {Report prepared for Tuberculosis Control Programme (TB CAP) September},
  title   = {A systematic review on the role of gender in tuberculosis control},
  year    = {2010},
}

@Article{Horton2020,
  author  = {Horton, Katherine C. and Hoey, Anne L. and Béraud, Guillaume and Corbett, Elizabeth L. and White, Richard G.},
  journal = {Emerging Infectious Diseases},
  title   = {Systematic review and meta-analysis of sex differences in social contact patterns and implications for tuberculosis transmission and control},
  year    = {2020},
  number  = {5},
  volume  = {26},
}

@Article{Ray2009,
  author    = {Ray, J. Christian J. and Flynn, JoAnne L. and Kirschner, Denise E.},
  journal   = {The Journal of Immunology},
  title     = {Synergy between individual TNF-dependent functions determines granuloma performance for controlling Mycobacterium tuberculosis infection},
  year      = {2009},
  number    = {6},
  volume    = {182},
  publisher = {Oxford University Press},
}

@Article{Doran2026,
  author  = {Doran, James W. G. and Mujuni, Dennis and Gallagher, Kit and Yates, Christian A. and Bowness, Ruth},
  journal = {arXiv preprint arXiv:2601.09813},
  title   = {An agent-based modelling approach to investigate the impact of gender on tuberculosis transmission in Uganda},
  year    = {2026},
}

@Article{Shorten2013,
  author    = {Shorten, Robert J. and McGregor, Alistair C. and Platt, S. and Jenkins, Claire and Lipman, M. C. I. and Gillespie, S. H. and Charalambous, B. M. and McHugh, T. D.},
  journal   = {Journal of Antimicrobial Chemotherapy},
  title     = {When is an outbreak not an outbreak? Fit, divergent strains of Mycobacterium tuberculosis display independent evolution of drug resistance in a large London outbreak},
  year      = {2013},
  number    = {3},
  volume    = {68},
  publisher = {Oxford University Press},
}

@Article{HendonDunn2016,
  author    = {Hendon-Dunn, Charlotte Louise and Doris, Kathryn Sarah and Thomas, Stephen Richard and Allnutt, Jonathan Charles and Marriott, Alice Ann Neville and Hatch, Kim Alexandra and Watson, Robert James and Bottley, Graham and Marsh, Philip David and Taylor, Stephen Charles},
  journal   = {Antimicrobial agents and chemotherapy},
  title     = {A flow cytometry method for rapidly assessing Mycobacterium tuberculosis responses to antibiotics with different modes of action},
  year      = {2016},
  number    = {7},
  volume    = {60},
  publisher = {American Society for Microbiology 1752 N St., NW, Washington, DC},
}

@Article{Doran2026a,
  author  = {Doran, James W. G. and Rowlatt, Christopher F. and Powathil, Gibin G. and Bowness, Ruth and Yates, Christian A.},
  journal = {arXiv preprint arXiv:2602.24258},
  title   = {A model of tuberculosis progression using CompuCell3D},
  year    = {2026},
}

@Article{Joslyn2022,
  author    = {Joslyn, Louis R. and Linderman, Jennifer J. and Kirschner, Denise E.},
  journal   = {Journal of theoretical biology},
  title     = {A virtual host model of Mycobacterium tuberculosis infection identifies early immune events as predictive of infection outcomes},
  year      = {2022},
  volume    = {539},
  publisher = {Elsevier},
}

@Article{Kiazyk2017,
  author  = {Kiazyk, S. and Ball, T. B.},
  journal = {Canada Communicable Disease Report},
  title   = {Latent tuberculosis infection: An overview},
  year    = {2017},
  number  = {3-4},
  volume  = {43},
}

@Article{VanFurth1973,
  author    = {Van Furth, Ralph and Diesselhoff-den Dulk, Martina M. C. and Mattie, Herman},
  journal   = {The Journal of Experimental Medicine},
  title     = {Quantitative study on the production and kinetics of mononuclear phagocytes during an acute inflammatory reaction},
  year      = {1973},
  number    = {6},
  volume    = {138},
  publisher = {Rockefeller University Press},
}

@Article{SegoviaJuarez2004,
  author    = {Segovia-Juarez, Jose L. and Ganguli, Suman and Kirschner, Denise},
  journal   = {Journal of Theoretical Biology},
  title     = {Identifying control mechanisms of granuloma formation during M. tuberculosis infection using an agent-based model},
  year      = {2004},
  number    = {3},
  volume    = {231},
  publisher = {Elsevier},
}

@Article{Sprent1993,
  author    = {Sprent, Jonathan},
  journal   = {Current Opinion in Immunology},
  title     = {Lifespans of naive, memory and effector lymphocytes},
  year      = {1993},
  number    = {3},
  volume    = {5},
  publisher = {Elsevier},
}

@Article{Melsew2018,
  author    = {Melsew, Y. A. and Doan, T. N. and Gambhir, M. and Cheng, A. C. and McBryde, E. and Trauer, J. M.},
  journal   = {Epidemiology \& Infection},
  title     = {Risk factors for infectiousness of patients with tuberculosis: a systematic review and meta-analysis},
  year      = {2018},
  number    = {3},
  volume    = {146},
  publisher = {Cambridge University Press},
}

@Article{Miller2021,
  author    = {Miller, Paige B. and Zalwango, Sarah and Galiwango, Ronald and Kakaire, Robert and Sekandi, Juliet and Steinbaum, Lauren and Drake, John M. and Whalen, Christopher C. and Kiwanuka, Noah},
  journal   = {BMC Infectious Diseases},
  title     = {Association between tuberculosis in men and social network structure in Kampala, Uganda},
  year      = {2021},
  volume    = {21},
  publisher = {Springer},
}

@Article{Suthar2016,
  author  = {Suthar, A. B. and Zachariah, R. and Harries, A. D.},
  journal = {International Journal of Tuberculosis and Lung Disease},
  title   = {Ending tuberculosis by 2030: can we do it?},
  year    = {2016},
  number  = {9},
  volume  = {20},
}

@Article{Prats2016,
  author    = {Prats, Clara and Montañola-Sales, Cristina and Gilabert-Navarro, Joan F. and Valls, Joaquim and Casanovas-Garcia, Josep and Vilaplana, Cristina and Cardona, Pere-Joan and López, Daniel},
  journal   = {Frontiers in Microbiology},
  title     = {Individual-based modeling of tuberculosis in a user-friendly interface: understanding the epidemiological role of population heterogeneity in a city},
  year      = {2016},
  volume    = {6},
  publisher = {Frontiers},
}

@Article{Rodriguez2023,
  author  = {Rodriguez, C. A. and Leavitt, S. V. and Bouton, T. C. and Horsburgh, C. R. and Zur Wiesch, P. Abel and Nichols, B. and Jenkins, H. E. and White, L. F.},
  journal = {The International Journal of Tuberculosis and Lung Disease},
  title   = {Survival of people with untreated TB: effects of time, geography and setting},
  year    = {2023},
  number  = {9},
  volume  = {27},
}

@Article{Zhang2016,
  author    = {Zhang, Liqun and Pang, Yu and Yu, Xia and Wang, Yufeng and Lu, Jie and Gao, Mengqiu and Huang, Hairong and Zhao, Yanlin},
  journal   = {Emerging microbes \& infections},
  title     = {Risk factors for pulmonary cavitation in tuberculosis patients from China},
  year      = {2016},
  number    = {1},
  volume    = {5},
  publisher = {Taylor & Francis},
}

@Article{Burrage2025,
  author    = {Burrage, Kevin and Burrage, Pamela M. and Kreikemeyer, Justin N. and Uhrmacher, Adelinde M. and Weerasinghe, Hasitha N.},
  journal   = {Frontiers in Applied Mathematics and Statistics},
  title     = {Learning surrogate equations for the analysis of an agent-based cancer model},
  year      = {2025},
  volume    = {11},
  publisher = {Frontiers Media SA},
}

@Article{Houben2016,
  author    = {Houben, Rein M. G. J. and Dodd, Peter J.},
  journal   = {PLOS Medicine},
  title     = {The global burden of latent tuberculosis infection: a re-estimation using mathematical modelling},
  year      = {2016},
  number    = {10},
  volume    = {13},
  publisher = {Public Library of Science San Francisco, CA USA},
}

@Article{Osei2015,
  author    = {Osei, Eric and Akweongo, Patricia and Binka, Fred},
  journal   = {BMC public health},
  title     = {Factors associated with DELAY in diagnosis among tuberculosis patients in Hohoe Municipality, Ghana},
  year      = {2015},
  number    = {1},
  volume    = {15},
  publisher = {Springer},
}

@Article{Asefa2014,
  author    = {Asefa, Anteneh and Teshome, Wondu},
  journal   = {PloS one},
  title     = {Total delay in treatment among smear positive pulmonary tuberculosis patients in five primary health centers, southern Ethiopia: a cross sectional study},
  year      = {2014},
  number    = {7},
  volume    = {9},
  publisher = {Public Library of Science San Francisco, USA},
}

@Book{WHO2010,
  author    = {World Health Organization, Stop T.B. Initiative (World Health Organization)},
  publisher = {{World Health Organization}},
  title     = {Treatment of tuberculosis: guidelines},
  year      = {2010},
}

@Article{HornaCampos2010,
  author  = {Horna-Campos, O. J. and Bedoya-Lama, A. and Romero-Sandoval, N. C. and Martin-Mateo, M.},
  journal = {The international journal of tuberculosis and lung disease},
  title   = {Risk of tuberculosis in public transport sector workers, Lima, Peru},
  year    = {2010},
  number  = {6},
  volume  = {14},
}

@Book{Tukey1977,
  author    = {Tukey, John Wilder},
  publisher = {Springer},
  title     = {Exploratory data analysis},
  year      = {1977},
  volume    = {2},
}

@Article{Gillespie2001,
  author    = {Gillespie, Daniel T.},
  journal   = {The Journal of Chemical Physics},
  title     = {Approximate accelerated stochastic simulation of chemically reacting systems},
  year      = {2001},
  number    = {4},
  volume    = {115},
  publisher = {AIP Publishing},
}
\newpage
\begin{appendices}
\section{Comparison of regression methods}
\label{appendix:regression comparison}
We used linear regression to infer the learnt equations for the seven cell types. We considered three methods. First, we considered the \texttt{lstsq} method in Python's \texttt{numpy.linalg} library. For example, the solution to Equation (\ref{eq:FGEB derivative}) found using this method, which minimises the mean squared error, is of the form:
\begin{equation}
    \hat{\xi} = \arg \min_{\xi \in \mathbb{R}^k}\left\{\frac{1}{n} \left\|\frac{d\bm{F_d}(t)}{dt}-\Theta \bm{\xi}\right\|_2^2\right\}.
\end{equation}
The second method we considered was the \texttt{Lasso} (least absolute shrinkage and selection operator) method from Python's \texttt{sklearn.linear\_model} library. For example, the solution to Equation (\ref{eq:FGEB derivative}) found using this method, which minimises the sum of the mean squared error and the regularisation term, is of the form:
\begin{equation}
    \hat{\xi} = \arg \min_{\xi \in \mathbb{R}^k}\left\{\frac{1}{2n}\left\|\frac{d\bm{F_d}(t)}{dt}-\Theta \bm{\xi}\right\|_2^2 + \lambda \left\|\bm{\xi}\right\|_1\right\}.
    \label{eq:lasso}
\end{equation}
The final method we considered, and ultimately ended up employing, was the \texttt{lsq\_linear} algorithm from Python's \texttt{scipy.optimize} library to solve the linear regression with upper and lower bounds imposed on the coefficients of the variables. For example, the solution to Equation (\ref{eq:FGEB derivative}) found using this method, which minimises the mean squared error subject to each coefficient being at least equal to a specified lower bound and at most equal to a specified upper bound, is of the form:
\begin{equation}
    \hat{\xi} = \arg \min_{\bm{b}_{l_i} \le \xi_i \le \bm{b}_{u_i}} \left\{\frac{1}{2n}\left\|\frac{d\bm{F_d}(t)}{dt}-\Theta \bm{\xi}\right\|_2^2\right\},
\end{equation}
where $\bm{b}_l,\bm{b}_u \in \mathbb{R}^k$ are vectors containing the lower and upper bounds for the coefficients to be learnt, respectively.\par
A comparison of the parameter values that each of the regression methods produced for the learned equations are shown below in Tables \ref{tab:regression comparison female} to \ref{tab:regression comparison neutral}. A comparison of the mean squared error for each of the cell types resulting from the parameters generated by each regression type are shown in Tables \ref{tab:regression MSE female} to \ref{tab:regression MSE neutral}. In the case of the Lasso regression, the penalisation term from Equation (\ref{eq:lasso}), $\lambda$, was set to $10^{-11}$, with a tolerance of $10^{-4}$, and the optimisation continued for a maximum of 30,000 iterations. In order to compute the mean squared errors, all ODE solutions were found using the \texttt{ode45} function on \texttt{MATLAB}. In the case of the standard least squares regression, the option to output non-negative solutions had to be specified for the solver to not produce an error.
\label{appendix:regression methods}
\begin{table}[H]
    \centering
    \begin{tabular}{|l|l|l|l|}
    \hline    
        \textbf{EQL parameter} & \textbf{Least squares} & \textbf{Lasso} & \textbf{Least squares with bounds}\\
        \hline
        $\xi_1$ & 0.63113 & 0.56517 & -0.65264\\
        $\xi_2$ & 0.0046 & 0.00567 & 0.02167\\
        $\xi_3$ & -0.09493 & -0.09354 & -0.02\\
        $\xi_4$ & -0.82003 & -0.55877 & -0.00001\\
        $\xi_5$ & 0.03774 & 0.03267 & 0.008\\
        $\xi_6$ & -0.1334 & -0.12123 & -0.00001\\
        $\xi_7$ & -0.00346 & -0.00342 & 0.0072\\
        $\xi_8$ & 0.66803 & 0 & 0.68043\\
        $\xi_9$ & 0.02407 & 0.01998 & 0.01645\\
        $\xi_{10}$ & -0.09084 & -0.23796 & -0.525\\
        $\xi_{11}$ & -0.01077 & -0.00819 & -0.00497\\
        $\xi_{12}$ & -88.04102 & -73.9505 & -58.5\\
        $\xi_{13}$ & 0.3018 & 0.36085 & 0.525\\
        $\xi_{14}$ & -0.02886 & -0.04526 & -0.002\\
        $\xi_{15}$ & 0.00166 & 0.00087 & -0.00164\\
        $\xi_{16}$ & -0.68095 & 0 & -0.1891\\
        $\xi_{17}$ & 0.00073 & 0.00109 & 0.00193\\
        $\xi_{18}$ & -0.13307 & 0 & -0.00642\\
        $\xi_{19}$ & 60.01821 & 0 & -0.8483\\
        $\xi_{20}$ & 70.04068 & 58.44115 & 58.08771\\
        $\xi_{21}$ & -0.00911 & -0.00675 & -0.00438\\
        $\xi_{22}$ & 0.01579 & 0.01576 & 0.01579\\
        $\xi_{23}$ & -0.02891 & -0.02884 & -0.02891\\
        \hline
    \end{tabular}
    \caption{Parameter values for ODEs learnt from the WHIDM simulations ran using the female parameter set, depending on the form of regression used.}
    \label{tab:regression comparison female}
\end{table}
\begin{table}
    \centering
    \begin{tabular}{|l|l|l|l|}
    \hline    
        \textbf{EQL parameter} & \textbf{Least squares} & \textbf{Lasso} & \textbf{Least squares with bounds}\\
        \hline
        $\xi_1$ & 0.59366 & 0.52235 & -0.51328\\
        $\xi_2$ & 0.00458 & 0.00579 & 0.02167\\
        $\xi_3$ & -0.0897 & -0.08991 & -0.07\\
        $\xi_4$ & -2.53529 & -1.43876 & -0.00001\\
        $\xi_5$ & 0.08517 & 0.05652 & 0.008\\
        $\xi_6$ & -0.37969 & -0.23064 & -0.0013\\
        $\xi_7$ & -0.00619 & -0.00375 & 0.01016\\
        $\xi_8$ & 28.54266 & 0 & 0.00001\\
        $\xi_9$ & 0.01966 & 0.01355 & 0.01368\\
        $\xi_{10}$ & -0.03833 & -0.26529 & -0.45\\
        $\xi_{11}$ & -0.00853 & -0.00481 & -0.00377\\
        $\xi_{12}$ & -87.63348 & -56.64864 & -46.5\\
        $\xi_{13}$ & -0.17708 & -0.06233 & 0.45\\
        $\xi_{14}$ & -0.022 & -0.09197 & -0.0015\\
        $\xi_{15}$ & 0.00903 & 0.00687 & -0.00216\\
        $\xi_{16}$ & -7.09691 & 0 & -1.9994\\
        $\xi_{17}$ & 0.00014 & 0.00009 & 0.00131\\
        $\xi_{18}$ & 0.69865 & 0 & -0.07891\\
        $\xi_{19}$ & -856.84049 & 0 & -23.71824\\
        $\xi_{20}$ & 49.22983 & 27.70516 & 46.4828\\
        $\xi_{21}$ & -0.00544 & -0.00074 & -0.00438\\
        $\xi_{22}$ & 0.00598 & 0.00589 & 0.00598\\
        $\xi_{23}$ & -0.03173 & -0.03116 & -0.03173\\
        \hline
    \end{tabular}
    \caption{Parameter values for ODEs learnt from the WHIDM simulations ran using the male parameter set, depending on the form of regression used.}
    \label{tab:regression comparison male}
\end{table}
\begin{table}
    \centering
    \begin{tabular}{|l|l|l|l|}
    \hline    
        \textbf{EQL parameter} & \textbf{Least squares} & \textbf{Lasso} & \textbf{Least squares with bounds}\\
        \hline
        $\xi_1$ & -2.20925 & -2.10029 & -0.69167\\
        $\xi_2$ & 0.04768 & 0.04589 & 0.02167\\
        $\xi_3$ & -0.01322 & -0.0147 & -0.02\\
        $\xi_4$ & 0.93185 & 0.70001 & -0.00001\\
        $\xi_5$ & 0.02496 & 0.02408 & 0.008\\
        $\xi_6$ & -0.43212 & -0.35524 & -0.01079\\
        $\xi_7$ & -0.00329 & -0.00266 & 0.00866\\
        $\xi_8$ & 17.37166 & 11.61823 & 0.00002\\
        $\xi_9$ & 0.02307 & 0.0181 & 0.01487\\
        $\xi_{10}$ & 0.04462 & -0.09442 & -0.4875\\
        $\xi_{11}$ & -0.01154 & -0.00863 & -0.00497\\
        $\xi_{12}$ & -84.32572 & -67.37425 & -47.5\\
        $\xi_{13}$ & -0.13453 & 0 & 0.4875\\
        $\xi_{14}$ & -0.01323 & -0.06749 & -0.002\\
        $\xi_{15}$ & 0.0075 & 0.00568 & -0.00138\\
        $\xi_{16}$ & -4.23744 & -1.40163 & -0.82445\\
        $\xi_{17}$ & 0.00018 & 0.00019 & 0.0015\\
        $\xi_{18}$ & -1.03023 & 0 & -0.03117\\
        $\xi_{19}$ & 697.5901 & 0 & -20.37359\\
        $\xi_{20}$ & 53.44479 & 39.48758 & 47.01748\\
        $\xi_{21}$ & -0.00703 & -0.0038 & -0.00438\\
        $\xi_{22}$ & 0.01228 & 0.01223 & 0.01228\\
        $\xi_{23}$ & -0.03631 & -0.03611 & -0.03631\\
        \hline
    \end{tabular}
    \caption{Parameter values for ODEs learnt from the WHIDM simulations ran using the neutral parameter set, depending on the form of regression used.}
    \label{tab:regression comparison neutral}
\end{table}
\begin{table}
    \centering
    \begin{tabular}{|l|l|l|l|}
    \hline    
        \textbf{Cell type} & \textbf{Least squares} & \textbf{Lasso} & \textbf{Least squares with bounds}\\
        \hline
        FGEB & $1.0155 \times 10^{-4}$ & $1.309 \times 10^{-4}$ & $\bm{2.9138 \times 10^{-6}}$\\
        SGEB & $2.1 \times 10^{-3}$ & $3.2 \times 10^{-3}$ & $\bm{1.1273 \times 10^{-6}}$\\
        MR & $\bm{5.8022 \times 10^{-7}}$ & $1.596 \times 10^{-6}$ & $7.6249 \times 10^{-7}$\\
        MI & $1.5568 \times 10^{-5}$ & $\bm{5.3393 \times 10^{-6}}$ & $5.5258 \times 10^{-6}$\\
        MCI & $2.6751 \times 10^{-10}$ & $4.4254 \times 10^{-9}$ & $\bm{9.2017 \times 10^{-11}}$\\
        MA & $7.7916 \times 10^{-7}$ & $2.933 \times 10^{-6}$ & $\bm{3.0754 \times 10^{-7}}$\\
        T & $2.6566 \times 10^{-8}$ & $2.8988 \times 10^{-8}$ & $\bm{1.2085 \times 10^{-8}}$\\
        \hline
    \end{tabular}
    \caption{Female mean squared errors for the three regression methods considered. The smallest mean squared error for each cell type is printed in bold font. Abbreviations: FGEB, fast-growing extracellular bacterium; SGEB, slow-growing extracellular bacterium; MR, resting macrophage; MI, infected macrophage; MCI, chronically infected macrophage; MA, activated amcrophage; T, T-cells.}
    \label{tab:regression MSE female}
\end{table}
\begin{table}
    \centering
    \begin{tabular}{|l|l|l|l|}
    \hline    
        \textbf{Cell type} & \textbf{Least squares} & \textbf{Lasso} & \textbf{Least squares with bounds}\\
        \hline
        FGEB & $3.2178 \times 10^{-4}$ & $3.2483 \times 10^{-4}$ & $\bm{1.4692 \times 10^{-6}}$\\
        SGEB & $1.68 \times 10^{-2}$ & $2.76 \times 10^{-2}$ & $\bm{1.0653 \times 10^{-5}}$\\
        MR & $3.3517 \times 10^{-6}$ & $2.0576 \times 10^{-5}$ & $\bm{6.8393 \times 10^{-7}}$\\
        MI & $4.4399 \times 10^{-5}$ & $6.1544 \times 10^{-5}$ & $\bm{3.8419 \times 10^{-6}}$\\
        MCI & $7.9227 \times 10^{-12}$ & $3.1399 \times 10^{-11}$ & $\bm{6.2013 \times 10^{-12}}$\\
        MA & $2.4094 \times 10^{-6}$ & $1.0351 \times 10^{-5}$ & $\bm{4.1857 \times 10^{-7}}$\\
        T & $4.1267 \times 10^{-9}$ & $8.2833 \times 10^{-9}$ & $\bm{4.2326 \times 10^{-10}}$\\
        \hline
    \end{tabular}
    \caption{Male mean squared errors for the three regression methods considered. The smallest mean squared error for each cell type is printed in bold font. Abbreviations: FGEB, fast-growing extracellular bacterium; SGEB, slow-growing extracellular bacterium; MR, resting macrophage; MI, infected macrophage; MCI, chronically infected macrophage; MA, activated amcrophage; T, T-cells.}
    \label{tab:regression MSE male}
\end{table}
\begin{table}
    \centering
    \begin{tabular}{|l|l|l|l|}
    \hline    
        \textbf{Cell type} & \textbf{Least squares} & \textbf{Lasso} & \textbf{Least squares with bounds}\\
        \hline
        FGEB & $3.214 \times 10^{-1}$ & 1.1235 & $\bm{2.6801 \times 10^{-6}}$\\
        SGEB & $3.375 \times 10^{-4}$ & $3.1474 \times 10^{-4}$ & $\bm{3.6045 \times 10^{-6}}$\\
        MR & $7.3267 \times 10^{-7}$ & $2.5132 \times 10^{-6}$ & $\bm{7.1188 \times 10^{-7}}$\\
        MI & $4.9971 \times 10^{-5}$ & $4.9971 \times 10^{-5}$ & $\bm{6.7768 \times 10^{-6}}$\\
        MCI & $3.312 \times 10^{-11}$ & $3.3126 \times 10^{-11}$ & $\bm{1.5174 \times 10^{-11}}$\\
        MA & $5.8621 \times 10^{-7}$ & $8.3219 \times 10^{-6}$ & $\bm{2.0675 \times 10^{-7}}$\\
        T & $2.8956 \times 10^{-7}$ & $1.5673 \times 10^{-6}$ & $\bm{7.97 \times 10^{-10}}$\\
        \hline
    \end{tabular}
    \caption{Neutral mean squared errors for the three regression methods considered. The smallest mean squared error for each cell type is printed in bold font. Abbreviations: FGEB, fast-growing extracellular bacterium; SGEB, slow-growing extracellular bacterium; MR, resting macrophage; MI, infected macrophage; MCI, chronically infected macrophage; MA, activated amcrophage; T, T-cells.}
    \label{tab:regression MSE neutral}
\end{table}
\newpage
\section{Multiscale model simulations with equal initial conditions for males and females}
\label{appendix:same IC}
Thirty additional simulations were conducted to determine how the average densities of the Males With Active TB and Females With Active TB compartments would evolve if an equal number of males and females were in each infection state initially. We simulated 10 years, with parameter values the same as those used in the Results section of the paper and the initial conditions in Table \ref{tab:same initial conditions}. A summary of the results of these simulations is shown in the following figures.\par
As observed in the simulations with unequal initial conditions for males and females in Section \ref{subsec:results}, the within-host dynamics appear to be more influential in producing differences between the number of males with active TB and the number of females with active TB. However, the figures also show that the between-host scale does have some impact on the male-to-female TB case ratio.
\begin{table}[H]
    \centering
    \begin{tabular}{|l|l|}
    \hline
    \textbf{Compartment} & \textbf{Value}\\
    \hline
        $S_F$ & 26,250\\
        $S_M$ & 26,250\\
        $L_F$ & 7,875\\
        $L_M$ & 7,875\\
        $I_F$ & 14\\
        $I_M$ & 14\\
        $T_F$ & 152\\
        $T_M$ & 152\\
        $R_F$ & 709\\
        $R_M$ & 709\\
        \hline
    \end{tabular}
    \caption{Initial conditions in additional simulations in which the number of males and females in each infection state were set to be equal initially. Abbreviations: $S_F$, susceptible females; $S_M$, susceptible males; $L_F$, females with latent TB; $L_M$, males with latent TB; $I_F$, females with active TB; $I_M$, males with active TB; $T_F$, females undergoing treatment; $T_M$, males undergoing treatment; $R_F$, females who have recovered but could relapse; $R_M$, males who have recovered but could relapse.}
    \label{tab:same initial conditions}
\end{table}
\begin{figure}[H]
    \centering
    \subfigure[Scenario 0.]{\includegraphics[width=0.45\linewidth]{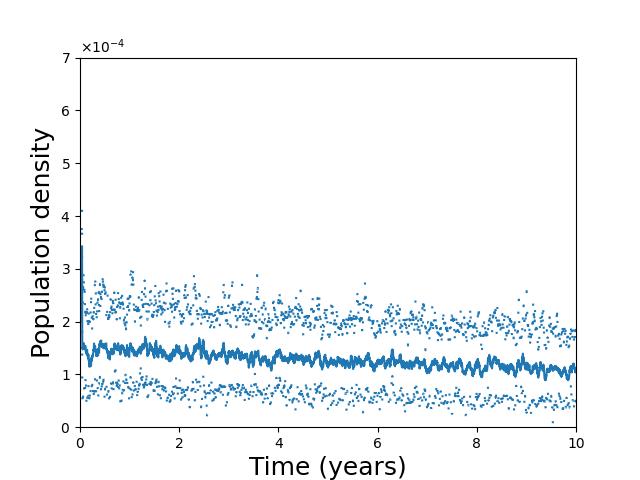}}
    \hfill
    \subfigure[Scenario 1.]{\includegraphics[width=0.45\linewidth]{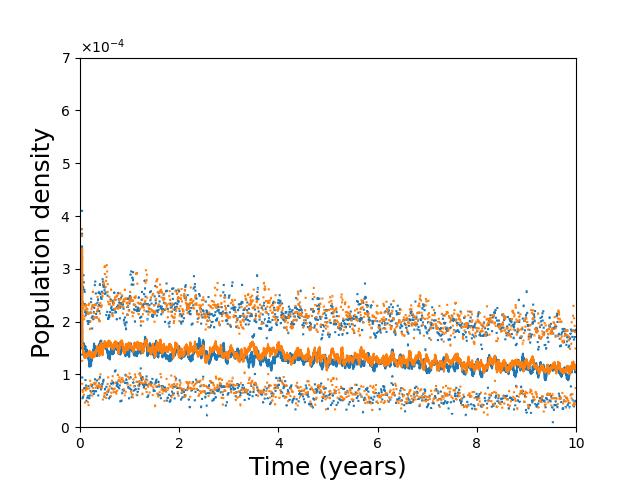}}
    \subfigure[Scenario 2.]{\includegraphics[width=0.45\linewidth]{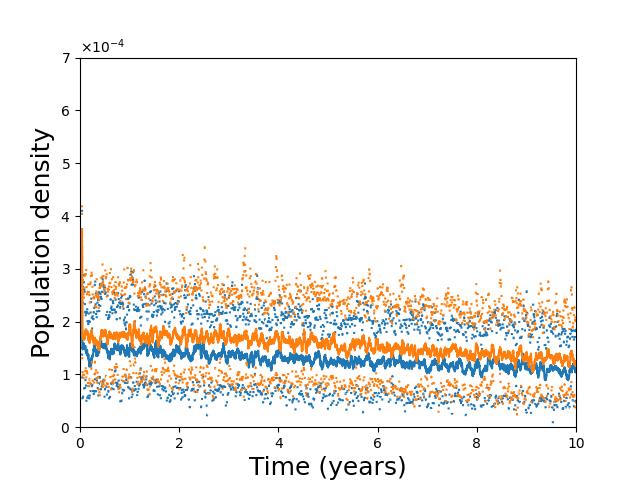}}
    \hfill
    \subfigure[Scenario 3.]{\includegraphics[width=0.45\linewidth]{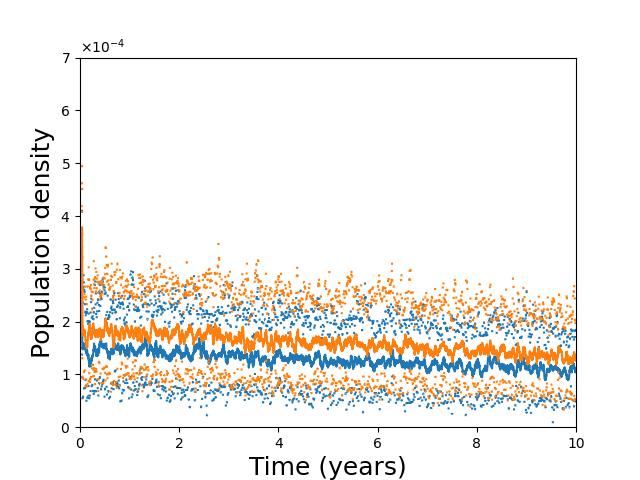}}
    \caption{Plots showing a summary of the average density of the Females With Active TB compartment over a 10-years simulated time period with equal initial densities of males and females in each infection state. The solid line indicates the mean, and the dotted lines indicate the upper and lower bounds of the 95\% confidence interval. The blue lines are the output when sex and gender differences at the within-host and between-host scales are considered (Scenario 0). The orange lines correspond to the output of the counterfactual scenario listed in the sub-figure caption.}
    \label{fig:IF_sameIC}
\end{figure}
\begin{figure}
    \centering
    \subfigure[Scenario 0.]{\includegraphics[width=0.45\linewidth]{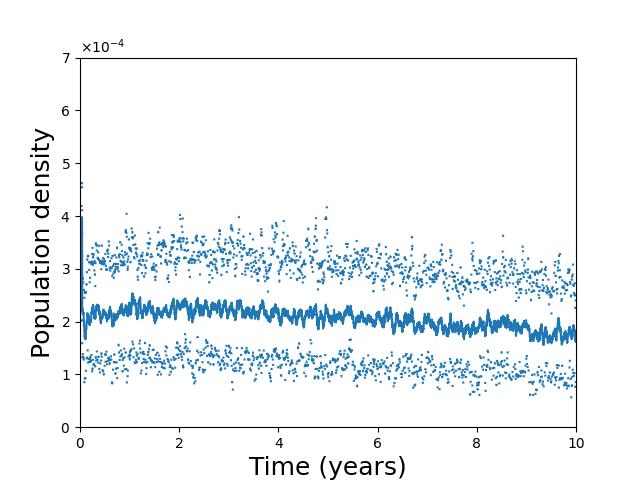}}
    \hfill
    \subfigure[Scenario 1.]{\includegraphics[width=0.45\linewidth]{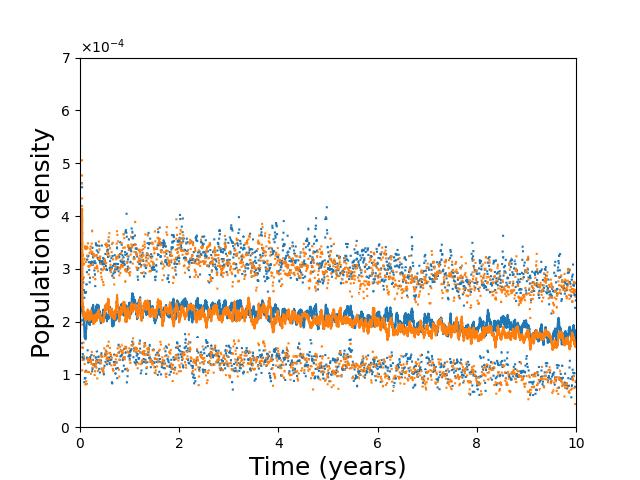}}
    \subfigure[Scenario 2.]{\includegraphics[width=0.45\linewidth]{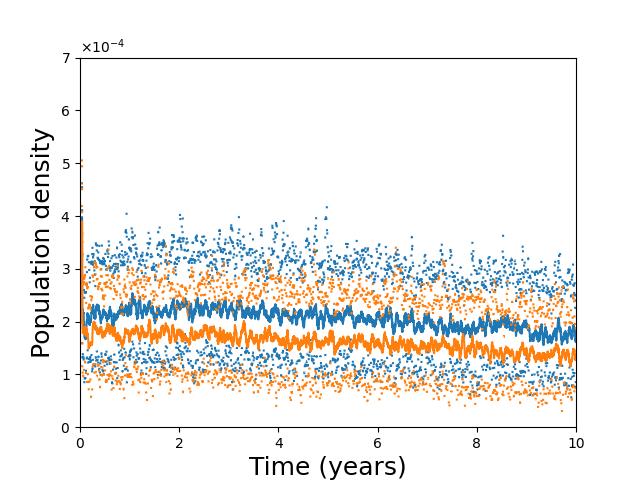}}    
    \hfill
    \subfigure[Scenario 3.]{\includegraphics[width=0.45\linewidth]{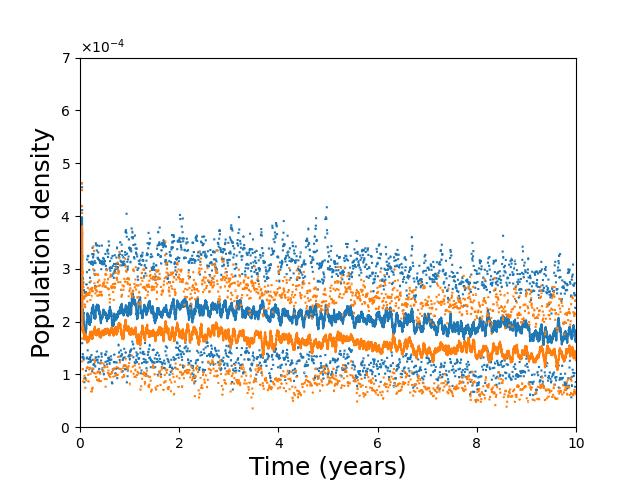}}
    \caption{Plots showing a summary of the average density of the Males With Active TB compartment over a 10-years simulated time period with equal initial densities of males and females in each infection state. The solid line indicates the mean, and the dotted lines indicate the upper and lower bounds of the 95\% confidence interval. The blue lines are the output when sex and gender differences at the within-host and between-host scales are considered (Scenario 0). The orange lines correspond to the output of the counterfactual scenario listed in the sub-figure caption.}
    \label{fig:IM_sameIC}
\end{figure}
\begin{figure}
    \centering
    \subfigure[Scenario 0.]{\includegraphics[width=0.45\linewidth]{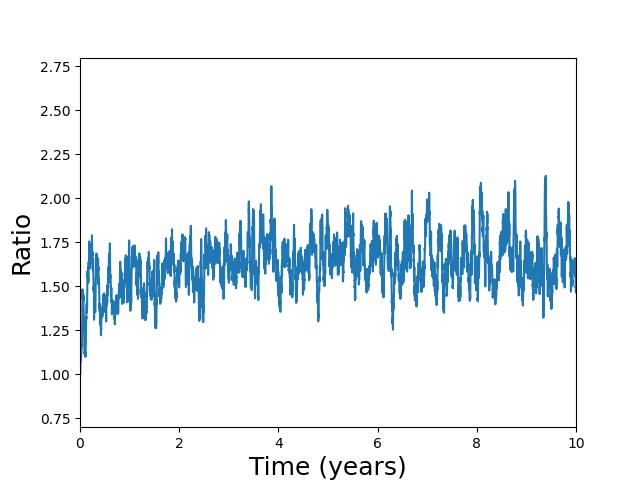}}
    \hfill
    \subfigure[Scenario 1.]{\includegraphics[width=0.45\linewidth]{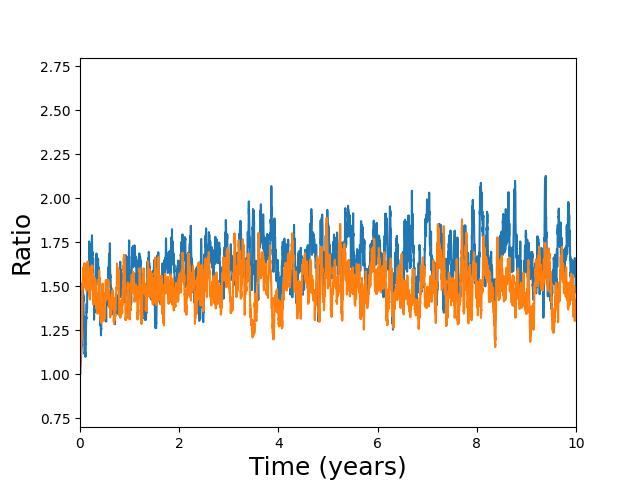}}
    \subfigure[Scenario 2.]{\includegraphics[width=0.45\linewidth]{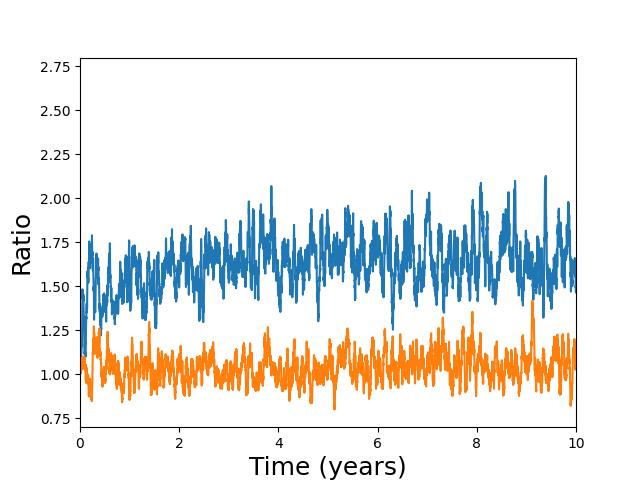}}
    \hfill
    \subfigure[Scenario 3.]{\includegraphics[width=0.45\linewidth]{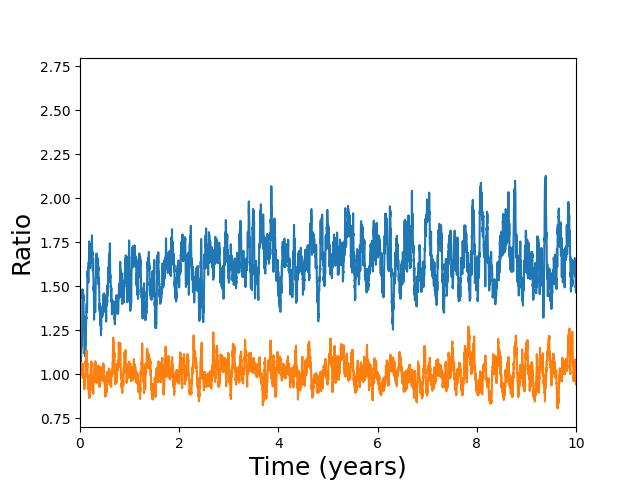}}
    \caption{Plots showing a summary of the average density of the Males With Active TB compartment divided by the average density of the Females With Active TB compartment over a 10-years simulated time period with equal initial densities of females and males in each infection state. The solid line indicates this ratio. The blue lines are the output when sex and gender differences at the within-host and between-host scales are considered (Scenario 0). The orange lines correspond to the output of the counterfactual scenario listed in the sub-figure caption.}
    \label{fig:Ratio_sameIC}
\end{figure}
\section{WHIDM parameters}
The following table lists all the parameter values used in the WHIDM simulations. Most parameter values were taken from previous research using this framework (\cite{Bowness2018,Doran2026a}). The parameter values for macrophage and T-cell recruitment rates were set to match those in the model presented by \citeauthor{Ray2009} \cite{Ray2009}. The parameter value for T-cell recruitment delay was set based on research by \citeauthor{Urdahl2011} \cite{Urdahl2011}. Different values were used for females, males and the neutral case for the probability of phagocytosis, the probability of T-cell recruitment, and the probability of macrophage activation per T-cell, based on research by \citeauthor{Nhamoyebonde2014} \cite{Nhamoyebonde2014}. These were found during calibration so that the proportion of simulations leading to dissemination aligned with real-world data of progression from latent TB to active TB \cite{Kiazyk2017}.
\label{appendix:WHIDM param sets}
\begin{table}
\centering
\resizebox{10.5cm}{!}{
\begin{tabular}{|l|l|}
\hline
\textbf{Parameter} & \textbf{Value}\\
\hline
Output interval (h) & 0.2\\
Simulation duration (h) & 360\\
Timestep (h) & 0.001\\
Grid cells & $100 \times 100$\\
Grid spacing (cm) & 0.002\\
Maximum neighbourhood size & 5\\
Granuloma neighbourhood size & 2\\
Oxygen diffusion coefficient ($\text{cm}^2\text{ h}^{-1}$) & 0.072\\
Oxygen decay ($\text{h}^{-1}$) & 0\\
Oxygen granuloma diffusion reduction & 2.7\\
Chemokine diffusion coefficient ($\text{cm}^2\text{ h}^{-1}$) & 0.0036\\
Chemokine decay ($\text{h}^{-1}$) & 0.347\\
Number of blood vessels & 49\\
Oxygen secretion rate ($\text{mol h}^{-1}$) & 29.52\\
Oxygen granuloma secretion rate reduction & 4\\
Resting macrophage recruitment probability per blood vessel per time step & $9.5 \times 10^{-6}$\\
T-cell recruitment delay (h) & 216\\
T-cell recruitment probability per blood vessel per time step & $1.4 \times 10^{-5}$ (female)\\
 & $9.5 \times 10^{-6}$ (neutral)\\
 & $5 \times 10^{-6}$ (male)\\
Macrophage area ($\text{cm}^2$) & $3.14 \times 10^{-6}$\\
Resting/infected/chronically infected macrophage lifespan (h) & $1200 \pm 1200$\\
Activated macrophage lifespan (h) & 240\\
Macrophage maximum neighbourhood size & 3\\
Initial number of resting macrophages & 105\\
Initial number of infected/chornically infected/activated macrophages & 0\\
Resting macrophage oxygen uptake ($\text{h}^{-1}$) & $1.15 \times 10^{-13}$\\
Infected macrophage oxygen uptake ($\text{h}^{-1}$) & $3.45 \times 10^{-13}$\\
Infected macrophage chemokine secretion rate ($\text{nM h}^{-1}$) & 0.0126\\
Chronically infected macrophage oxygen uptake ($\text{h}^{-1}$) & $4.6 \times 10^{-13}$\\
Chronically infected macrophage chemokine secretion rate ($\text{nM h}^{-1}$) & 0.0126\\
Activated macrophage oxygen uptake ($\text{h}^{-1}$) & $2.3 \times 10^{-13}$\\
Activated macrophage chemokine secretion rate ($\text{nM h}^{-1}$) & 0.0126\\
Time interval for resting macrophage movement (h) & 0.333\\
Time interval for infected macrophage movement (h) & 24\\
Time interval for chronically infected macrophage movement (h) & 24\\
Time interval for activated macrophage movement (h) & 7.8\\
Macrophage chemotactic migration bias & 35\\
Macrophage chemotactic migration weight & 1\\
Probability of macrophage activation per T-cell & 0.11 (female)\\
 & 0.09 (neutral)\\
 & 0.07 (male)\\
 Number of T-cells needed for $\texttt{RM} \rightarrow \texttt{AM}$ & 1\\
Phagocytosis probability & 1 (female)\\
 & 0.8125 (neutral)\\
 & 0.625 (male)\\
Number of bacteria needed for $\texttt{RM} \rightarrow \texttt{IM}$ & 1\\
Probability of $\texttt{RM} \rightarrow \texttt{IM}$ after threshold reached & 1\\
Number of bacteria needed for $\texttt{IM} \rightarrow \texttt{CIM}$ & 10\\
Probability of $\texttt{IM} \rightarrow \texttt{CIM}$ after threshold reached & 1\\
Number of bacteria needed for $\texttt{CIM}$ to burst & 20\\
Probability of $\texttt{CIM}$ bursting after threshold reached & 1\\
Maximum neighbourhood size for distributing bacteria after $\texttt{CIM}$ burst & 3\\
T-cell area ($\text{cm}^2$) & $3.14 \times 10^{-6}$\\
T-cell lifespan (h) & $36 \pm 36$\\
Initial number of T-cells & 0\\
T-cell oxygen uptake ($\text{h}^{-1}$) & $1.4375 \times 10^{-14}$\\
Time interval for T-cell movement (h) & 0.167\\
T-cell chemotactic migration bias & 35\\
T-cell chemotactic migration weight & 1\\
Bacterium area ($\text{cm}^2$) & $3.14 \times 10^{-6}$\\
Bacterium lifespan (h) & 360\\
Initial number of fast-growing extracellular bacteria & 6\\
Initial number of slow-growing extracellular bacteria & 6\\
Fast-growing extracellular bacterium oxygen uptake rate ($\text{h}^{-1}$) & $2.08 \times 10^{-11}$\\
Slow-growing extracellular bacterium oxygen uptake rate ($\text{h}^{-1}$) & $2.08 \times 10^{-11}$\\
Extracellular bacteria state switch delay (h) & 2\\
Oxygen threshold for $\texttt{FGEB} \rightarrow \texttt{SGEB}$ (\%) & 6\\
Oxygen threshold for $\texttt{SGEB} \rightarrow \texttt{FGEB}$ (\%) & 65\\
Fast-growing extracellular bacterium replication rate (h) & 15-32\\
Slow-growing extracellular bacterium replication rate (h) & 48-96\\
Extracellular bacterium replication maximum neighbourhood size & 3\\
\hline
\end{tabular}}
\caption{List of the parameters and their assigned values in WHIDM simulations. Abbreviations: RM, resting macrophage; AM, activated macrophage; IM, infected macrophage; CIM, chronically infected macrophage; FGEB, fast-growing extracellular bacterium; SGEB, slow-growing extracellular bacterium.}
\label{tab:WHIDM params multiscale}
\end{table}
\section{WHIDM cell populations in simulations ending in containment}
\label{appendix:WHIDM containment}
A summary of the cell numbers in the simulations of WHIDM that led to containment (that is, fewer than 10 extracellular bacteria remaining at the end of a simulation) is shown in this section. Unlike simulations of WHIDM that resulted in dissemination (that is, 10 or more extracellular bacteria remaining at the end of a simulation), the numbers of fast-growing extracellular bacteria and slow-growing extracellular bacteria did not grow exponentially over time. Instead, such simulations saw both cell populations reduce to zero in under 50 hours on average. The number of infected macrophages is also reduced in simulations leading to containment compared to simulations leading to dissemination; other cell populations appear to be very similar in magnitude, regardless of the simulation outcome.
\begin{figure}[H]
    \centering
    \subfigure[Female cell numbers]{\includegraphics[width=0.49\linewidth]{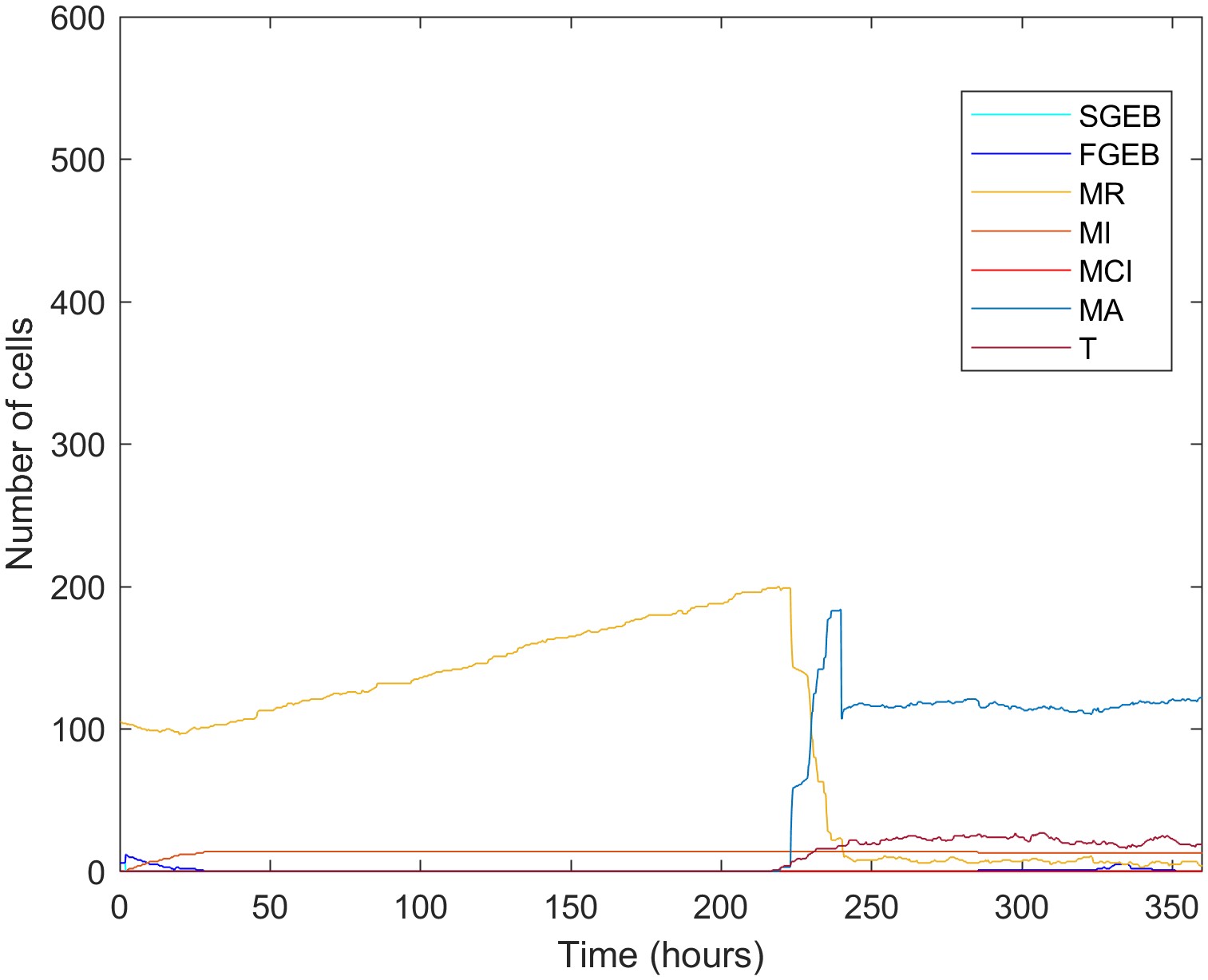}}
    \subfigure[Male cell numbers]{\includegraphics[width=0.49\linewidth]{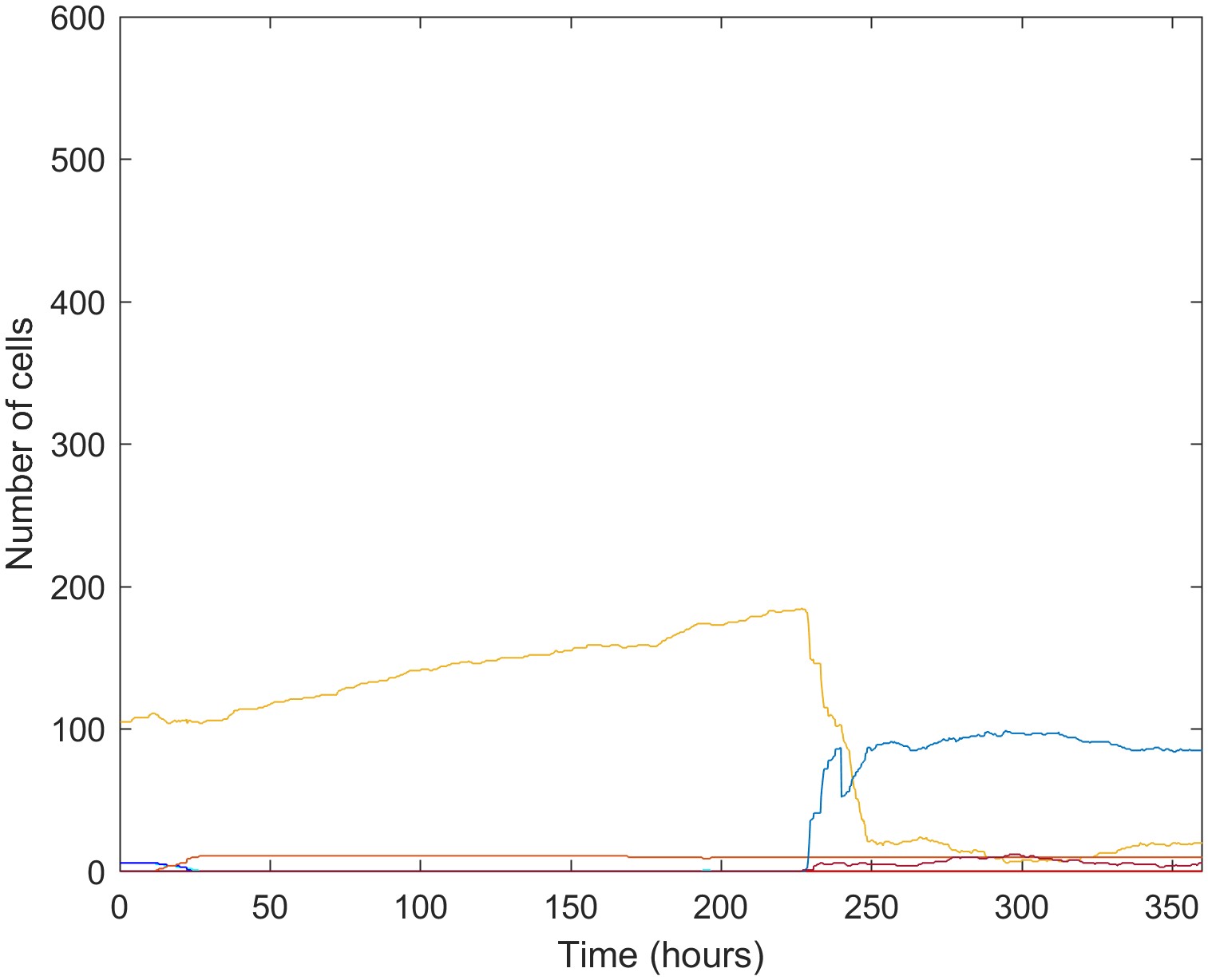}}
    \subfigure[Neutral cell numbers]{\includegraphics[width=0.49\linewidth]{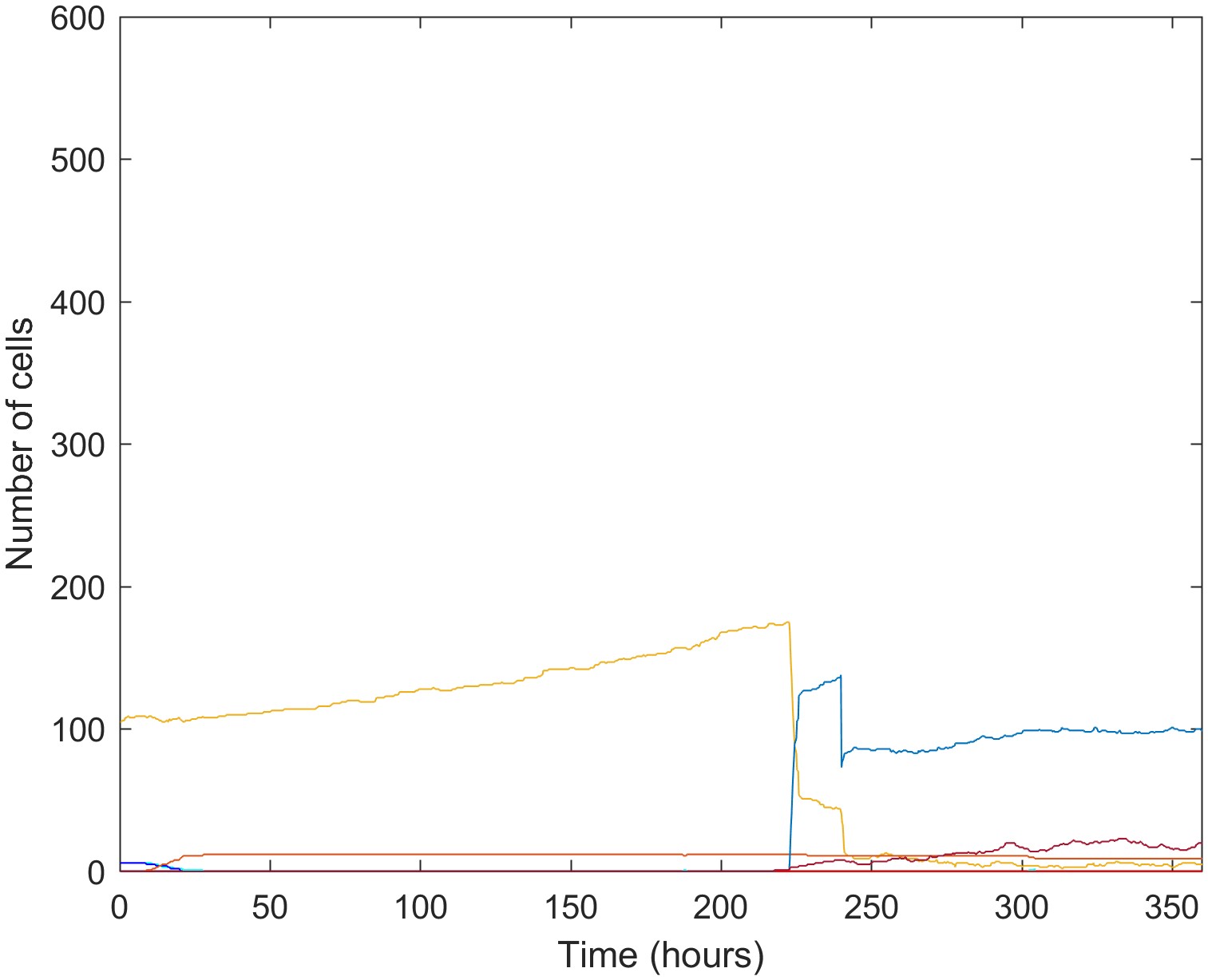}}
    \caption{Summary plots of the evolution of cell population numbers over time in WHIDM simulations that ended in containment. Abbreviations: SGEB, slow-growing extracellular bacteria; FGEB, fast-growing extracellular bacteria; MR, resting macrophages; MI, infected macrophages; MCI, chronically infected macrophages; MA, activated macrophages; T, T-cells.}
    \label{fig:placeholder}
\end{figure}
\newpage
\section{Multiscale model 20-year simulations}
\label{appendix:20 years}
Thirty additional simulations were conducted to determine how the average densities of the Males With Active TB and Females With Active TB compartments would evolve over longer time-frames than the time period considered in our main results. We simulated 20 years, with parameter values and initial conditions the same as those used in the Results section of the paper. A summary of the results of these simulations is shown in the following figures.
\begin{figure}[H]
    \centering
    \subfigure[Scenario 0.]{\includegraphics[width=0.45\linewidth]{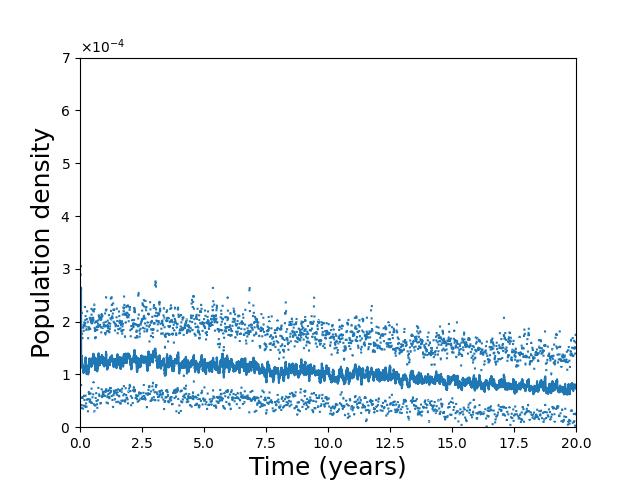}}
    \hfill
    \subfigure[Scenario 1.]{\includegraphics[width=0.45\linewidth]{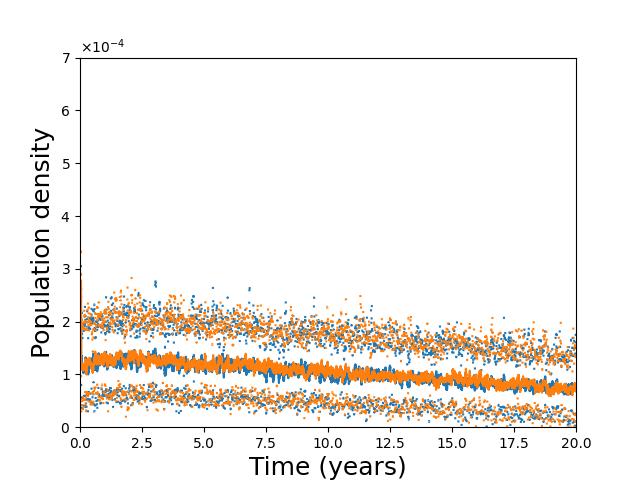}}
    \subfigure[Scenario 2.]{\includegraphics[width=0.45\linewidth]{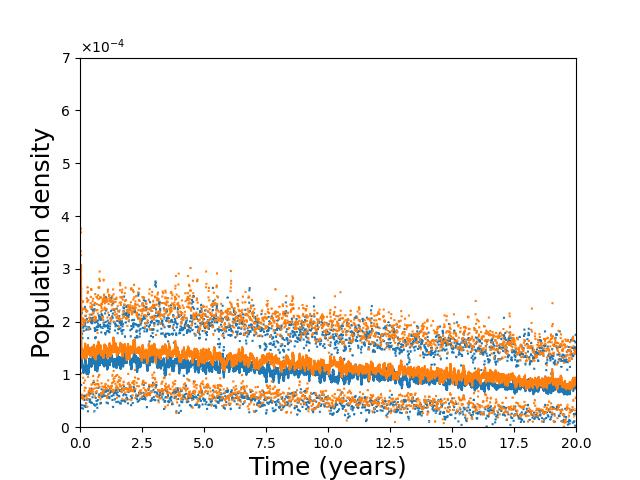}}    
    \hfill
    \subfigure[Scenario 3.]{\includegraphics[width=0.45\linewidth]{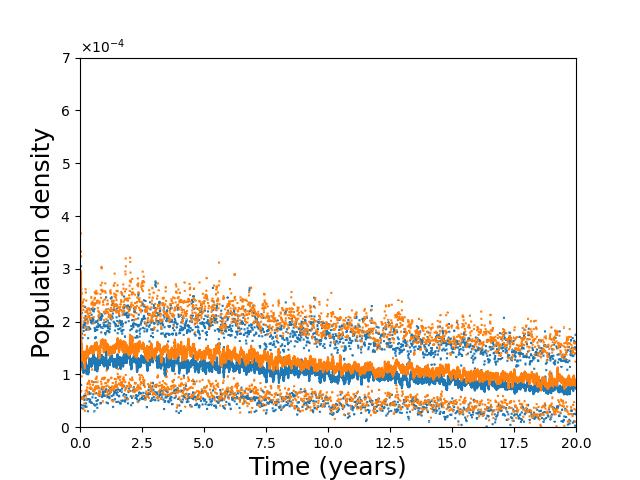}}
    \caption{Plots showing a summary of the average density of the Females With Active TB compartment over a 20-years simulated time period. The solid line indicates the mean, and the dotted lines indicate the upper and lower bounds of the 95\% confidence interval. The blue lines are the output when sex and gender differences at the within-host and between-host scales are considered (Scenario 0). The orange lines correspond to the output of the counterfactual scenario listed in the sub-figure caption.}
    \label{fig:IF_20years}
\end{figure}
\begin{figure}
    \centering
    \subfigure[Scenario 0.]{\includegraphics[width=0.45\linewidth]{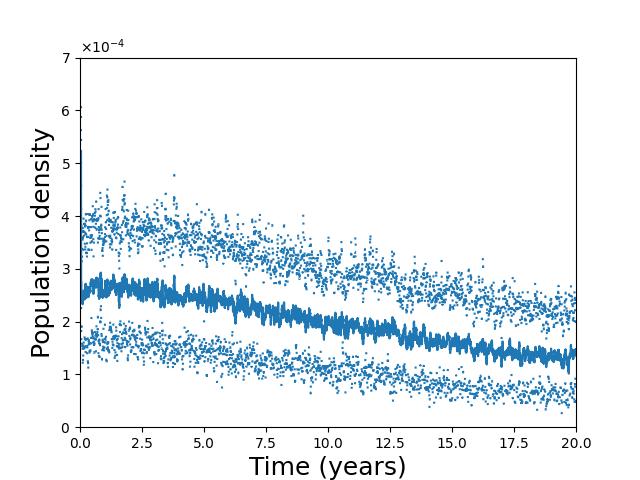}}
    \hfill
    \subfigure[Scenario 1.]{\includegraphics[width=0.45\linewidth]{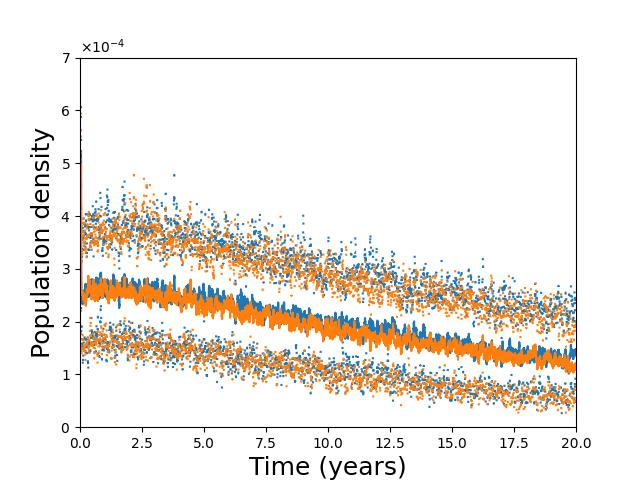}}
    \subfigure[Scenario 2.]{\includegraphics[width=0.45\linewidth]{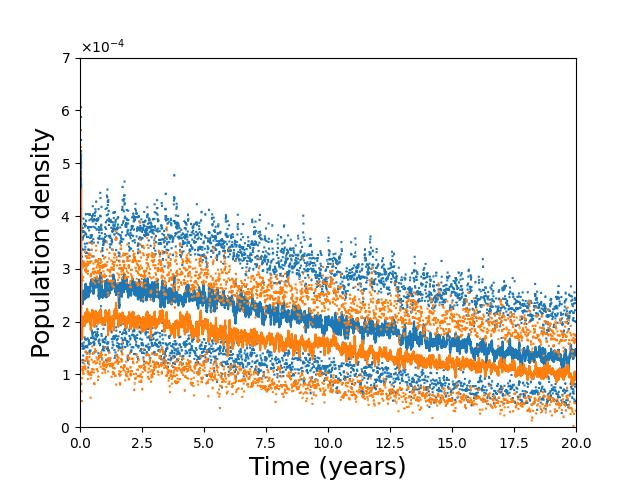}}    
    \hfill
    \subfigure[Scenario 3.]{\includegraphics[width=0.45\linewidth]{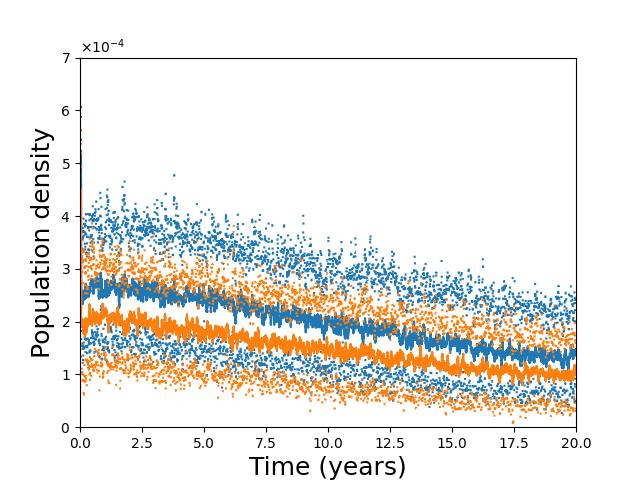}}
    \caption{Plots showing a summary of the average density of the Males With Active TB compartment over a 20-years simulated time period. The solid line indicates the mean, and the dotted lines indicate the upper and lower bounds of the 95\% confidence interval. The blue lines are the output when sex and gender differences at the within-host and between-host scales are considered (Scenario 0). The orange lines correspond to the output of the counterfactual scenario listed in the sub-figure caption.}
    \label{fig:IM_20years}
\end{figure}
\begin{figure}
    \centering
    \subfigure[Scenario 0.]{\includegraphics[width=0.45\linewidth]{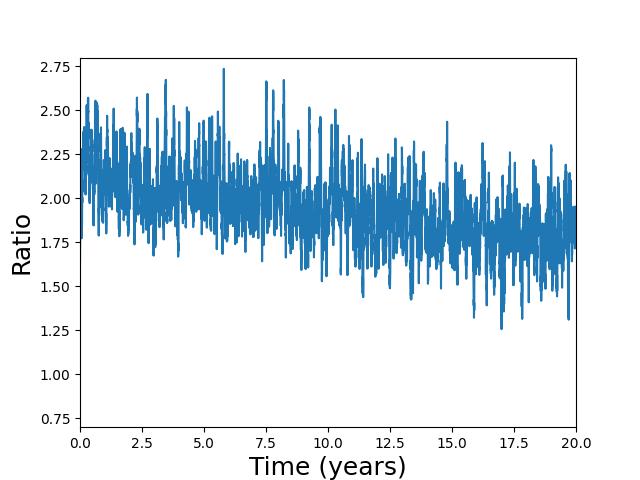}}
    \hfill
    \subfigure[Scenario 1.]{\includegraphics[width=0.45\linewidth]{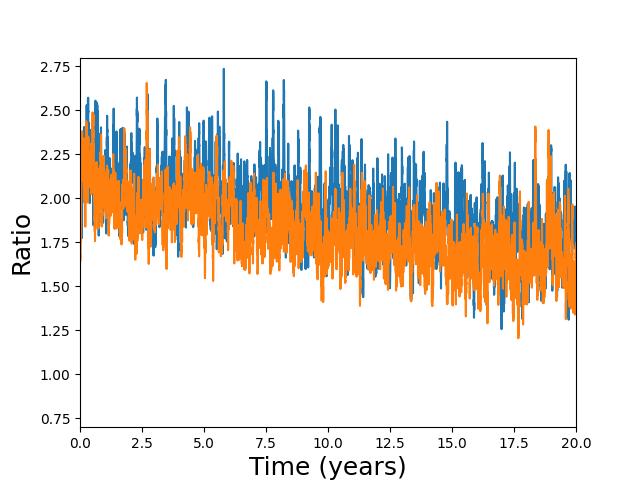}}
    \subfigure[Scenario 2.]{\includegraphics[width=0.45\linewidth]{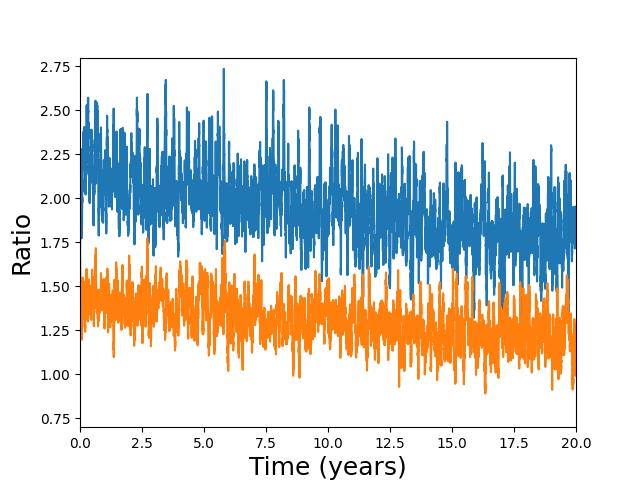}}    
    \hfill
    \subfigure[Scenario 3.]{\includegraphics[width=0.45\linewidth]{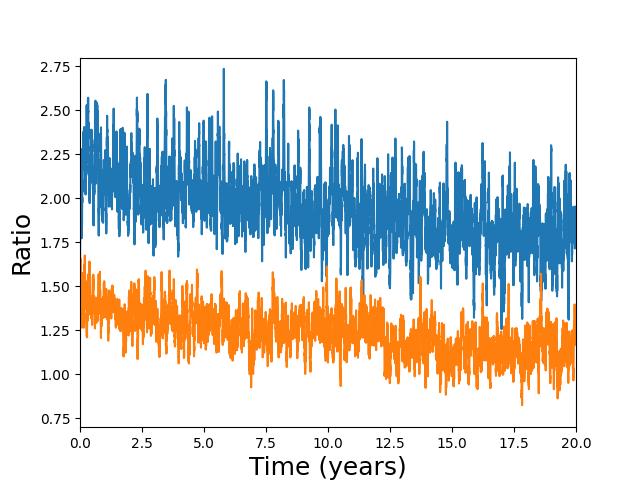}}
    \caption{Plots showing a summary of the average density of the  Males With Active TB compartment divided by the average density of the Females With Active TB compartment over a 20-years simulated time period. The blue lines are the output when sex and gender differences at the within-host and between-host scales are considered (Scenario 0). The orange lines correspond to the output of the counterfactual scenario listed in the sub-figure caption.}
    \label{fig:Ratio_20years}
\end{figure}
\section{Multiscale model 40-year simulations}
\label{appendix:40 years}
Thirty additional simulations were conducted to determine how the average densities of the Males With Active TB and Females With Active TB compartments would evolve over longer time-frames than the time period considered in our main results. We simulated 40 years, with parameter values and initial conditions the same as those used in the Results section of the paper. A summary of the results of these simulations is shown in the following figures.
\begin{figure}[H]
    \centering
    \subfigure[Scenario 0.]{\includegraphics[width=0.45\linewidth]{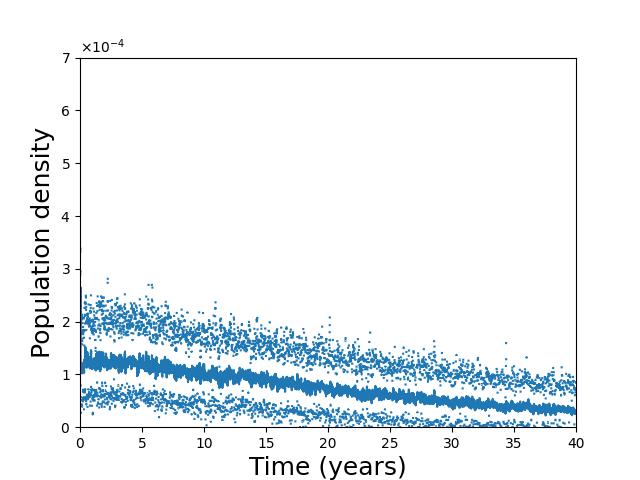}}
    \hfill
    \subfigure[Scenario 1.]{\includegraphics[width=0.45\linewidth]{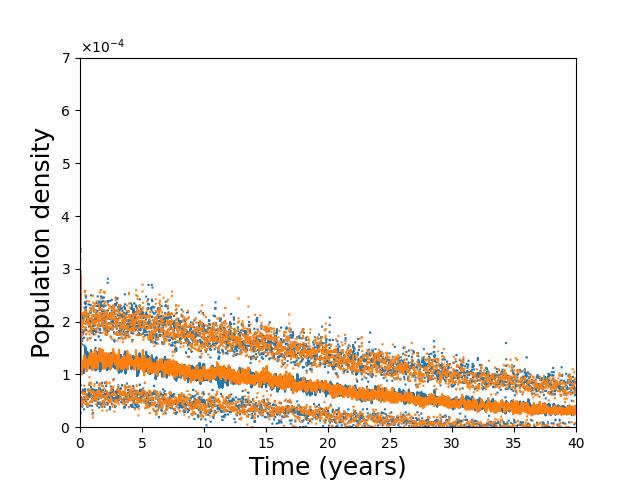}}
    \subfigure[Scenario 2.]{\includegraphics[width=0.45\linewidth]{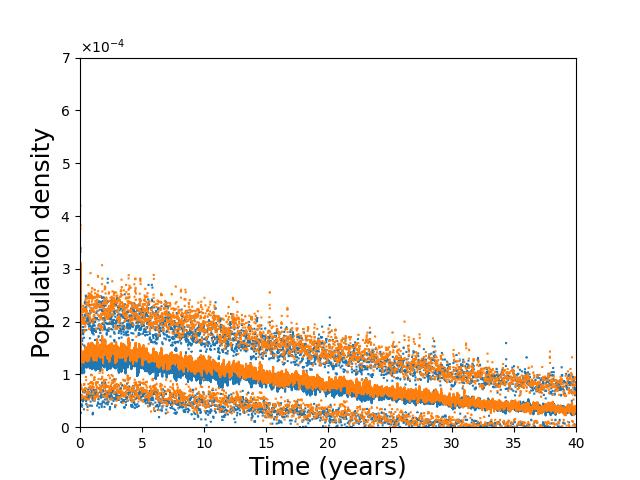}}    
    \hfill
    \subfigure[Scenario 3.]{\includegraphics[width=0.45\linewidth]{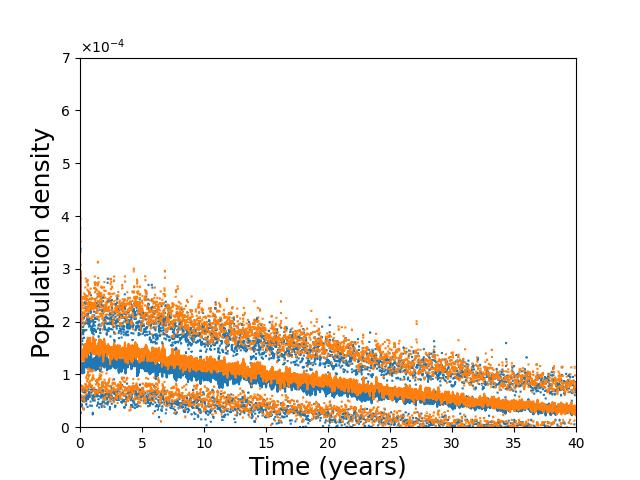}}
    \caption{Plots showing a summary of the average density of the Females With Active TB compartment over a 40-years simulated time period. The solid line indicates the mean, and the dotted lines indicate the upper and lower bounds of the 95\% confidence interval. The blue lines are the output when sex and gender differences at the within-host and between-host scales are considered (Scenario 0). The orange lines correspond to the output of the counterfactual scenario listed in the sub-figure caption.}
    \label{fig:IF_40years}
\end{figure}
\begin{figure}
    \centering
    \subfigure[Scenario 0.]{\includegraphics[width=0.45\linewidth]{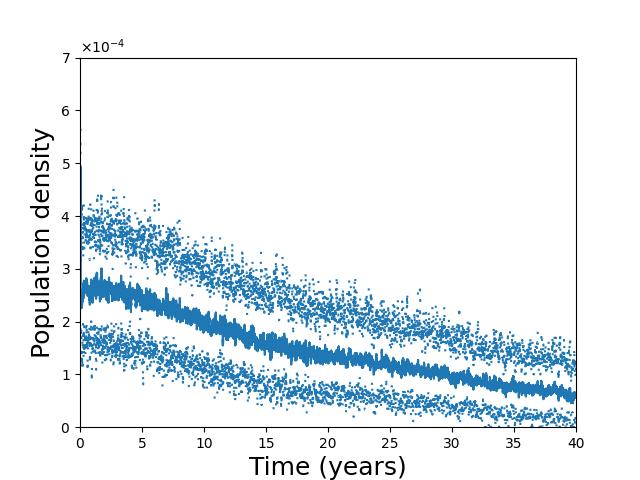}}
    \hfill
    \subfigure[Scenario 1.]{\includegraphics[width=0.45\linewidth]{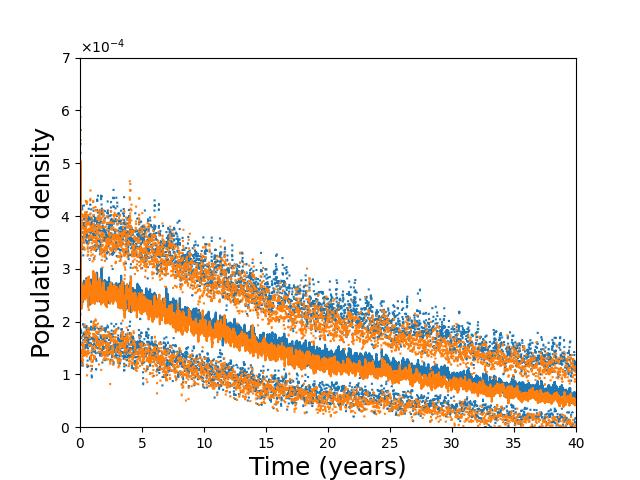}}
    \subfigure[Scenario 2.]{\includegraphics[width=0.45\linewidth]{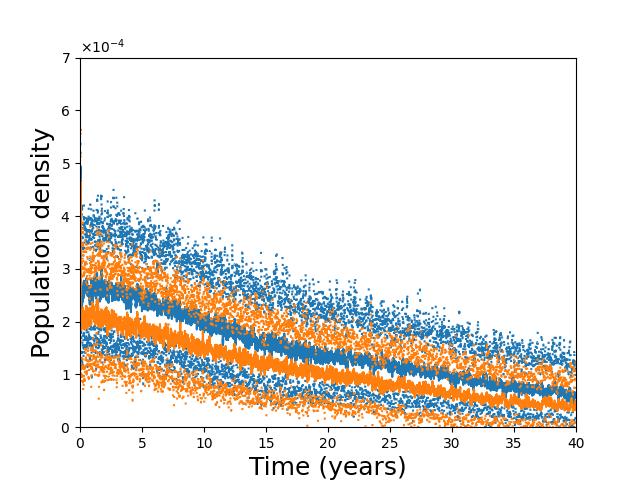}}    
    \hfill
    \subfigure[Scenario 3.]{\includegraphics[width=0.45\linewidth]{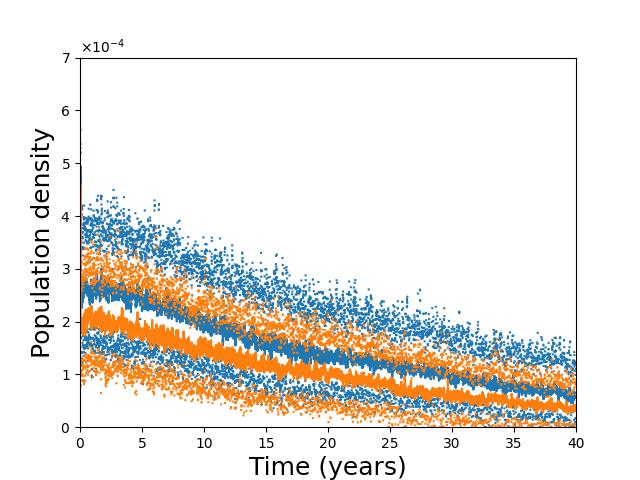}}
    \caption{Plots showing a summary of the average density of the Males With Active TB compartment over a 40-years simulated time period. The solid line indicates the mean, and the dotted lines indicate the upper and lower bounds of the 95\% confidence interval. The blue lines are the output when sex and gender differences at the within-host and between-host scales are considered (Scenario 0). The orange lines correspond to the output of the counterfactual scenario listed in the sub-figure caption.}
    \label{fig:IM_40years}
\end{figure}
\begin{figure}
    \centering
    \subfigure[Scenario 0.]{\includegraphics[width=0.45\linewidth]{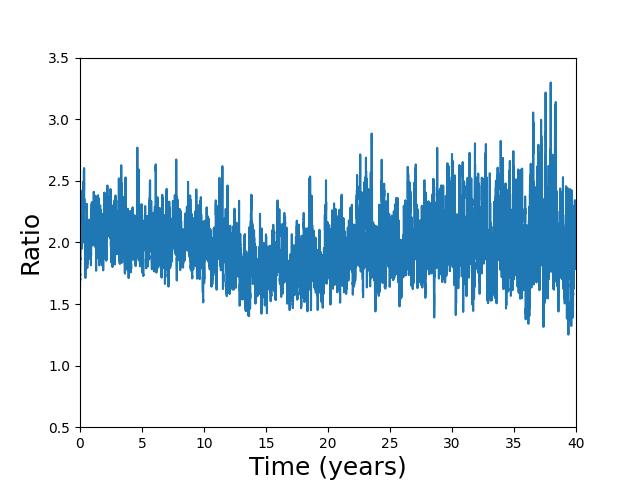}}
    \hfill
    \subfigure[Scenario 1.]{\includegraphics[width=0.45\linewidth]{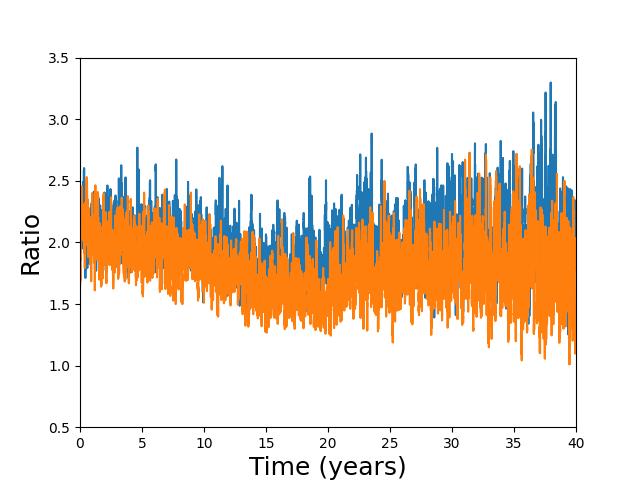}}
    \subfigure[Scenario 2.]{\includegraphics[width=0.45\linewidth]{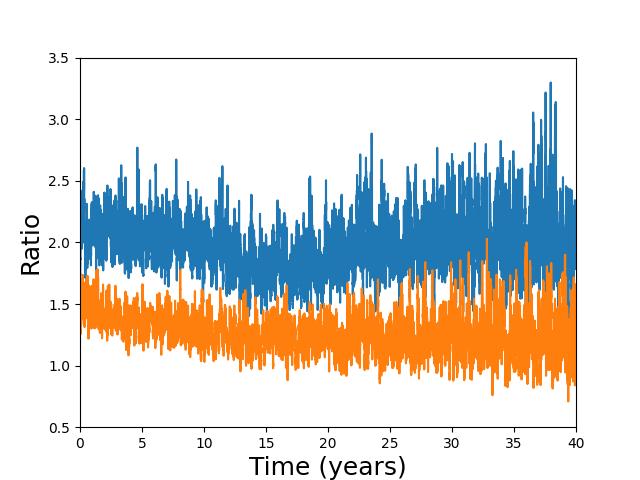}}    
    \hfill
    \subfigure[Scenario 3.]{\includegraphics[width=0.45\linewidth]{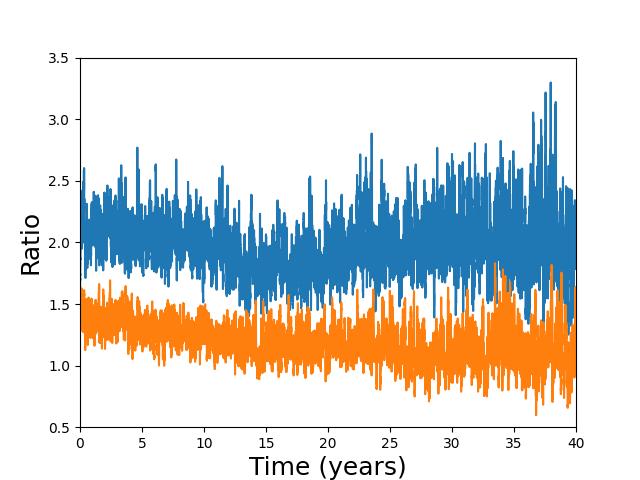}}
    \caption{Plots showing a summary of the average density of the  Males With Active TB compartment divided by the average density of the Females With Active TB compartment over a 40-years simulated time period. The blue lines are the output when sex and gender differences at the within-host and between-host scales are considered (Scenario 0). The orange lines correspond to the output of the counterfactual scenario listed in the sub-figure caption.}
    \label{fig:Ratio_40years}
\end{figure}
\end{appendices}
\end{document}